\documentclass[usenatbib]{mnras}
\usepackage{newtxtext,newtxmath}
\usepackage{graphicx}	% Including figure files
\usepackage{amsmath}	% Advanced maths commands
\usepackage{amssymb}	% Extra maths symbols
\usepackage{tabularx}
\usepackage{subfigure}
\usepackage[flushleft]{threeparttable}
\usepackage{hyperref}
\usepackage{gensymb}
\usepackage{epstopdf}
\usepackage{longtable}
\usepackage{afterpage}

\newcommand{\thetaop}{\theta_{\rm op}}
\newcommand{\betah}{\beta_{\rm h}}
\newcommand{\betac}{\beta_{\rm c}}
\newcommand{\Vc}{V_{\rm c}}
\newcommand{\Pc}{P_{\rm c}}
\newcommand{\rc}{r_{\rm c}}
\newcommand{\zc}{z_{\rm c}}
\newcommand{\thetab}{\theta_{\rm b}}
\newcommand{\thetabo}{\theta_{\rm b0}}
\newcommand{\thetabomin}{\theta_{\rm b0,min}}
\newcommand{\thetabinf}{\theta_{\rm b \infty}}
\newcommand{\Rb}{R_{\rm b}}
\newcommand{\Rbo}{R_{\rm b0}}
\newcommand{\Rbomin}{R_{\rm b0,min}}
\newcommand{\chibo}{\chi_{\rm b0}}
\newcommand{\chibomin}{R_{\rm b0,min}}
\newcommand{\omegabo}{\omega_{\rm b0}}
\newcommand{\Lj}{L_{\rm j}}
\newcommand{\tj}{t_{\rm j}}
\newcommand{\epsiloneff}{\epsilon_{\rm eff}}
\newcommand{\deriv}{{\rm d}}
\newcommand{\req}{r_{\rm eq}}
\newcommand{\rb}{r_{\rm b}}
\newcommand{\zb}{z_{\rm b}}
\newcommand{\thetaeqo}{\theta_{\rm eq0}}
\newcommand{\Reqo}{R_{\rm eq0}}
\newcommand{\chieqo}{\chi_{\rm eq0}}
\newcommand{\omegaeqo}{\omega_{\rm eq0}}
\newcommand{\tsph}{t_{\rm sph}}
\newcommand{\tbo}{t_{\rm bo}}
\newcommand{\tw}{t_{\rm w}}
\newcommand{\teq}{t_{\rm eq}}
\newcommand{\etabo}{\eta_{\rm bo}}
\newcommand{\etaw}{\eta_{\rm w}}
\newcommand{\etaeq}{\eta_{\rm eq}}
\newcommand{\vw}{v_{\rm w}}
\newcommand{\Deltab}{\Delta_{\rm b}}
\newcommand{\Deltaeq}{\Delta_{\rm eq}}

\title[The propagation of choked jets]{The propagation of choked jet outflows in power-law external media}

\author[C. M. Irwin et al.]{
Christopher M. Irwin,$^{1}$\thanks{E-mail: christopheri@mail.tau.ac.il} 
Ehud Nakar,$^{1}$
and Tsvi Piran$^{2}$
\\
$^{1}$The Raymond and Beverly Sackler School of Physics and Astronomy, Tel Aviv University, Tel Aviv 69978, Israel \\
$^{2}$The Racah Institute of Physics, The Hebrew University of Jerusalem, Jerusalem 91904, Israel
}

\date{Accepted XXX. Received YYY; in original form ZZZ}

\pubyear{2019}

\begin{document}
\label{firstpage}
\pagerange{\pageref{firstpage}--\pageref{lastpage}}
\maketitle

\begin{abstract}
Observations of both gamma-ray bursts (GRBs) and active galactic nuclei (AGNs) point to the idea that some relativistic jets are suffocated by their environment before we observe them.  In these `choked' jets, all the jet's kinetic energy is transferred into a hot and narrow cocoon of near-uniform pressure.  We consider the evolution of an elongated, axisymmetric cocoon formed by a choked jet as it expands into a cold power-law ambient medium $\rho \propto R^{-\alpha}$, in the case where the shock is decelerating ($\alpha<3$).  The evolution proceeds in three stages, with two breaks in behaviour: the first occurs once the outflow has doubled its initial width, and the second once it has doubled its initial height.  Using the Kompaneets approximation, we derive analytical formulae for the shape of the cocoon shock, and obtain approximate expressions for the height and width of the outflow versus time in each of the three dynamical regimes.  The asymptotic behaviour is different for flat ($\alpha \le 2$) and steep ($2 < \alpha < 3$) density profiles. Comparing the analytical model to numerical simulations, we find agreement to within $\sim 15$ per cent out to 45 degrees from the axis, but discrepancies of a factor of 2--3 near the equator.  The shape of the cocoon shock can be measured directly in AGNs, and is also expected to affect the early light from failed GRB jets.  Observational constraints on the shock geometry provide a useful diagnostic of the jet properties, even long after jet activity ceases.
\end{abstract}


\section{Introduction}

Relativistic jets are ubiquitous in high-energy astrophysics and essential to our understanding of phenomena such as short and long gamma-ray bursts (GRBs) and active galactic nuclei (AGNs).  Although these systems have markedly different  origins---AGNs are powered by accretion on to a supermassive black hole \citep[e.g.,][and references therein]{rees84,blandford19}; long GRBs (LGRBs) are connected to the death of high-mass stars \citep[see, e.g., ][for a review]{wb06}; and, thanks to GW 170817, short GRBs (SGRBs) have at long last been unambiguously linked to the merger of two neutron stars \citep[e.g.,][]{abbott17}---the underlying physics is none the less similar, as ultimately they all involve a compact object driving a jet into an external medium.   The relevant medium is the intracluster medium in the case of AGNs, the host star and surrounding circumstellar gas in the case of LGRBs, or the ejecta thrown out by the merger in the case of SGRBs.

While the jet-launching process remains murky, and the emission mechanism of relativistic jets is a long-standing problem, the dynamics of jet propagation are comparatively well-understood, both analytically \citep[e.g.,][]{bc89,mw01,matzner03,lb05,b11} and numerically \citep[e.g.,][]{marti97,aloy00,mwh01,zwh04,mizuta06,mlb07,waz08,mizuta09,lmb09,nagakura11,lopez13,mizuta13,ito15,harrison18}.  Whether describing an AGN or a GRB, the dynamics are captured by the same physical quantities: the kinetic power $\Lj$ and Lorentz factor $\Gamma_j \gg 1$ of the jet, the opening angle $\thetaop$ into which the jet is injected, and the external density $\rho$.  In AGN environments, the ambient pressure $P_a$ and magnetic field may also be important.  As the jet drills through its surroundings, a forward and reverse shock structure known as the jet head develops at the interface between the jet ejecta and the ambient gas.  Due to a strong pressure gradient in the jet head, matter flowing in through the forward and reverse shocks is squeezed out to the sides, forming a hot `cocoon' full of shocked ejecta.  This cocoon, which is an unavoidable consequence of jet propagation, surrounds and exerts pressure on the jet.  Because this pressure can be significant enough to collimate the jet, the dynamics of the jet-cocoon system must be solved self-consistently, as described by \citet{b11}.

The cocoon contains both an inner part comprised of relativistic jet ejecta that entered through the reverse shock, and an outer part containing non-relativistic ambient matter swept up by the forward shock.  The two components have different densities and temperatures---the relativistic gas is light and hot, whereas the non-relativistic gas is heavy and cold---and are likely mixed together to some degree \citep{np17,gottlieb18}.  Importantly, however, the different parts of the cocoon are roughly in pressure equilibrium, provided that the cocoon expands non-relativistically, because the near-relativistic sound speed in the lighter regions acts to smooth out pressure differences on time-scales shorter than a dynamical time \citep{b11}.  The assumption of near-uniform pressure, which is crucial to our model, is supported by numerical jet simulations \citep[e.g.,][]{mizuta09,mizuta13,harrison18}.

The above description is valid while the jet remains active and matter continues to flow into the reverse shock.  However, there is reason to believe that some GRB jets fail to penetrate their immediate surroundings.  In these `choked' jets, the reverse shock crosses the flow before the jet breaks out, and all of the jet energy is dumped into the cocoon.  Several lines of evidence point towards the existence of choked jets.  First, the duration distribution of LGRBs has a plateau, with few objects having an observed duration much shorter than the typical breakout time, indicating a significant population of failed jets \citep{bromberg12}.  A similar plateau has been observed in the duration distribution of SGRBs, as well \citep{moharana17}.  Second, in low-luminosity GRBs ($ll$GRBs), a peculiar faint and long-lived class of long GRBs, early optical emission suggests the presence of an extended, optically thick envelope \citep{nakar15,ic16}.  The inferred mass ($\sim 10^{-2} M_\odot$) and radius ($\ga 100 R_\odot$) of the envelope are more than sufficient to choke a GRB of typical luminosity and duration \citep{nakar15,ic16}.  Mildly relativistic shock breakout from such an extended envelope could explain the unusual prompt emission of $ll$GRBs \citep{kulkarni98,campana06,ns12,nakar15}.  Interestingly, although they are more difficult to observe, $ll$GRBs may be more common per cosmic volume than standard GRBs \citep{soderberg06}, again hinting that choked jets may be common.  Third, early spectroscopy of Type Ib and Ic supernovae (SNe) has unveiled a distinct high-velocity component in several cases \citep[e.g.,][]{piran19,izzo19}.  The inferred energy ($\sim 10^{51}$\,erg\,s$^{-1}$) and velocity ($\sim 0.1c$) of this component are consistent with the expectations for a GRB jet's cocoon.  Finally, in AGNs, the evidence for choked jets is even more direct: some galaxies contain `relic bubbles' that were likely formed by past jet activity \citep[e.g.][]{churazov00,mcnamara00,fabian02,mcnamara05,tc17}. In recently quenched systems where the bubbles have not yet been deformed by buoyancy, the shape of the bubbles can be used to constrain the jet properties (Irwin et al., in preparation).

Compared to the case where the jet is active, the evolution of the outflow after the jet is choked is less clear.  The expectation is that the outflow will eventually become self-similar, but our understanding of the transition to this scale-free regime is lacking.  While several authors have explored the process of jet choking through numerical simulations \citep[e.g.][]{lazzati12,gottlieb18}, there is not yet a firm analytic theory underpinning the results.  We aim to rectify this situation by developing an analytical model for the propagation of a choked jet outflow in a power-law external density profile, thereby extending the solutions of \citet{b11} to beyond the moment of choking.  To do so, we employ the well-known Kompaneets approximation \citep[hereafter, KA;][]{kompaneets60}, which takes advantage of the fact that the cocoon pressure is nearly uniform (as discussed above).  

Before further discussing aspherical outflows, we briefly review important results in the spherical case.  In the scale-free limit, a spherical explosion expands according to the well-known Sedov--Taylor (ST) blast wave solution, with the radius growing in time as $R \propto t^{2/(5-\alpha)}$ if the density obeys $\rho \propto R^{-\alpha}$ \citep{taylor50,sedov59}.  The ST solution applies when the outflow is decelerating, which is true for $\alpha < 3$.  In media of finite mass ($\alpha > 3$), the shock accelerates with time and eventually becomes causally disconnected from the swept up mass \citep{koo90}.  The case $3<\alpha<5$ can be described by a second type of self-similar solution \citep{koo90,waxman93}, although the ST formula remains a valid approximation until causal connection is lost.  For $\alpha>5$, however, the ST solution breaks down completely, since the radius goes to infinity in a finite time, and a different class of self-similar solution applies \citep{waxman93}.   

While aspherical explosions have been considered in the past, the literature has mainly focused on point explosions.  Early work on this topic was carried out by \citet{kompaneets60}, who developed the eponymous approximation and obtained a solution for the case of an exponentially stratified density.  Later authors applied the KA to a wide variety of other problems, as reviewed by \citet{bs95} \citep[see also][and references therein]{lyutikov11,bannikova12,rimoldi15}.  The work of \citet{k92} (hereafter K92), who investigated off-centre point explosions in a power-law ambient medium, is of particular relevance to our study.  Also closely related is the case of an initially spherical explosion with a velocity depending on polar angle, which was treated by \citet{bb82} for a homogeneous medium.

We consider a different generalization of the Kompaneets problem to power-law ambient media, in which the explosion is centred at the origin but elongated along the symmetry axis.  This configuration arises naturally when the energy is injected by a relativistic jet.  We restrict our discussion to the $\alpha<3$ case where the shock is decelerating and the flow evolves like an ST blast wave at late times; the case of an accelerating shock will be treated in a separate paper.  Applying the KA, we derive general analytical expressions for the evolution of the shock's shape, and examine how the outflow transitions from an initially elongated shape to a quasi-spherical one.  As we will show, although the evolution eventually comes to resemble a point explosion, deviations from spherical symmetry persist long after the jet has been quenched, especially when the initial shape is narrow.  Meaningful information about the jet can therefore be extracted by measuring the shape of the shock, even when jet activity has already ended (as in AGN with relic bubbles) or was hidden from view (as in failed GRB jets).
 
The KA is powerful, but our idealized model is not without its limitations.  First, as mentioned above, the model depends on the cocoon pressure being uniform, and will not give reliable results if this is not the case.  Second, we only consider the case where the cocoon propagates non-relativistically, and the speed of the ambient gas is negligible.  This assumption is accurate for LGRBs and AGNs, and is somewhat reasonable for $ll$GRBs, which are at most mildly relativistic, but is not suitable for SGRBs, where the ejecta may have an expansion speed comparable to the cocoon's speed.  Third, we ignore gravity.  This is fine for GRBs where the gravitational time-scale is much longer than a dynamical time; however, in AGNs, our model breaks down once buoyancy becomes important.  In spite of these drawbacks, the idealized model presented here is an important first step towards understanding the ultimate fate of choked jet explosions.

The paper has a somewhat unconventional structure in that we begin with a summary of key results in Section~\ref{overview}, where we overview the important time-scales and the role of the density profile in shaping the evolution of the choked jet outflow.   Then, in Section~\ref{analytical}, we use the KA to derive analytical solutions for the shape of the shock.  We start by considering the case of uniform density in Section~\ref{alpha0}, then generalize to a power-law external density in Section~\ref{alphax}.  We find different behaviour in each of the cases $\alpha<2$, $\alpha=2$, and $2< \alpha < 3$, which are discussed respectively in subsections~\ref{alpha0-2}--\ref{alpha2-3}.  We calculate the volume of the expanding cocoon in Section~\ref{volume}, and provide approximate formulae for the cocoon's height and width as functions of time in Section~\ref{convert}.  We then wrap up this section with a discussion of the long-term breakdown of the solution in Section~\ref{breakdown}.   Next, in Section~\ref{jet}, we consider how the initial conditions of the cocoon problem depend on the properties of the injected jet, and conversely how observationally-inferred cocoon properties can constrain the underlying jet.  After that, we compare our analytical results to numerical simulations of choked jets in Section~\ref{numerical}.  Finally, we present our conclusions and discuss some possible applications in Section~\ref{conclusions}.
\\

\section{Overview}
\label{overview}

Consider a relativistic jet propagating in a power-law external medium, $\rho \propto R^{-\alpha}$, with a head velocity $c \betah$.  As the jet drills into the ambient gas, shocked material is pushed to the side to form a hot cocoon around the jet.  Assuming that the ambient pressure is negligible, this cocoon expands sideways supersonically, driving a shock into the ambient medium at a speed $c \betac$.  Since the sideways expansion is driven by the cocoon pressure alone, while the jet head is pushed forward by both internal pressure and ram pressure from the jet, we necessarily have $\betah > \betac$.

After the central engine shuts off, jet material continues to flow into the cocoon until the last-emitted material catches up with the head.  Once all of the jet ejecta have entered the cocoon, at the time $t_0$, we consider the jet to be `choked.'  The time $t_0$ is the endpoint of the \citet{b11} jet propagation model, and is the starting point for our choked jet model.  The evolution of the system after the jet is choked can be described by three parameters: the height $a$ and width $b$ of the outflow upon choking, and the total energy $E_0$ injected into the cocoon.  (These quantities are directly related to the jet properties in the \citet{b11} model, as discussed in Section~\ref{jet}).  From the condition $\betah > \betac$, it follows that $a>b$.  

Elongated, axisymmetric explosions are best understood through analogy with the familiar spherical case.  In the case of a spherical explosion of energy $E_0$, with initial size $R_0$, detonated at time $t_0$, there is one length scale ($R_0$), and one characteristic time-scale, ${t_{\rm ST} \sim (\rho R_0^5/E_0)^{1/2}}$, which is roughly the time for the outflow to double its initial radius.  This divides the evolution into two dynamical regimes: the planar phase ($t-t_0 \ll t_{\rm ST}$), and the self-similar phase ($t-t_0 \gg t_{\rm ST}$).  During the planar phase, outflow properties such as the volume, pressure, and expansion speed are nearly constant in time, with values that depend on $R_0$.  Conversely, in the self-similar phase, these quantities vary in time, but the dependence on the initial conditions is lost.

The axisymmetric case, on the other hand, has two inherent length scales: the initial width ($b$), and the initial height ($a$).  Consequently, there are two characteristic time-scales, and three dynamical regimes.  The relevant time-scales are the time for the initial width to double ($t_b$), and the time for the initial height to double ($t_a$).  The early phase of evolution ($t-t_0 \ll t_b$) is analogous to the planar phase of a spherical explosion.  Also, like in the spherical case, the expansion becomes self-similar when $t - t_0 \gg t_a$.  However, in the axisymmetric solution there is a transitional regime ($t_b \ll t - t_0 \ll t_a$) which does not appear in the spherical case.  During this period, the outflow's width changes significantly, while its height remains about the same.   The more elongated the explosion, the more pronounced is the transitional regime.

To put it another way, spherical evolution corresponds to a special case of the axisymmetric solution, with $a=b$ and $t_b=t_a$, where pressure changes become important at around the same time that initial conditions become unimportant.  In the more general case of an elongated axisymmetric explosion, however, changes in pressure start to affect the outflow while the initial size is still relevant.

\begin{figure*} 
\centering
\includegraphics[width=\linewidth,keepaspectratio]{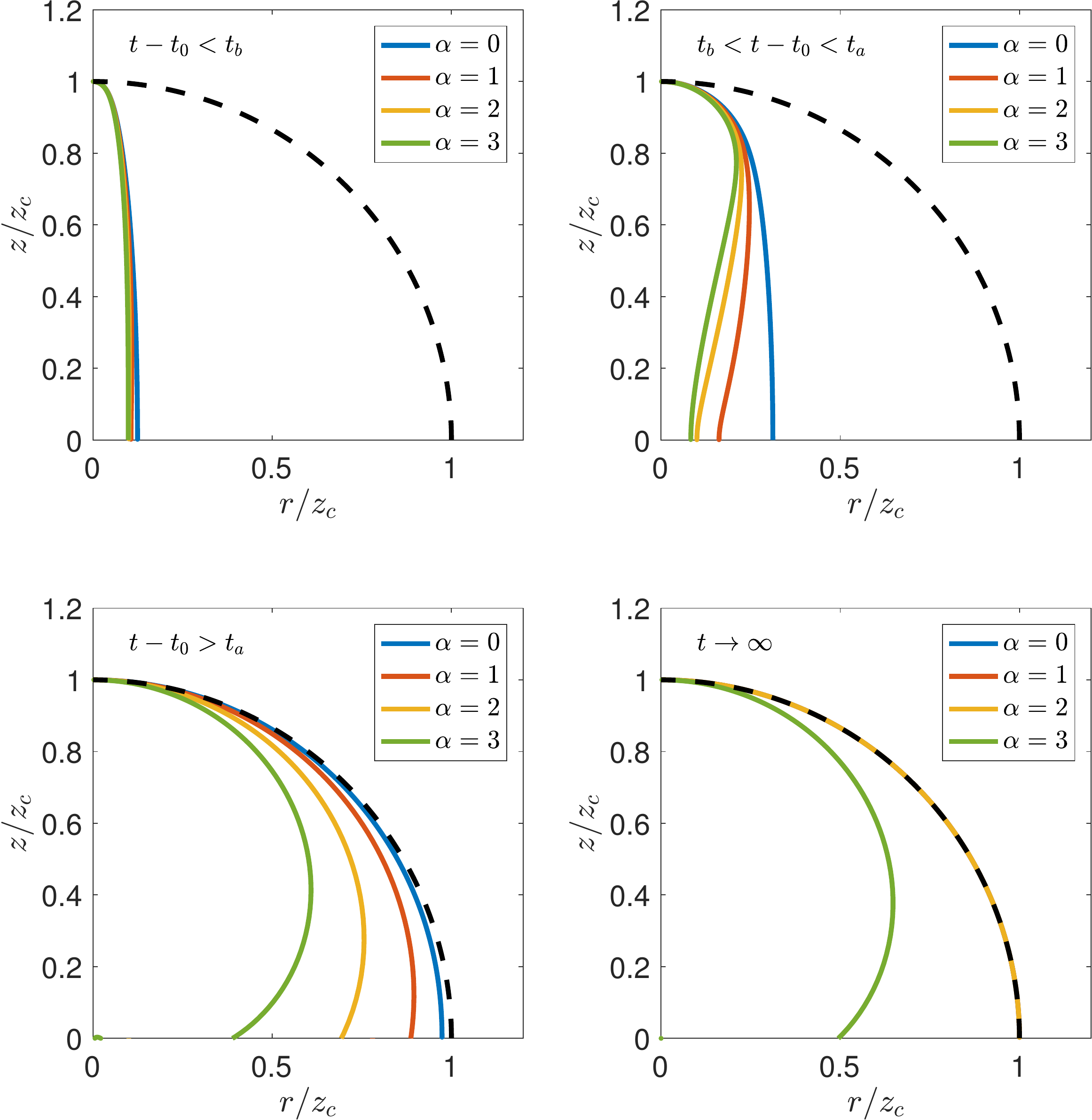}
\caption{Comparison of the shape of the shock in each phase of the evolution, for a density profile $\rho \propto R^{-\alpha}$ with $\alpha=0$, 1, 2, and 3.  The axes are scaled to the height of the cocoon, $\zc$, to allow for an easy comparison.  The black dashed line is a sphere of radius $\zc$. \textit{Upper left:} The planar phase ($t-t_0 \ll t_b$).  \textit{Upper right:} The sideways expansion phase ($t_b \ll t-t_0 \ll t_a$).   \textit{Lower left:} The quasi-spherical phase ($t_a \ll t-t_0$).   \textit{Lower right:} The asymptotic shape as $t \rightarrow \infty$.}
\label{comparison}
\end{figure*}

In Fig.~\ref{comparison}, we illustrate the shape of the shock in each dynamical regime for several different density profiles, taking as an initial condition an ellipsoidal cocoon shock with $b/a=0.1$.  The system is initially in the \textit{planar phase}, with $t-t_0 \ll t_b$ (upper left panel).  In this regime, the shape remains roughly ellipsoidal.  The evolution is not very sensitive to the density profile, although minor differences appear towards the equator.  Once the width has roughly doubled, the system transitions to the \textit{sideways expansion} phase (upper right panel), for which $t_b \ll t-t_0 \ll t_a$.  The shock shape in this case is strongly influenced by the initial height $a$, but only weakly dependent on the initial width $b$.   The effect of the density profile also starts to become apparent, with the cocoon being wider towards the base in shallower density gradients, and wider towards the tip in steeper gradients.  However, the aspect ratio of the shock is similar regardless of density profile.  Over a time-scale $\sim t_a$, the height and width of the shock become comparable, and the outflow enters the \textit{quasi-spherical} regime (lower left panel).  In this phase, $t-t_0 \gg t_a$, and the outflow's size is much larger than its initial size, so the evolution is approximately scale-free.  The differences in shape are mainly due to the density profile.  Different density profiles no longer have a similar aspect ratio; outflows in shallow density profiles are closer to becoming spherical.  Finally, as $t \rightarrow \infty$, the system approaches an asymptotic self-similar solution, as shown in the lower right panel.  

The asymptotic behaviour can be divided into two different regimes depending on $\alpha$.  For $\alpha \le 2$, the shock becomes fully spherical and the system is described by the classical spherical ST blast wave solution at late times.  Interestingly, however, non-sphericity persists indefinitely in the KA when $\alpha > 2$.  Although the shock still becomes self-similar in this case, it does not become spherical: the shock radius remains smaller at the equator than at the poles, resulting in a peanut-like shape.  The reasons for the different shapes will be explored further in Section~\ref{alpha2-3}.

Our model assumes that the pressure in the ambient medium is negligible compared to the cocoon pressure, and that the ambient medium is effectively static, with an expansion velocity much smaller than the shock velocity.  If either of these conditions are violated, the solution breaks down.  The range of validity of the model is discussed further in Section~\ref{breakdown}.

Because the sideways expansion is driven by pressure in the cocoon, both before and after choking, the width of the outflow doubles over a time-scale comparable to the choking time, i.e. $t_b \sim t_0$.  Alternatively, we can write $t_b \sim b/c\betac$, with the sideways shock expansion speed $\betac$ determined by balancing the upstream and downstream momentum flux at the shock.  Applying the shock jump conditions gives $\betac = [(\gamma+1)P_0/2\rho c^2]^{1/2}$, where $P_{0} = (\gamma-1)E_0/V_{0}$ is the initial pressure in the cocoon, $V_{0} \approx \frac{4}{3} \pi a b^2$ is the cocoon's initial volume, and $\gamma$ is the adiabatic index.  Adopting $\gamma=4/3$, we arrive at an expression for $t_b$ in terms of the cocoon parameters $a$, $b$, and $E_0$:
\begin{equation}
\label{tb}
t_b \sim t_0 \sim \left(\dfrac{24 \pi \rho b^4 a}{7E_0}\right)^{1/2}.
\end{equation}

The time-scale $t_a$ is also straightforward to derive.  After the jet is choked, both the forward and sideways expansion are driven by the cocoon's pressure.  Most of the early expansion occurs near the tip of the jet where the density is lowest.  Forward-moving material near the tip encounters nearly the same density as sideways-moving material, so the expansion speed is comparable in both directions.  Therefore, in the time it takes for the jet's height to increase from $a$ to $2a$, its width also increases from $b$ to $\sim b+a$.  The volume at that time is roughly $\Vc \sim \frac{8}{3} \pi a^3$, so the cocoon pressure is $\Pc \sim E_0/8\pi a^3$.  The time-scale $t_a$ is then given by $\sim a/v_z$, where $v_z \sim [(\gamma+1)\Pc/2\rho]^{1/2} $ is the on-axis expansion speed, leading to
\begin{equation}
\label{ta}
t_a \sim \left(\dfrac{48 \pi \rho a^5}{7E_0}\right)^{1/2}.
\end{equation}

The fact that $t_a \sim (a/b)^2 t_0$ implies that an outflow takes much longer to become quasi-spherical if it is initially narrow (i.e., if $b \ll a$).  This is due to a combination of two effects which cause narrow jets to be more strongly decelerated.  First, narrow jets undergo a larger drop in velocity when the jet ram pressure vanishes.  After the jet is choked, the on-axis velocity decreases from its initial value, $\betah$, to a new value, $\sim  \betac$, which is comparable to the sideways expansion speed.  Since $a\sim c\betah t_0$ and $b \sim c \betac t_0$, the velocity is reduced by a factor of $\betac/\betah \sim b/a$.   Second, narrow jets are more strongly affected by sideways expansion.  As discussed above, the width of the outflow grows from $b$ to $\sim a$ in the time it takes the height to double.  This increases the volume by a factor of $\sim (a/b)^2$, and decreases the pressure by a factor of $\sim(b/a)^2$.  Since $v_z \propto \Pc^{1/2}$, the sideways expansion reduces the velocity by an additional factor of $(\Pc/P_{0})^{1/2} \sim b/a$, so by the time the shock reaches a height of $2a$, the velocity is only $\sim (b/a) \betac \sim (b/a)^2 \betah$.

For $b \ll a$, the cocoon's height $\zc$ and width $\rc$ in the three dynamical regimes evolve approximately as 
\begin{equation}
\label{zsimple}
\zc \sim 
\begin{cases}
a, & t-t_0 \ll t_b \\
a\left[1+ \left(\dfrac{t}{t_a}\right)^{1/2} \right] & t_b \ll t-t_0 \ll t_a \\
2a\left(\dfrac{t}{t_a}\right)^{2/(5-\alpha)}, & t-t_0 \gg t_a 
\end{cases}
\end{equation}
and
\begin{equation}
\label{rsimple}
\rc \sim 
\begin{cases}
b, & t-t_0 \ll t_b \\
a\left(\dfrac{t}{t_a}\right)^{1/2}, & t_b \ll t-t_0 \ll t_a \\
a\left(\dfrac{t}{t_a}\right)^{2/(5-\alpha)}, & t-t_0 \gg t_a 
\end{cases}.
\end{equation}
Once the cocoon's width has increased significantly (i.e., for $t \gg t_b$), the angle $\thetab$ between the axis and the point where the cocoon is widest is of order
\begin{equation}
\label{thetasimple}
\thetab \sim
\begin{cases}
\left(\dfrac{t}{t_a}\right)^{1/2}, & t_b \ll t-t_0 \ll t_a \\
1, & t-t_0 \gg t_a 
\end{cases}
\end{equation}
(For the derivation of equations~\ref{zsimple}--\ref{thetasimple}, including the accurate determination of order-unity prefactors, see Section~\ref{analytical}.)  If $\alpha \le 2$, the ratio $\rc/\zc$ approaches 1 as $t \rightarrow \infty$, and the angle $\thetab$ approaches $\pi/2$.   Otherwise, $\rc/\zc \rightarrow [\sin(\pi/\alpha)]^{\alpha/(\alpha-2)}$ and $\thetab \rightarrow \pi/\alpha$.

In cases where the outflow becomes spherical, we find that the convergence to a sphere is rather gradual: as $t\rightarrow \infty$, the difference between the height and width goes to zero as (see Section~\ref{analytical})
\begin{equation}
\label{difference}
\zc-\rc \propto \zc^{(\alpha-2)/2} \propto t^{(\alpha-2)/(5-\alpha)}
\end{equation}
for $\alpha < 2$, or as $ (\ln \zc)^{-1} \propto (\ln t)^{-1}$ for $\alpha=2$.  Because this quantity decreases so slowly, the difference $\zc-\rc$ can be used to constrain the radius at which the jet was choked, even at times considerably larger than $t_a$. If the both the height and width of the outflow are measured through observations, the choking radius can be estimated via
\begin{equation}
a \sim 
\begin{cases} 
\zc - \rc, & \zc - \rc \ga \zc/2 \\
[\zc^{-\alpha} (\zc-\rc)^2]^{1/{(2-\alpha)}}, & \zc - \rc \ll \zc/2.
\end{cases}
\end{equation}
Knowing the choking radius places strong constraints on the properties of the jet, as we discuss in Section~\ref{jet}.  In particular, if the explosion energy and the density at $z=\zc$ are also known, then the jet's luminosity $\Lj$, opening angle $\thetaop$ and duration $\tj$ satisfy
\begin{equation}
\label{jetconstraints}
\Lj \tj  =   E_0
\end{equation}
and
\begin{equation}
\tj \thetaop^{-2}  \sim  \left(\dfrac{\rho(\zc) \zc^5}{E_0}\right)^{1/2}  \left(\dfrac{\zc-\rc}{\zc}\right)^{q_\alpha},
\end{equation}
where we have defined
\begin{equation}
q_\alpha \equiv \begin{cases}
(5-\alpha)/2, & \zc - \rc \ga \zc/2 \\
(5-\alpha)/(2-\alpha), & \zc - \rc \ll \zc/2
\end{cases}.
\end{equation}

\section{Analytical Solutions for Cocoon Evolution}
\label{analytical}

As discussed above, the choking of a relativistic jet produces a narrow cocoon filled with hot gas.  We suppose that the cocoon propagates in an ambient medium with a negligible pressure and a radially stratified density, 
\begin{equation}
\label{rho}
\rho=\rho_0(R/a)^{-\alpha}.
\end{equation}
The forward shock bounding the cocoon can be described by a curve $\theta=\theta(R,t)$ at time $t$, where $\theta$ and $R$ are the usual polar coordinates.  

To help keep track of the many definitions employed throughout this paper, we provide a list of the symbols used in Appendix~\ref{appendix0}, along with their meaning and the place in the text where they are defined.   In naming variables, we make use of the following subscripts to simplify our notation:
\begin{itemize}
\item The subscript `0' refers to to values measured at the choking time, $t=t_0$.   
\item Global properties of the cocoon (such as its height, volume, and pressure) which depend only on time are marked with the subscript `c'.  
\item Quantities pertaining to the `bulge' (a special point on the cocoon surface where the instantaneous velocity is parallel to the equator; see Section~\ref{alphax}) are denoted by the subscript `b'.  
\item Parameters characterizing the jet prior to choking are subscripted with a `j'.  
\item Finally, various constants that depend on the density profile are given the subscript `$\alpha$'.
\end{itemize}

\begin{figure} 
\centering
\includegraphics[width=\columnwidth]{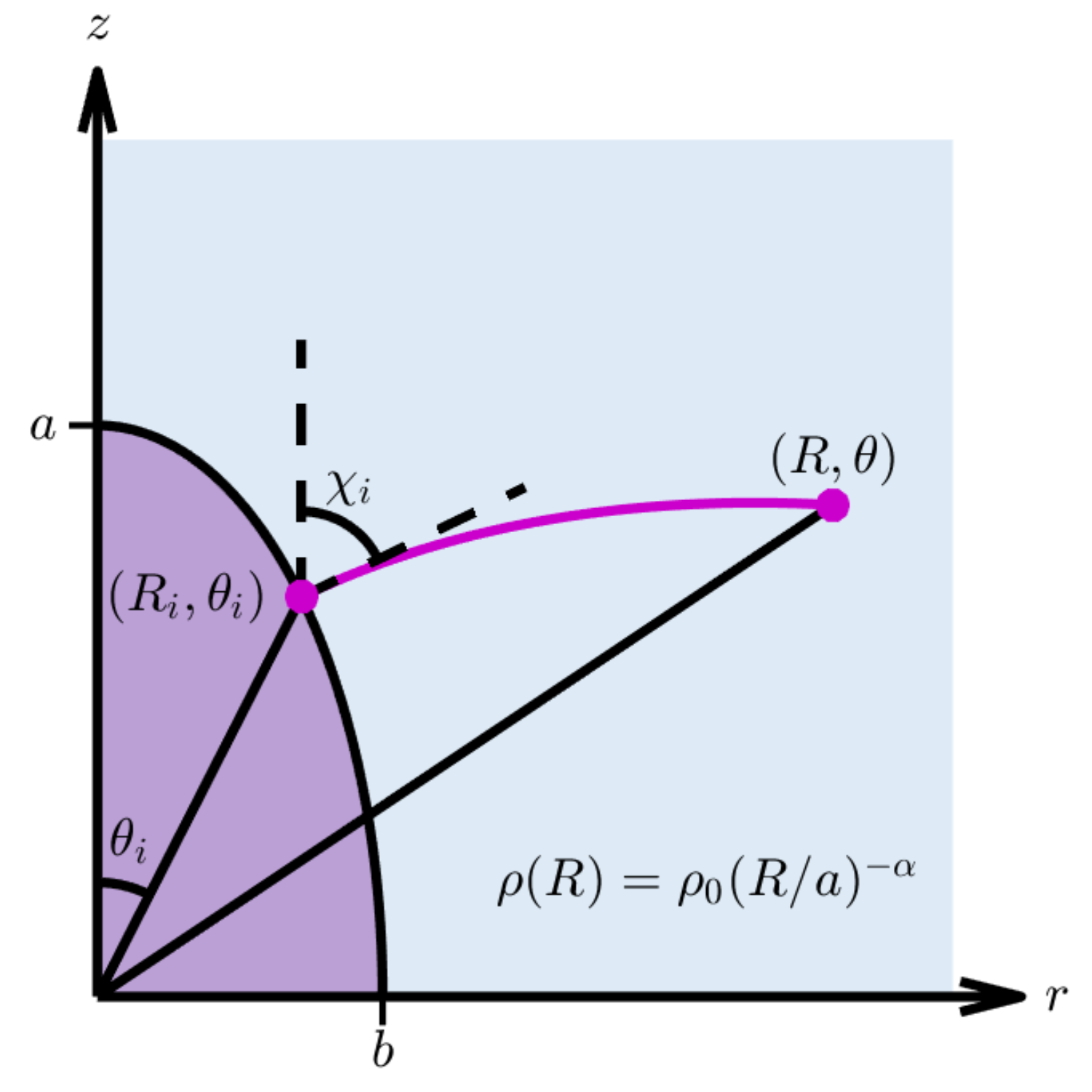}
\caption{Problem setup.  A high-pressure cocoon (purple) expands into an external medium (blue).  In a power-law density profile, particles at $(R_i,\theta_i)$ travel along the curved pink path to reach the point $(R,\theta)$.  The initial direction of motion is perpendicular to the shock surface; the angle between the axis and the surface normal is $\chi_i$.}
\label{illustration1}
\end{figure}

Let $(R_i,\theta_i)$ be the set of coordinates describing the cocoon surface at the initial time $t=t_0$ when the jet was choked.  We can then parametrize the initial shape of the forward shock by a curve $R_i(\theta_i)$.  Additionally, we define $\chi_i(\theta_i)$ as the angle between the surface normal at $(R_i,\theta_i)$ and the axis, which is given by
\begin{equation}
\label{chi_i}
\chi_i(\theta_i) =  \tan^{-1}\left(\dfrac{\tan \theta_i - \frac{\deriv \ln R_i}{\deriv \theta_i}}{1+\frac{\deriv \ln R_i}{\deriv \theta_i} \tan \theta_i} \right).
\end{equation}
A schematic of the initial conditions is shown in Fig.~\ref{illustration1}.  Particles starting with different values of $\theta_i$ each follow a unique trajectory as the outflow expands, such that any point along the shock surface can be associated with a particular value of $\theta_i$.

For concreteness, we take the initial shape of the cocoon to be a prolate ellipsoid of semi-major axis $a$ and semi-minor axis $b$, centred at the origin and oriented along the $z$-axis. In that case, geometry dictates
\begin{equation}
\label{R_i}
R_i(\theta_i) = \dfrac{ab}{\sqrt{a^2 \sin^2 \theta_i+b^2 \cos^2 \theta_i }}
\end{equation}
and
\begin{equation}
\label{thetaN}
\chi_i(\theta_i) = \tan^{-1}\left(\dfrac{a^2}{b^2} \tan \theta_i\right).
\end{equation}
However, the model is general and applicable to arbitrary initial shapes, as long as 1) the shape has axisymmetry and reflective symmetry about the equator; 2) $R_i$\footnote{Hereafter, to reduce clutter, we refer to $R_i(\theta_i)$ and $\chi_i(\theta_i)$ as simply $R_i$ and $\theta_i$.  These and all other quantities subscripted with `$i$' should be understood as having an implicit dependence on $\theta_i$.} decreases smoothly with $\theta_i$; and 3) there is no cusp on the axis (cusps at the equator are permissible since, as we will see, they form even when not present initially). The choice of shape mainly affects the early evolution of the outflow, and does not significantly impact our conclusions.  

Now, suppose that at $t=t_0$ the gas within the cocoon has a pressure $P_0$, distributed uniformly over a volume $V_0$.  At each point along the shock surface, the internal pressure applies a force in the direction normal to the surface.  The speed $v$ of a surface element is set by balancing the mass and momentum flux in a frame comoving with the shock, leading to $v=[(\gamma+1)P'/2\rho]^{1/2}$, where $P'$ is the postshock gas pressure.  The ambient pressure is assumed to be negligible compared to the ambient ram pressure.  The speed in the direction normal to the surface can alternatively be found by direct differentiation of $\theta(R,t)$, yielding $v=|\partial \theta/\partial t|/|\nabla \theta|$. Equating the two expressions for $v$ results in a partial differential equation for $\theta(R,t)$,\footnote{Eq.~\ref{EoM} is undefined for a spherical flow since in that case $\partial \theta/\partial t \rightarrow \infty$ and $\partial \theta/\partial R \rightarrow \infty$.  The equation is more well-behaved in the spherical limit when $\theta$ is chosen as the independent variable, i.e. when the shock is described by a curve $R=R(\theta,t)$.  However, in the general case of an aspherical shock in a spherically symmetric density, it is more convenient to use $R$ as the independent variable, because the density is a function of $R$, not $\theta$. }
\begin{equation}
\label{EoM}
\sqrt{\dfrac{\gamma+1}{2} \dfrac{P'(t)}{\rho(R)}} = v(R,t) = \dfrac{\left| \partial\theta/\partial t \right|}{\sqrt{1/R^2+(\partial \theta/\partial R)^2 }},
\end{equation}
where we have assumed a strong, non-relativistic shock and applied the jump condition $\rho'/\rho=v/v'={(\gamma+1)/(\gamma-1)}$ for adiabatic index $\gamma$.

The KA allows us to replace the postshock pressure $P'(t)$ in equation~\ref{EoM} by the volume-averaged pressure of the cocoon $\Pc(t)$.    If losses are unimportant, the total cocoon energy, $E_0$, can be treated as constant, so that $\Pc(t) \propto E_0/\Vc(t)$.  We can then write
\begin{equation}
\label{P}
P'(t) = \dfrac{\lambda (\gamma-1)E_0}{\Vc(t)},
\end{equation}
where $\Vc(t)$ is the volume of the cocoon at time $t$, and 
\begin{equation}
\label{lambda}
\lambda \equiv \left(\dfrac{E_{\rm th}}{E_0}\right)\left(\dfrac{P'}{\Pc}\right)
\end{equation}
is an order-unity dimensionless parameter which depends on the ratio of the thermal energy $E_{\rm th}$ to the total energy $E_0$, as well as the ratio of the postshock pressure to the average pressure.  The essential assumption of the KA is that $\lambda$ is constant in time.  In this case the dynamics are governed solely by the change in pressure due to the expanding volume, and the evolution of the cocoon can be determined to a reasonable approximation without the need to calculate the structure of the postshock region in detail.  For simplicity, we adopt $\lambda =1$ and $P' = \Pc=E_0/(3\Vc)$ in subsequent calculations.  The actual value of $\lambda$ can be estimated from numerical simulations (see Section~\ref{numerical}).

Equation~\ref{EoM} can be simplified by noting that $P'$ depends only on $t$, while $\rho$ depends only on $R$.  Following K92, we define a dimensionless parameter $x=x(t)$:
\begin{equation}
\label{x}
x(t) \equiv  \int_{t_0}^t \sqrt{\dfrac{\gamma+1}{2} \dfrac{\Pc(t')}{\rho_0 a^2}} \deriv t'.
\end{equation}
The parameter $x$ serves as a dimensionless replacement for the time.  It is initialized to $x=0$ at $t=t_0$ and thereafter increases monotonically with $t$.  Using the fact that the initial shock velocity along the axis is $v_0=[(\gamma+1)P_0/2\rho_0]^{1/2}$, equation~\ref{x} can be rewritten as
\begin{equation}
\label{xsimple}
x(t) = \int_{t_0}^t \sqrt{\dfrac{\Pc(t')}{P_0}} \dfrac{v_0 \deriv t'}{a}.
\end{equation}
We then see that $x$ is essentially the product of two dimensionless ratios: the ratio $[\Pc(t)/P_0]^{1/2}$ comparing the pressure at time $t$ to the initial pressure, and the ratio $t/(a/v_0)$ comparing the age to the initial lengthwise sound-crossing time, $a/v_0 \sim \sqrt{t_b t_a}$.

It is possible for $x$ to be bounded or unbounded, depending on whether integral in eq.~\ref{x} converges or diverges as ${t \rightarrow \infty}$ (see also Section~\ref{convert}).  Since we expect to recover the ST solution at late times, with $R \propto t^{2/(5-\alpha)}$ and $\Pc \propto R^{-3} \propto t^{-6/(5-\alpha)}$, we find that the integral diverges for $\alpha \le 2$.  The transition to the sideways expansion phase discussed in Section~\ref{overview} occurs at $x \sim b/a$, while the transition to the quasi-spherical phase occurs when $x\sim 1$.

Substituting equations~\ref{rho} and \ref{x} into equation~\ref{EoM}, we obtain
\begin{equation}
\label{diffeq}
\dfrac{1}{a^2} \left( \dfrac{\partial \theta}{\partial x} \right)^2 = \left(\dfrac{R}{a}\right)^\alpha \left[ \left(\dfrac{\partial \theta}{\partial R} \right)^2 + \dfrac{1}{R^2} \right],
\end{equation}
which can be solved to find the cocoon shape as a function of $R$ and $x$ (see Sections~\ref{alpha0} and~\ref{alphax}).  Once the shape is known, the volume can be calculated (see Section~\ref{volume}), and the pressure can be found via ${\Pc=P_0(V_0/\Vc)}$.  The time can then be determined by inverting equation~\ref{x} and integrating (see Section~\ref{convert}).

As we will see, the evolution of the cocoon depends strongly on $\alpha$, the power-law index of the external density profile.  We begin by discussing the simple case of a constant ambient density as an illustrative example.  Then, we examine how the evolution changes as the external density profile steepens.

\subsection{A constant external density}
\label{alpha0}

\begin{figure}
\centering
\includegraphics[width=\columnwidth]{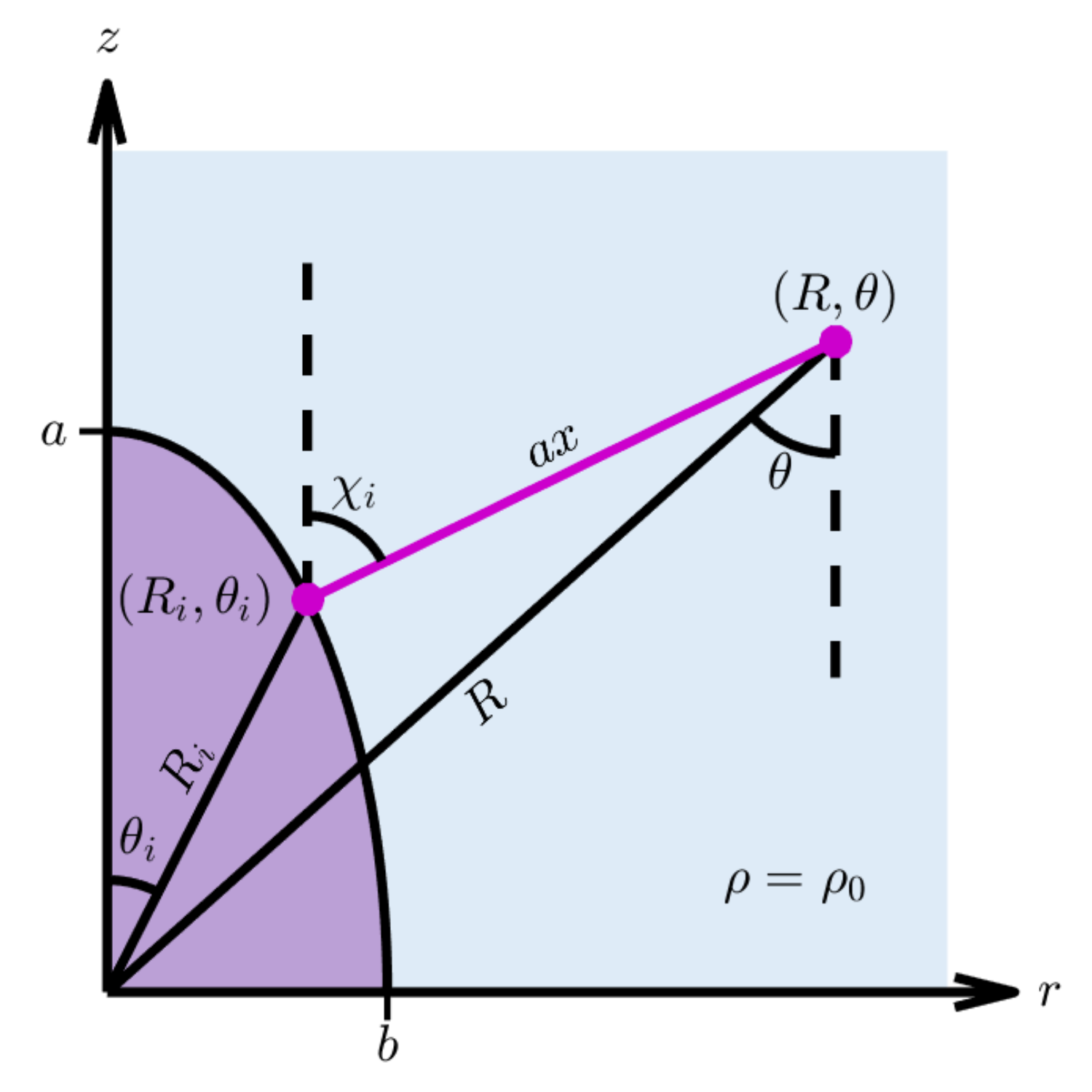}
\caption{As in Fig.~\ref{illustration1}, but for a uniform density ${\rho=\rho_0}$.  Particles follow a straight-line trajectory from ($R_i,\theta_i$) to ($R,\theta$), covering a distance $ax$.  A triangle of side lengths $R_i$, $R$, and $ax$ is formed by connecting the starting location, the new location, and the origin.  The interior angle opposite $R$ is given by $\pi-(\chi_i-\theta_i)$.}
\label{illustration2}
\end{figure}

When the external density is uniform with $\rho(R)=\rho_0$, each point along the shock surface travels with the same velocity, $v(t) = [(\gamma+1)\Pc(t)/2 \rho_0]^{1/2}$, since all points are driven by the same uniform pressure into the same density.  Furthermore, since there is no differential velocity between adjacent points on the surface, each small patch of surface maintains its original orientation.  Therefore, a particle on the surface travels to infinity along a straight line in the direction normal to the surface.  Examining equation~\ref{xsimple}, we see that it reduces to $x(t) =\int v(t) \deriv t/a$ in the constant density case.  In other words, when $\alpha=0$, the distance traveled along each trajectory at time $t$ is simply $ax(t)$.    Thus, a particle beginning at $(R_i,\theta_i)$ at time $t_0$ travels a distance $ax$ along a straight line to reach a new position $(R,\theta)$ at time $t$, as illustrated in Fig.~\ref{illustration2}.

Because the trajectories are straight lines in this case, $R$ and $\theta$ can be determined simply from geometry.  A triangle with side lengths $R$, $R_i$, and $ax$ is formed by connecting the points $(0,0)$, $(R_i,\theta_i)$, and $(R,\theta)$, as seen in Fig.~\ref{illustration2}.    
From the Law of Cosines, 
\begin{equation}
\label{R0}
R(x,\theta_i) = \left(R_i^2 +a^2 x^2 + 2 a x R_i \cos\omega_i\right)^{1/2},
\end{equation}
where we have defined $\omega_i$ as the angle between the initial direction of motion and the radial direction, i.e.
\begin{equation}
\label{omega}
\omega_i \equiv \chi_i-\theta_i.
\end{equation}
The cylindrical coordinates $r$ and $z$ are obtained via ${r=R_i \sin\theta_i + ax \sin \chi_i}$ and ${z = R_i \cos \theta_i + ax \cos \chi_i}$.  After some manipulation using angle sum and difference identities,  ${\theta= \tan^{-1}(r/z)}$ becomes
\begin{equation}
\label{theta0}
\theta(x,\theta_i)  = \theta_i + \tan^{-1}\left(\dfrac{x \sin\omega_i}{R_i/a + x \cos\omega_i}\right).
\end{equation}
If $\theta_i$ is held fixed, equations~\ref{R0} and \ref{theta0} give the trajectory of a particle that started at $\theta_i$, while if $x$ is fixed, the equations describe the shape of the cocoon surface at the time $t=t(x)$.  It is also possible to combine equations~\ref{R0} and \ref{theta0} to eliminate $x$, which gives $R$ as a function of $\theta$ along a trajectory stemming from $\theta_i$:
\begin{equation}
R(\theta,\theta_i) = \dfrac{R_i}{\cos(\theta-\theta_i)-\sin(\theta-\theta_i)\cot\omega_i}.
\end{equation}
As expected, this describes a straight line that passes through $(R_i,\theta_i)$ and forms an angle $\chi_i$ with the axis.

In order to understand the long term evolution of the outflow, it is informative to compare the expansion along the axis to the expansion perpendicular to the axis.    We denote the height and width of the cocoon as $\zc$ and $\rc$, respectively.  The height is defined as the radius at the pole,
\begin{equation}
\label{zcdef}
\zc(x) \equiv R(x,\theta_i=0),
\end{equation}
while the width is defined as the maximum distance between the shock and the axis, i.e. 
\begin{equation}
\label{rcdef}
\rc(x) \equiv \max\{R (x, \theta_i) \sin [\theta(x,\theta_i)]\}
\end{equation}  
for a given $x$.  In the constant-density case, since the trajectories are linear and the expansion velocity is equal in magnitude at each point on the surface, the width is ultimately set by the ejecta which have an initial velocity perpendicular to the axis, with $\chi_i = \pi/2$.

We now turn to the specific case of an initially ellipsoidal shape.  For an ellipsoid with semi-major axis $a$ and semi-minor axis $b$, we have $(R_i,\chi_i)=(a,0)$ at $\theta_i=0$, and $(R_i,\chi_i)=(b,\pi/2)$ at $\theta_i=\pi/2$.   Then, from equation~\ref{R0}, $\zc = a(x+1)$ and $\rc=a(x+b/a)$, so that the ratio of the cocoon's width to its height is
\begin{equation}
\label{ratio0}
\dfrac{\rc}{\zc} = \dfrac{x+b/a}{x+1} = 1-\dfrac{a}{\zc}\left(1-b/a\right).
\end{equation}
Defining an effective eccentricity, $\epsiloneff = \left(1-\rc^2/\zc^2\right)^{1/2}$, we have
\begin{equation}
\label{ecc0}
\epsiloneff \approx \left[2 \left(1-b/a\right) a/\zc\right]^{1/2}
\end{equation}
to leading order in $a/\zc$.

As $x \rightarrow \infty$, the outflow becomes gradually wider and ${\rc/\zc \rightarrow 1}$ while $\epsiloneff \rightarrow 0$.   In fact, we see from equation~\ref{R0} that for large $x$, $R\approx ax$ regardless of $\theta_i$, indicating that the cocoon becomes spherical.  Taking $x \propto R$ and $\Pc \propto  \Vc \propto R^{-3}$ in equation~\ref{x}, we see that the usual $R \propto t^{2/5}$ behaviour for a blast wave in a constant density is recovered.

\begin{figure} 
\centering
\includegraphics[width=\columnwidth]{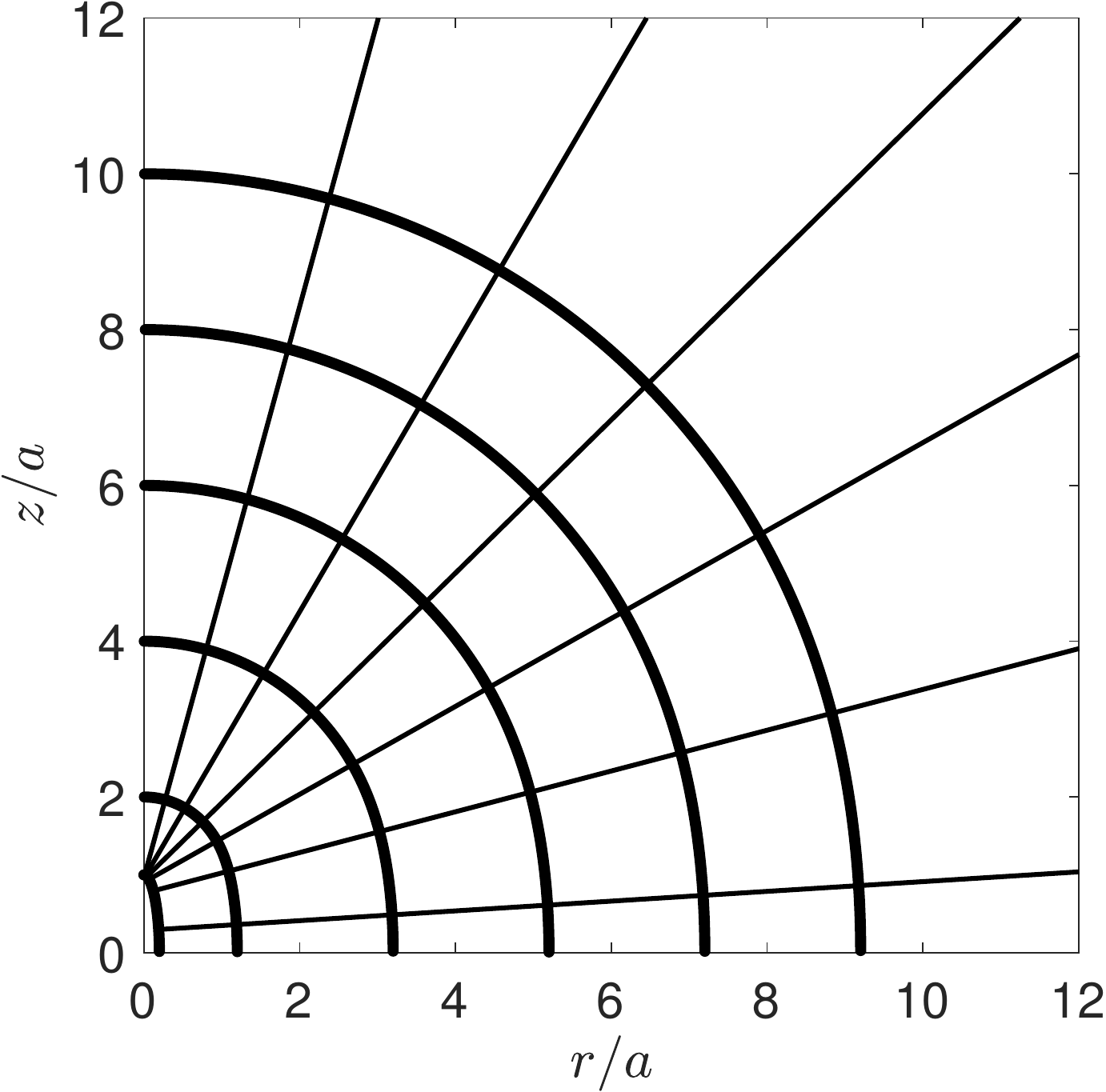}
\caption{The shape of the cocoon (thick lines) when $\zc/a= 1$, $2$, $4$, $6$, $8$ and $10$, for a constant density profile with $\alpha=0$ and $a/b=5$.  Several trajectories for different starting points on the cocoon surface are also shown as thin lines.}
\label{web0}
\end{figure}

The solution for $\alpha=0$ is plotted in Fig.~\ref{web0} at the times when the cocoon's height has reached 2, 4, 6, 8, and 10 times its initial value.  The initial shape was taken to be an ellipsoid with aspect ratio $b/a=0.2$.  Several particle paths are also indicated.  A few key properties of the constant-density solution are worth emphasizing.  First, all of the trajectories are straight lines that do not cross.  There are no self-intersections of the cocoon surface.  Second, the width of the outflow is always a maximum in the equatorial plane.  Finally, the solution gradually approaches a self-similar spherical blast wave as $t$ becomes large.   The process of becoming spherical is relatively rapid: by the time the cocoon's height has doubled to $\zc=2a$, eq.~\ref{ratio0} shows that its width is already $\rc>\zc/2$, regardless of how narrow it was to start.  As we will see, these basic features do not necessarily hold true at higher $\alpha$.

\subsection{A power-law external density}
\label{alphax}

In a non-uniform, spherically symmetric density profile, there is a density gradient in the radial direction that is not aligned with the surface normal.  Although each small patch of the surface is driven from behind by a uniform pressure, adjacent points encounter different densities, feel different ram pressures, and therefore move with different speeds.  As a result, elements of the surface change their orientation as they travel outwards, following trajectories that curve towards higher density.   The strength of this effect depends on the misalignment of the density gradient and the surface normal, which is expressed by $|\chi_i-\theta_i|$.  The cocoon becomes more deformed in regions where $|\chi_i-\theta_i|$ is larger, eventually developing a bulge near the location where $|\chi_i-\theta_i|$ is maximal. 

\begin{figure} 
\centering
\includegraphics[width=\columnwidth]{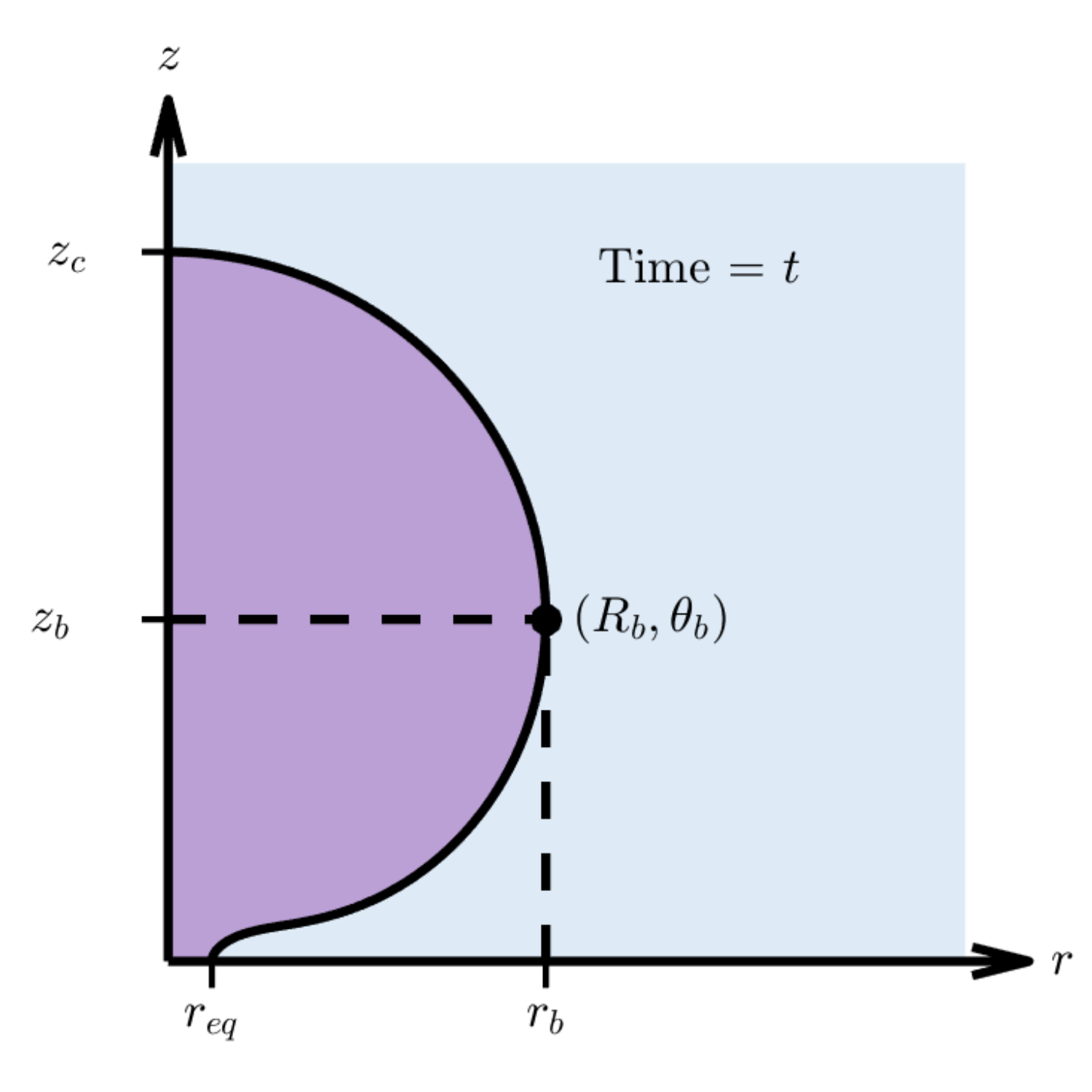}
\caption{The quantities $\zc$, $\rb$, $z_{b}$, and $\req$ that characterize the cocoon shape.  We define the cocoon width $\rc$ to be the larger of $\rb$ and $\req$ (in this case, $\rc=\rb$).}
\label{illustration}
\end{figure}

Because the bulge forms at a larger radius where the density is lower, it expands sideways faster than material in the equatorial plane, and eventually overtakes it.  Thus, unlike the constant-density case, in decreasing density profiles the cocoon's width is ultimately set by the behaviour of the bulge, not the expansion near the equator.  Once the bulge has formed, we can describe the cocoon's shape with four quantities: the extent along the $z$-axis, $\zc$; the width in the equatorial plane, $\req$; and the coordinates $\Rb$ and $\thetab$ indicating the location of the bulge, which we define to be the point where the cocoon surface is locally parallel to the axis.  This point is located at a height $\zb = \Rb \cos \thetab$ above the equatorial plane and at a distance $\rb=\Rb \sin \thetab$ from the axis.  The cocoon width is then $\rc = \text{max}(\rb, \req)$.  This configuration is illustrated in Fig.~\ref{illustration}.    

The determination of $\req$ comes with an important caveat.  Because the points on the surface follow curved paths, self-intersections can occur in a non-uniform density profile: when any part of the cocoon reaches the equator, it encounters material coming from the opposite side due to the reflective symmetry of the system.  Collisions at the equator are expected to generate a high-pressure region and drive a shock outwards in the equatorial plane.  Our simple KA model does not take equatorial interactions into account, and therefore underestimates the true equatorial width.  In addition, there is a cusp at the equator in the idealized model which is unlikely to exist in reality.  These discrepancies are clearly seen in a comparison with numerical simulations (see Section~\ref{numerical}).  While the analytically-obtained value of $\req$ is only accurate to within a factor of a few, it may none the less be useful as a lower bound.  

For $\alpha \ne 2$, K92 showed that equation~\ref{diffeq} can be solved by means of a conformal coordinate transformation.  We define
\begin{equation}
\label{kalpha}
k_\alpha \equiv \dfrac{2-\alpha}{2},
\end{equation} 
and then introduce new dimensionless coordinates ${\xi=(R/a)^{k_\alpha}}$ and $\phi=k_\alpha \theta$.  Applying this transformation to equation~\ref{diffeq} results in 
\begin{equation}
\label{transformeddiffeq}
\dfrac{1}{k_\alpha^2}\left( \dfrac{\partial \phi}{\partial x} \right)^2 = \left(\dfrac{\partial \phi}{\partial \xi} \right)^2 + \dfrac{1}{\xi^2},
\end{equation}
which has the same form as the Kompaneets equation~\ref{diffeq} for a constant density medium.  Essentially, this converts the problem of solving equation~\ref{diffeq} for an arbitrary density power-law to the much easier problem of solving the constant-density case for different initial conditions.  For convenience, we also define $\xi_i=(R_i/a)^{k_\alpha}$ and $\phi_i=k_\alpha \theta_i$.

The angle $\psi_i$ between the axis and the initial direction of motion in $\xi$--$\phi$ coordinates can be derived as follows.  We first observe that the quantity $\deriv \ln R/\deriv \theta=\deriv \ln \xi/\deriv\phi$ is preserved by the coordinate transformation.  Then, we note that in order for the slope of a trajectory to be $\cot\chi_i$ at the point $(R_i,\theta_i)$, the initial condition $\deriv \ln R/\deriv\theta=\cot(\chi_i-\theta_i)$ must be satisfied at $\theta=\theta_i$.  By the same logic, $\deriv \ln \xi/\deriv \phi=\cot(\psi_i-\phi_i)$ when evaluated at $\phi=\phi_i$.  Thus, we have $\cot(\psi_i-\phi_i)=\cot(\chi_i-\theta_i)=\cot \omega_i$, and therefore $\psi_i= \phi_i+\omega_i$ (up to a phase that we choose to be zero).

Having determined the initial conditions in $\xi$--$\phi$ space, the solution of equation~\ref{transformeddiffeq} now proceeds in the same fashion as in the $\alpha=0$ case.  Following the procedure outlined in the previous section yields
\begin{equation}
\label{xi1}
\xi(x,\phi_i) = \left[ \xi_i^2 + k_\alpha^2 x^2 + 2 k_\alpha x \xi_i \cos(\psi_i-\phi_i) \right]^{1/2}
\end{equation}
and
\begin{equation}
\label{phi1}
\phi(x,\phi_i) = \phi_i + \tan^{-1}\left( \dfrac{x \sin(\psi_i-\phi_i)}{\xi_i + x \cos (\psi_i-\phi_i)}\right).
\end{equation}
Transforming back to $R$--$\theta$ coordinates, and using the fact that ${\psi_i-\phi_i=\omega_i}$ (see above) we obtain:
\begin{equation}
\label{R1}
R(x,\theta_i) = \left[R_i^{2k_\alpha}+a^{2k_\alpha} k_\alpha^2 x^2 + 2 k_\alpha x (R_i a)^{k_\alpha} \cos \omega_i \right]^{1/2k_\alpha}
\end{equation}
\begin{equation}
\label{theta1}
\theta(x,\theta_i) = \theta_i + \dfrac{1}{k_\alpha} \tan^{-1} \left( \dfrac{  k_\alpha x \sin \omega_i}{ \left(R_i/a\right)^{k_\alpha} + k_\alpha x \cos\omega_i} \right).
\end{equation}
Just as before, we can combine the parametric equations~\ref{R1} and \ref{theta1} to eliminate $x$ and obtain an expression describing the trajectory of a particle that began at $\theta_i$:
\begin{equation}
\label{traj1}
R(\theta,\theta_i) = \dfrac{R_i}{\left\{ \cos\left[k_\alpha(\theta-\theta_i)\right] -  \cot \omega_i \sin\left[k_\alpha(\theta-\theta_i)\right]\right\}^{1/k_\alpha}}.
\end{equation}
Note that when $R_i=a$, $\theta_i=0$, and $\omega_i=\chi_i$, equations~\ref{R1}--\ref{traj1} reduce to the expressions given in K92 for an on-axis point explosion. 

\begin{figure} 
\centering
\includegraphics[width=\columnwidth,keepaspectratio]{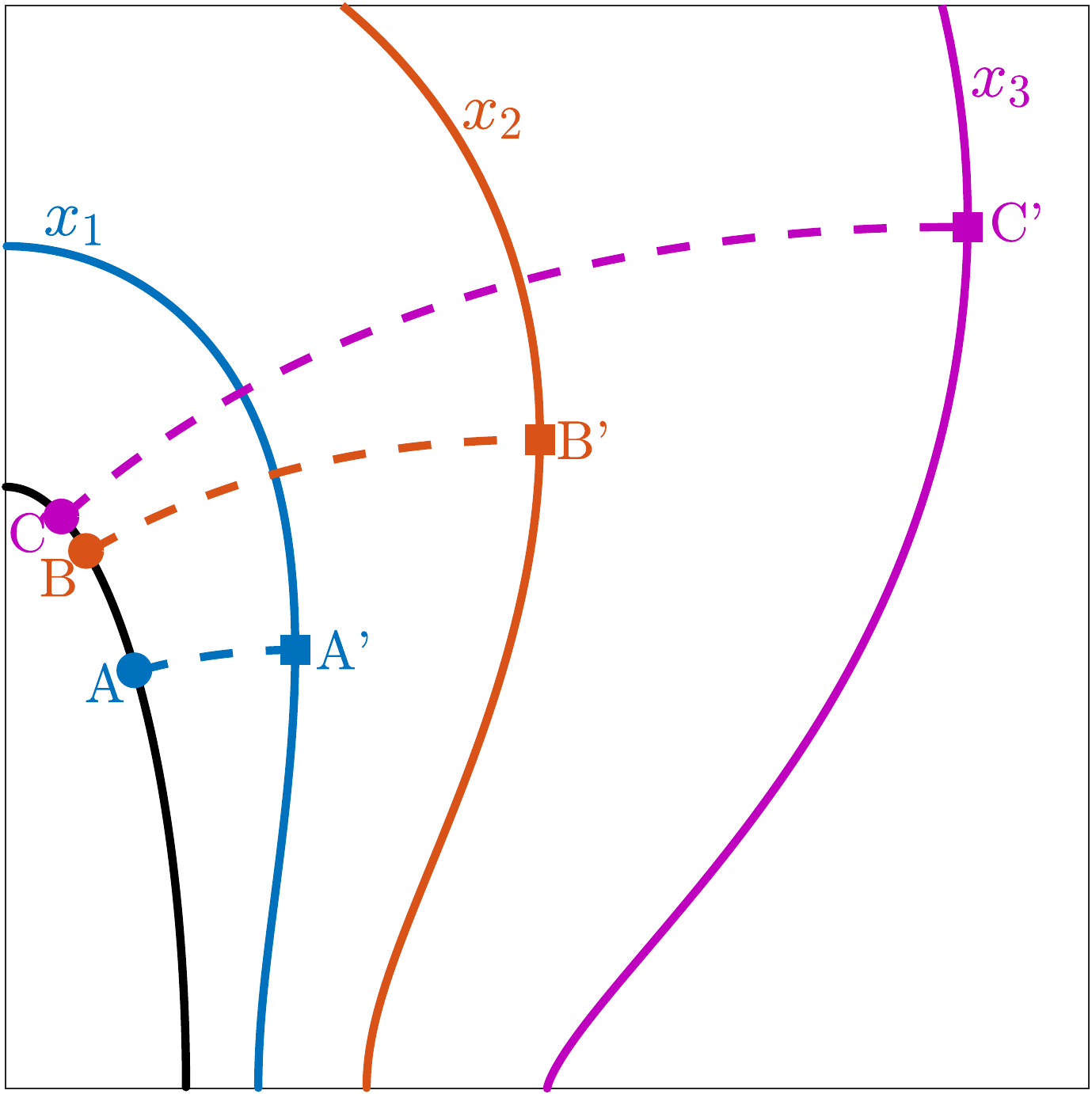}
\caption{The shape of the shock at three times--$x_1$ (blue), $x_2$ (red), and $x_3$ (magenta)--taking $\alpha=2$ and $b/a=0.3$ as an illustrative example.  The coloured squares  mark the location of the bulge where the velocity is parallel to the equator.  The angular coordinates of the points A', B', and C' are, respectively, $\thetab(x_1)$, $\thetab(x_2)$, and $\thetab(x_3)$.  At each time, the dashed trajectory passing through the bulge can be traced to a unique point on the initial surface, as indicated by the filled coloured circles along the black curve.  The points labeled A, B, and C have initial angular positions of $\theta_i=\thetabo(x_1)$, $\thetabo(x_2)$, and $\thetabo(x_3)$ respectively.}
\label{bulgeparams}
\end{figure}

Equation~\ref{traj1} can be used to derive parametric equations for the coordinates of the bulge.  To do so, we first note that since $\Rb$ and $\thetab$ describe a particular point on the cocoon's surface, they must be functions of time alone, i.e. $\Rb=\Rb(x)$ and $\thetab=\thetab(x)$.  The trajectory passing through the point ($\Rb(x)$, $\thetab(x)$) at a given time $x$ can be traced back to a specific initial angular position $\thetabo(x)$, as shown in Fig.~\ref{bulgeparams}.  That is to say, for each time $x$, there exists a corresponding initial angle $\thetabo(x)$ such that $R(x,\thetabo(x)) = \Rb(x)$ and $\theta(x,\thetabo(x))=\thetab(x)$.  The relationship between $x$ and $\thetabo$ is derived in Appendix~\ref{appendixa}, but the exact functional form is not important for the present discussion.  For now, it suffices to say that there is a one-to-one mapping between $\thetabo$ and $t$.  Therefore, $\thetabo$ can be thought of as an alternative dimensionless time parameter, which has an initial value of $\thetabo=\pi/2$ and decreases monotonically to a final value $\thetabomin$ as $t\rightarrow \infty$.

Now, since every trajectory meets the surface at a right angle, the trajectory which intersects the surface at ($\Rb(x), \thetab(x)$) must be parallel to the equator at that point (see Fig.~\ref{bulgeparams}).  This means that $d \ln R/d\theta = \tan \thetab$ must be satisfied when evaluated at $R=\Rb$ and $\theta=\thetab$.  Differentiating equation~\ref{traj1} and setting $\theta_i=\thetabo(x)$ leads to the equations
\begin{equation}
\label{thetab}
\thetab(x) = \thetabo(x) + \dfrac{2}{\alpha}\left[\pi/2-\chibo(x)\right]
\end{equation}
\begin{equation}
\label{Rb}
\Rb(x) = \Rbo(x) \left(\dfrac{\sin[\chibo(x)-\thetabo(x)]}{\cos[\thetab(x)]}\right)^{1/k_\alpha},
\end{equation}
where we defined $\Rbo(x) \equiv R_i(\theta_i=\thetabo(x))$ and $\chibo(x) \equiv \chi_i(\theta_i=\thetabo(x))$.  As the cocoon expands, the position of the bulge traces out a path described by the parametric curve ($\Rb(\thetabo),\thetab(\thetabo))$.\footnote{From here on, we write $\thetabo(x)$ as simply $\thetabo$.}

To understand the transition to spherical flow, our ultimate goal is to derive an expression akin to equation~\ref{ratio0} which relates the width of the cocoon $\rc$ to its height $\zc$.  To make a direct comparison between $\rc$ and $\zc$, we must first express them in terms of the same parameter.  For example, in the constant density case, we derived expressions for both $\rc$ and $\zc$ in terms of $x$, which we then solved to obtain $\rc$ as an explicit function of $\zc$.  In the general case, however, things are not so straightforward.
While it is still possible to determine the height in terms of $x$ by setting $\theta_i=\chi_i=0$ and $R_i=a$ in equation~\ref{R1}, 
\begin{equation}
\label{z1}
\zc(x) = a (1+k_\alpha x)^{1/k_\alpha},
\end{equation}
in general there is not a way to express $\rc(x)$ explicitly in a similar fashion.
The underlying reason that $\zc$ can be described by a simple function of $x$, but $\rc$ can not, is that the location where the height is greatest always occurs along the axis, while the place where the width is greatest changes its relative position on the surface as the cocoon evolves.  Put another way, whereas the \textit{height} is always greatest \textit{along a particular trajectory}, the point where the \textit{width} is greatest is associated with \textit{different trajectories at different times}.  (This also explains why an explicit form of $\rc(x)$ exists in the constant-density case, since in that case the width is always determined by the ejecta traveling along a certain trajectory parallel to the equator.)

We thus conclude that $x$ is not, in general, a suitable choice of parameter for comparing $\rc$ and $\zc$ in a non-uniform density.  What, then, would be a better choice? The discussion surrounding equations~\ref{thetab}--\ref{Rb} suggests another option: parametrizing the system in terms of $\thetabo$.  This turns out to be a good choice, in the sense that the important quantities characterizing the cocoon's shape ($\zc$, $\rb$, and $\zb$; see Fig.~\ref{illustration}) can be written explicitly in terms of $\thetabo$ and the quantities $\Rbo$ and $\chibo$ derived from the initial shape.  For clarity of presentation, we reserve the derivation of the full system of parametric equations describing the shape for Appendix~\ref{appendixa}.  

The new parametrization allows a direct comparison between the height $\zc(\thetabo)$ and the width $\rc(\thetabo)$ of the cocoon at a given time $x$.  However, a drawback compared to the constant-density case is that the equations cannot be solved to obtain an explicit function $\rc(\zc)$ relating the height and width.  None the less, it is possible to simplify the exact equations to obtain an approximate relation between $\rc$ and $\zc$ in each of the dynamical regimes discussed in Section~\ref{overview} (see Appendix~\ref{appendixb}).  Expressions relating $\req$ and $\zc$ can also be derived in a similar manner (see Appendix~\ref{appendixc}), although there are issues with the KA near the equatorial plane as discussed above.

Here, we give only a qualitative overview of the results, assuming an initially ellipsoidal shape as given by equation~\ref{R_i}.  In this case, the cocoon width is initially set by the ejecta in the equatorial plane, with $\rc=\req$.  We find that the evolution of $\rc$ depends on the relationship between $\alpha$ and the outflow's initial aspect ratio, $a/b$. If $\alpha \le 2b^2/a^2$, no bulge develops, and the outflow remains widest at the equator indefinitely, similar to the constant-density case.  In this scenario, there are no trajectories which pass through the equator, so there is no interaction between ejecta from opposite sides of the equator.   On the other hand, if $\alpha > 2b^2/a^2$, the bulge eventually forms and overtakes the equatorial ejecta, and collisions at the equator occur as well.

We are mainly interested in the scenario where ${a/b \gg 1}$, in which case $\alpha \gg 2b^2/a^2$ also holds unless the density profile is very flat.  In this limit, the bulge overtakes the equatorial material early in the planar phase, once the width has increased by $\rc -b \approx (2b^2/\alpha a^2)b \ll b$.  The cocoon's width is therefore governed by the behaviour of the bulge, with $\rc=\rb$ for the majority of the evolution.  The motion of the bulge during each dynamical phase can be summarized as follows:
\begin{enumerate}
\item During the planar phase, the bulge initially remains close to the equator.  At first, particles have not had sufficient time to significantly change their direction of motion, so the cocoon width is determined by the ejecta from $\thetabo \gg b/a$, which had an initial velocity almost perpendicular to the axis, with $\chibo \approx \pi/2$.  As the cocoon expands, material at larger $z$ (which started closer to the axis, but with a higher speed due to the lower density) begins to overtake material at smaller $z$, causing the bulge to march along the surface towards smaller $\theta$.  By the time the cocoon's width has doubled, the bulge has crossed most of the surface, with $\thetab$ decreasing from $ \sim \pi/2$ to $ \sim b/a$.
\item In the sideways expansion phase, the particles at the location of the bulge originated from ${b^2/a^2 \ll \thetabo \ll b/a}$.  On this part of the surface, the $r$- and $z$-components of the initial velocity are comparable, with $\chibo \ga \pi/4$.  The particles in the bulge have had to change their direction of motion by $\la 45$ degrees, moving farther from the axis in the process.  By the time a particle starting from $\thetabo$ passes through the bulge, its angular position has increased to $\thetab \gg \thetabo$.  Because the width of the outflow increases in this phase, while the height stays about the same (see Section~\ref{overview}), $\thetab$ now grows over time.  This phase ends when the height and width of the cocoon become comparable, at which point $\thetab \sim 1$.
\item Finally, during the quasi-spherical phase, the width is governed by material that came from $\thetabo \ll b^2/a^2$.   The material in the bulge was initially moving almost parallel to the axis, with $\chibo \ll \pi/4$, and had its direction of motion changed by $\sim 90$ degrees.  As in the previous phase, $\thetab \gg \thetabo$.   As time goes on, the value of $\thetab$ continues to gradually increase.  Eventually, the evolution becomes scale-free, with all length scales $\propto \zc$.  However, the outflow does not necessarily become spherical in the KA, as we discuss in Section~\ref{alpha2-3}. To capture the shape at infinity, we define two constants of proportionality depending on $\alpha$, $f_\alpha$ and $g_\alpha$, such that
\begin{equation}
\label{fdefine}
\begin{array}{ll}
f_\alpha \equiv \lim_{t \rightarrow \infty} (\rc/\zc)  \\
g_\alpha \equiv \lim_{t \rightarrow \infty} (\req/\zc)
\end{array}
\end{equation}
In addition, we define the maximum angle reached by the bulge as
\begin{equation}
\label{thetabmax}
\thetabinf \equiv \lim_{t\rightarrow \infty} \thetab.
\end{equation}
If the outflow becomes spherical, then ${f_\alpha=g_\alpha=1}$, and the bulge moves to the equator so $\thetabinf= \pi/2$.  Otherwise, we have $f_\alpha < 1$ and $g_\alpha < 1$, with $f_\alpha \ne g_\alpha$, and $\thetabinf < \pi/2$.
\end{enumerate}

In both the sideways expansion and quasi-spherical phases, we find that $\thetabo$ is negligible compared to $\chibo$ and $\thetab$.  This property is not unique to the bulge point.  In fact, as long as $b \ll a$ and $\alpha \gg 2b^2/a^2$, the conditions $\theta_i \ll \theta$ and $\theta_i \ll \chi_i$ hold over all parts of the surface where $\theta \la x$.  Particles with $\theta_i \ll 1$ also satisfy $R_i \approx a$, since they started out near the tip of the cocoon.  Therefore, we can neglect $\theta_i$ and set $R_i/a \approx 1$ in equations~\ref{R1} and~\ref{theta1} to obtain an approximate formula for the shape,
\begin{equation}
\label{shape1}
R(\theta,x) \approx a\left[\cos \theta + k_\alpha x \sqrt{1- \left(\dfrac{\sin k_\alpha \theta}{k_\alpha x}\right)^2}\right]^{1/k_\alpha},
\end{equation}
which is valid for initially narrow cocoons out to angles of $\theta \la x$. During the sideways expansion phase, it can be shown that $x \sim \thetab$ (see Appendix B); therefore, in this regime, eq.~\ref{shape1} describes the shape between the axis and the bulge, for ${0 \le \theta \la \thetab}$.  On the other hand, during the quasi-spherical phase $x \ga 1$ and eq.~\ref{shape1} is approximately valid over most of the cocoon surface.  

The approximation used to generate equation~\ref{shape1} is equivalent to assuming that the shock was generated by a point explosion placed at $z=a$.  Although K97 did not explicitly give an equation for $R(\theta,x)$ in the point explosion case, equation~\ref{shape1} can be reproduced by combining the two expressions given in their equation 23.  As with other results derived from the KA, equation~\ref{shape1} is expected to be less accurate towards the equator.

While the evolution during the planar and sideways expansion phases is not affected much by the density gradient, the asymptotic behaviour of the outflow depends strongly on the density profile.  Two different scenarios are possible, depending on whether the density profile is flat ($\alpha<2$) or steep ($\alpha>2$).  We discuss each regime in turn below, along with the special case $\alpha=2$.  As before, we treat only the specific case of an initially ellipsoidal shock.

\subsubsection{A flat external density profile $(0<\alpha<2)$}
\label{alpha0-2}

For $\alpha<2$, $k$ is positive, and the integral in equation~\ref{x} diverges so $x\rightarrow \infty$ as $t\rightarrow \infty$.  Inspecting equation~\ref{z1}, we see that $\zc \propto x^{1/k_\alpha}\propto x^{2/(2-\alpha)}$ for large $x$.  Assuming that the pressure drops as $\Pc \propto \Vc \propto \zc^{-3}$ asymptotically, equation~\ref{x} implies $x \propto t^{(2-\alpha)/(5-\alpha)}$.  Thus, as expected, we recover a blast wave evolution with $\zc  \propto t^{2/(5-\alpha)}$ at late times.

As discussed above, interactions at the equator occur in a non-uniform external density when $\alpha > 2b^2/a^2$.  Which regions on the cocoon surface eventually experience a collision?  For $2b^2/a^2<\alpha<2$, it turns out that $\thetabo$ has a minimum value $\thetabomin>0$.  Trajectories stemming from $\thetabomin$ are asymptotically parallel to the equator, i.e. they have $\theta \rightarrow \pi/2$ as $x \rightarrow \infty$.  Particles originating from $\theta_i < \thetabomin$ follow paths that never become parallel to the equator, so they can never be located at the bulge's location.  These particles stream freely to infinity, without interacting with any other ejecta.  On the other hand, particles with $\theta_i >\thetabomin$ eventually undergo a collision at the equator.  The value of $\thetabomin$ is derived in Appendix~\ref{appendixb}.

As time goes on, $\thetabo$ decreases (see Fig.~\ref{bulgeparams}) and becomes progressively closer to $\thetabomin$. To understand the sideways spreading at late times, we can therefore study the behaviour of equations~\ref{thetab} and \ref{Rb} as $\thetabo \rightarrow \thetabomin$.  
For $\zc \gg a$, we find (see Appendix~\ref{appendixb}) that the ratio of the cocoon's width to its height is approximated by 
\begin{equation}
\label{ratio1}
\dfrac{\rc}{\zc} \approx 1-\dfrac{A_\alpha}{\zeta(\zc)},
\end{equation}
where 
\begin{equation}
\label{Z}
\zeta(\zc) \equiv \dfrac{(\zc/a)^{|k_\alpha|}-1}{|k_\alpha|},
\end{equation}
and
\begin{equation}
\label{Aalpha1}
A_\alpha =  \dfrac{2}{k_\alpha^2} \sin^2(k_\alpha \pi/4)
\end{equation}
is an order-unity factor that depends on the density power-law.   The effective eccentricity in this case is
\begin{equation}
\label{ecc1}
\epsiloneff \approx \left(2 A_\alpha/\zeta\right)^{1/2}.
\end{equation}
Since $\rc/\zc \rightarrow 1$ and $\thetab \rightarrow \pi/2$ as $\zc \rightarrow \infty$, we infer that the outflow  becomes spherical in the $0<\alpha<2$ case, and therefore $f_\alpha = g_\alpha =1$.  However, the process of spherization is slower than in the constant-density case, with $\rc/\zc$ approaching unity as $\zeta^{-1}\propto \zc^{-k_\alpha}$, rather than as $\zc^{-1}$ as in equation~\ref{ratio0}.  

\begin{figure} 
\centering
\includegraphics[width=\columnwidth]{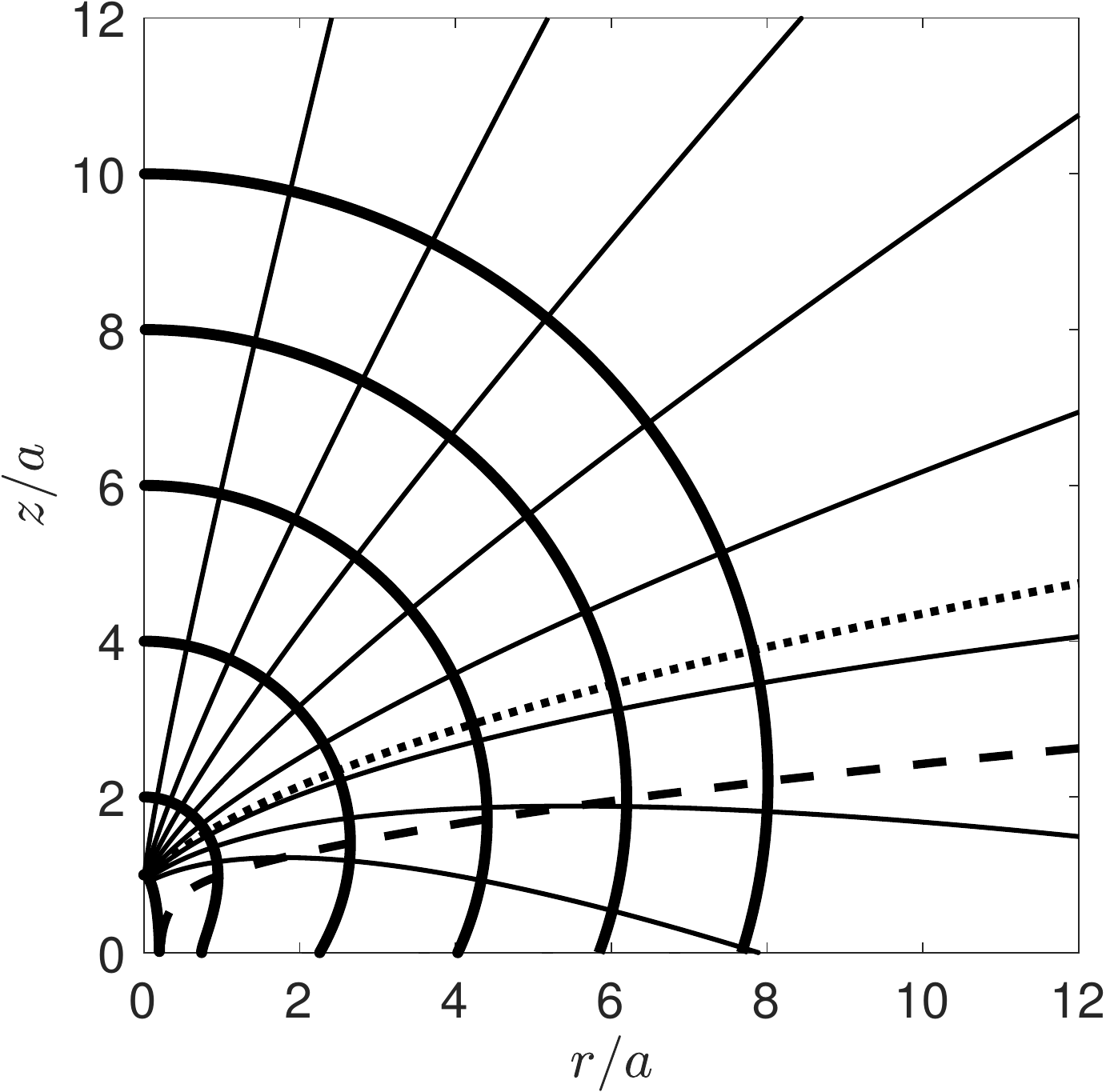}
\caption{As in Fig.~\ref{web0}, but for $\alpha=1$.  The dotted line separates trajectories that eventually reach the equator and collide with cocoon material from the other side, from those that proceed freely to infinity without interacting.  The dashed line shows the curve $\Rb(\theta_{b})$, which always passes through the cocoon surface at its widest point.}
\label{web1}
\end{figure}

The shape of the cocoon at $z/a=2$, 4, 6, 8, and 10 is shown in Fig.~\ref{web1}, for $\alpha=1$ and $b/a=0.2$.  The trajectory stemming from $\thetabomin$ and the evolution of the bulge's location are also shown, respectively as dotted and dashed lines.  To summarize, as in the constant density solution, the flow becomes more spherical with time, eventually becoming a self-similar ST blast wave.  However, in contrast with the constant-density case, if $\alpha > 2b^2/a^2$ a portion of cocoon material with $\theta_i>\thetabomin$ eventually reaches the equator and experiences a collision.  Furthermore, whereas for a constant density the cocoon is always widest at the equator, for a decreasing density profile, the cocoon bulges out and its width is maximum at some height $z_{b}>0$.  But, because $z_{b}$ grows more slowly than $\zc$, $z_{b}/\zc$ steadily decreases and the relative position of the bulge gradually approaches the equator at late times.

\subsubsection{A wind-like external density profile $(\alpha=2)$}
\label{alpha2}

When $\alpha=2$, equation~\ref{EoM} cannot be solved by the coordinate transformation used in Section~\ref{alpha0-2}, because $k_\alpha=0$.  Instead, we go back to the original differential equation~\ref{diffeq}, which for $\alpha=2$ becomes
\begin{equation}
\label{diffeq2}
\left(\dfrac{\partial \theta}{\partial x}\right)^2 = R^2\left(\dfrac{\partial \theta}{\partial R}\right)^2+1.
\end{equation}
This equation admits a solution of the form $\theta^2(x,R)=\theta_x(x)+\theta_R(R)$ \citep{kompaneets60}.  By separation of variables, we find 
\begin{equation}
\theta^2 = {(x+\kappa_1)^2 - (\ln R+\kappa_2)^2 + \kappa_1^2-\kappa_2^2},
\end{equation} 
where $\kappa_1=\kappa_1(\theta_i)$ and $\kappa_2=\kappa_2(\theta_i)$ depend on the original shape.  The initial conditions $\theta(x=0,\theta_i) = \theta_i$, $R(x=0,\theta_i)=R_i$, and $\deriv \ln R_i/\deriv \theta_i= -\tan\omega_i$ give $\kappa_2=\theta_i \cot\omega_i - \ln R_i$ and $\kappa_1^2 = [\kappa_2^2+\theta_i^2 \csc^2\omega_i]/2$.  Rewriting, we have
\begin{equation}
\label{surf2}
(x+\kappa_1)^2 = \theta^2 + [\ln(R/R_i)+\theta_i \cot\omega_i]^2 -\kappa_1^2+ \kappa_2^2.
\end{equation}
To derive an equation for the trajectories, we note that since the shock is a surface of constant $x$, $\nabla x$ is normal to the cocoon surface and parallel to the particle paths.  Therefore, the trajectories satisfy
\begin{equation}
\label{slope2}
R \dfrac{d\theta}{dR} = \dfrac{(1/R)(\partial x/\partial \theta)}{\partial x/\partial R} = \dfrac{\theta}{\ln(R/R_i) + \theta_i \cot\omega_i}.
\end{equation}
Integrating subject to the initial condition $R=R_i$ at $\theta=\theta_i$ yields
\begin{equation}
\label{traj2}
R(\theta,\theta_i) = R_i {\rm e}^{(\theta-\theta_i) \cot\omega_i}.
\end{equation}
The $\alpha=2$ case is unique in the sense that $d \ln R/d \theta=\cot\omega_i$ is constant along particle paths.

As in Section~\ref{alphax}, $R$ and $\theta$ can be written as parametric functions of $x$ and $\theta_i$.  After substitution of equation~\ref{surf2} into \ref{traj2}, we eliminate either $\theta$ or $R$ to obtain, respectively,
\begin{equation}
\label{R2}
R(x,\theta_i) = R_i {\rm e}^{x \cos \omega_i}
\end{equation}
and
\begin{equation}
\label{theta2}
\theta(x,\theta_i) = \theta_i + x \sin \omega_i.
\end{equation}
As before, we obtain an approximate expression for the shape of the shock in the sideways expansion and quasi-spherical regimes by neglecting $\theta_i$ and adopting $R_i \approx a$.  This gives
\begin{equation}
\label{shape2}
R(\theta,x) \approx a {\rm e}^{\sqrt{x^2-\theta^2}}.
\end{equation}
Setting $\theta_i=\chi_i=0$ and $R_i=a$ in equation~\ref{R2} gives the cocoon height, 
\begin{equation}
\label{z2}
\zc = a {\rm e}^{x}.
\end{equation}
Repeating the procedure of Section~\ref{alphax}, we find that the bulge is located at
\begin{equation}
\label{thetab2}
\theta_{b}=\pi/2-(\chibo-\thetabo)
\end{equation}
\begin{equation}
\label{Rb2}
\Rb = \Rbo {\rm e}^{(\thetab-\thetabo) \cot(\chibo-\thetabo)}   .
\end{equation}
Equations~\ref{traj2}--\ref{Rb2} agree with equations~\ref{R1}--\ref{Rb} and~\ref{shape1} in the limit of $\alpha \rightarrow 2$ and $k_\alpha \rightarrow 0$.  

Overall, the late-time evolution for $\alpha=2$ is similar to the $\alpha<2$ case.  One notable difference is that  ${\thetabomin=0}$ for $k_\alpha=0$, so $\theta_i>\thetabomin$ holds everywhere except on the axis.  Therefore, all of the cocoon material eventually reaches the equator and interacts.  The asymptotic ratio of width to height in this case is
\begin{equation}
\label{ratio2}
\dfrac{\rc}{\zc} \approx 1 - \dfrac{\pi^2}{8\ln (\zc/a)},
\end{equation}
and the eccentricity is 
\begin{equation}
\label{ecc2}
\epsiloneff \approx \left[\pi^2/4\ln(\zc/a)\right]^{1/2}.
\end{equation}
Equations~\ref{ratio2} and \ref{ecc2} can also be derived by taking the limit of equations~\ref{ratio1}--\ref{ecc1} as $k_\alpha \rightarrow 0$.  Although the cocoon once again becomes spherical, the logarithmic dependence on $\zc$ makes the transition to sphericity relatively slow in a wind-like medium.  As such, significant deviations from spherical symmetry are expected even after the cocoon has expanded to much larger than its original size.

\begin{figure} 
\centering
\includegraphics[width=\columnwidth]{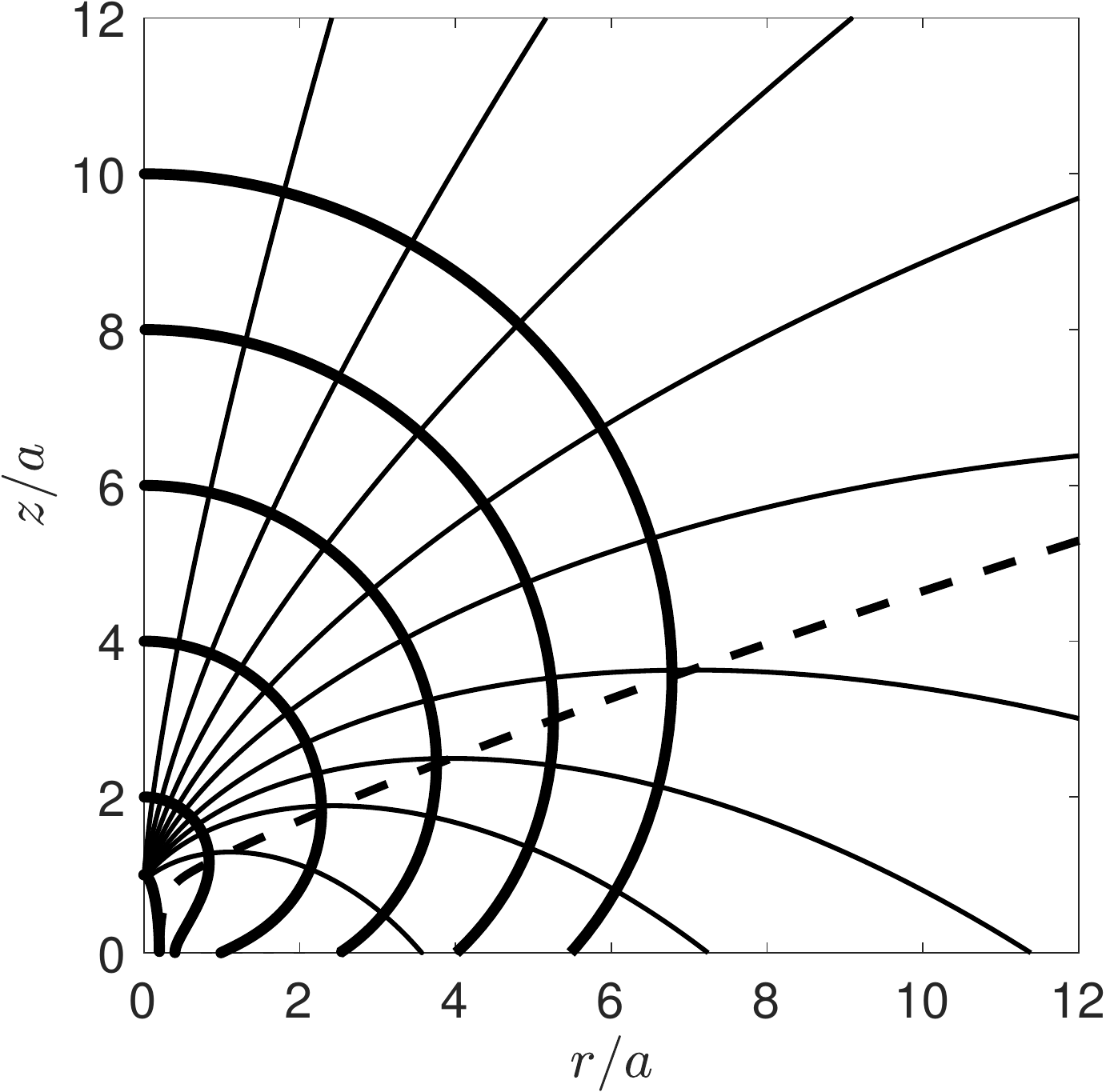}
\caption{As in Fig.~\ref{web0}, but for $\alpha=2$.}
\label{web2}
\end{figure}

Cocoon shapes and particle trajectories for $\alpha=2$ are shown in Fig.~\ref{web2}, using the same values as before for $\zc$ and $b/a$.  The dashed line, which tracks the position of the bulge in space, would become horizontal for a spherical flow.  The extremely gradual change in the slope of this line captures the slow transition to sphericity.  

\subsubsection{A steep external density profile $(2<\alpha < 3)$}
\label{alpha2-3}

When $2<\alpha<3$ ($k_\alpha <0$), the integral~\ref{x} converges, implying that $x$ has a finite value as $t \rightarrow \infty$.   From equation~\ref{z1}, we see that $\zc \propto (1+k_\alpha x)^{1/k_\alpha}$.  Thus, in order to have $\zc \rightarrow \infty$ as $t \rightarrow \infty$, we require $x \rightarrow 1/|k_\alpha|$.  Evaluating both sides of equation~\ref{x} as $t\rightarrow \infty$ gives
\begin{equation}
\label{x3}
\dfrac{1}{|k_\alpha|} = \int_{t_0}^{\infty}\sqrt{\dfrac{(\gamma+1)\Pc(t')}{2\rho_0 a^2}} \deriv t'
\end{equation}
Then, subtracting equation~\ref{x} from equation~\ref{x3} and multiplying through by $|k_\alpha|$, we obtain
\begin{equation}
\label{xscaling}
1 + k_\alpha x = |k_\alpha| \int_t^{\infty}\sqrt{\dfrac{(\gamma+1)\Pc(t')}{2\rho_0 a^2}} \deriv t'.
\end{equation}
Now, if we now choose $t$ to be very large, we must have $\Pc \propto \zc^{-3} \propto (1+k_\alpha x)^{-3/k_\alpha}$ since the flow will be self-similar.  Plugging this into equation~\ref{xscaling}, we find ${(1+k_\alpha x)} \propto t^{(2-\alpha)/(5-\alpha)}$.  Therefore the usual blast wave scaling $\zc \propto t^{2/(5-\alpha)}$ is also recovered for $\alpha >2$.

A peculiar feature of the solution for $\alpha > 2$ is that, while the flow eventually becomes self-similar, it does \textit{not} become spherical.  This is most easily seen from equation~\ref{thetab}.  Recall that as $t \rightarrow \infty$, we have ${\thetabo \rightarrow \thetabomin}$ and $\thetab \rightarrow \thetabinf$.  Since $\thetabomin=0$ in this case, we take $\thetabo \rightarrow 0$ in equation~\ref{thetab} to obtain $\thetabinf = \pi/\alpha < \pi/2$, implying that the bulge stops short of reaching the equator.  As we show in Appendix~\ref{appendixb}, the width and the height in this case are related by
\begin{equation}
\label{ratio3}
\dfrac{\rc}{\zc} \approx f_\alpha \left(1-\dfrac{A_\alpha}{\zeta} \right),
\end{equation}
where $\zeta$ is defined as before, and
\begin{equation}
\label{falpha3}
f_\alpha =  \left[\sin\left(\dfrac{\pi}{\alpha}\right)\right]^{\alpha/(\alpha-2)}.
\end{equation}
In this regime the parameter $A_\alpha$ takes on a different form,
\begin{equation}
\label{Aalpha3}
A_\alpha = \dfrac{1}{2k_\alpha^2} \cot^2(\pi/\alpha).
\end{equation}
In the limit $\alpha \rightarrow 2$ and $k\rightarrow 0$, $A_\alpha \rightarrow \pi^2/8$ and $f_\alpha \rightarrow 1$, so we recover equation~\ref{ratio2}.  The effective eccentricity is now given by
\begin{equation}
\label{ecc3}
\epsiloneff \approx \ (1-f_\alpha^2 +2 f_\alpha^2 A_\alpha/\zeta)^{1/2}.
\end{equation}
Unlike the previous cases, $\epsiloneff$ no longer goes to zero, but instead approaches $(1-f_\alpha^2)^{1/2}$ as $\zc$ becomes large.

Considering equation~\ref{shape1} in the limit $x\rightarrow 1/|k_\alpha|$ leads to an asymptotic expression for the cocoon's shape: as $\zc$ tends to infinity, the curve bounding the cocoon is increasingly well-approximated by
\begin{equation}
\label{shape3}
R(\theta) \approx \zc \left[\cos(k_\alpha \theta)\right]^{-1/k_\alpha}.
\end{equation}
The overall shape of the shock front in the scale-free limit is the same as the shape that would be obtained from setting off two point explosions at $z=\pm a$ (see, e.g., K92).\footnote{Interestingly, the asymptotic shape often takes a familiar mathematical form when $k_\alpha$ is a rational fraction.  For example, when $k_\alpha=-1/2$, the asymptotic shape is a cardioid.}
Setting $\theta=\pi/2$ in equation~\ref{shape3} and comparing with formula~\ref{fdefine}, we find
\begin{equation}
\label{galpha}
g_\alpha =\left[\cos(k_\alpha \pi/2)\right]^{-1/k_\alpha}.
\end{equation}
Whereas $g_\alpha=f_\alpha=1$ for density profiles flatter than $r^{-2}$ because the cocoon ultimately becomes a sphere, steeper density profiles have $g_\alpha < f_\alpha <1$ instead.  Therefore, $\req<\rc<\zc$ at late times; rather than a sphere, the outflow becomes a self-similar `peanut.' 

\begin{figure} 
\centering
\includegraphics[width=\columnwidth]{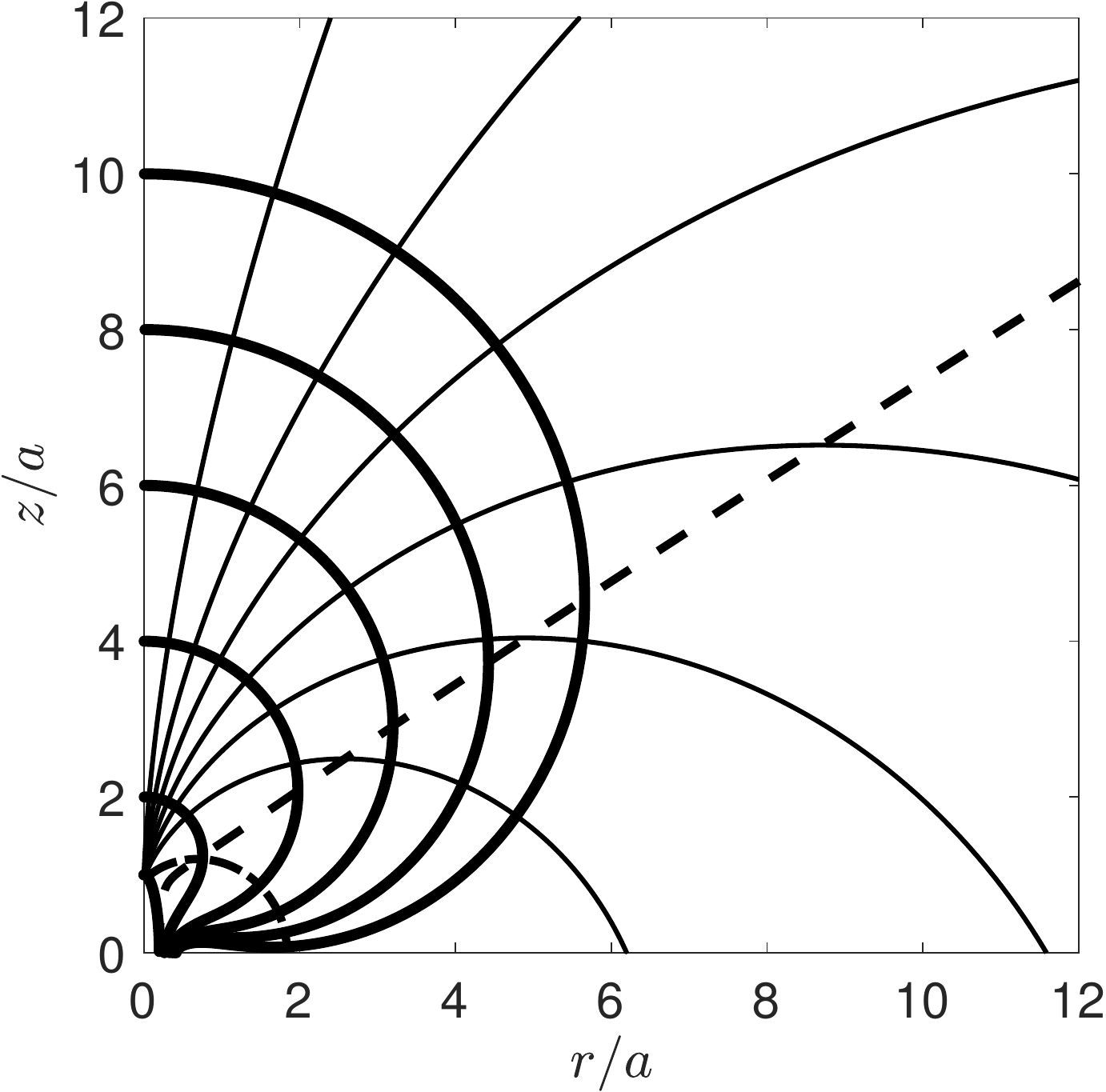}
\caption{As in Fig.~\ref{web0}, but for $\alpha=3$.  The dash-dotted line marks the trajectory of the first particles to reach the equator.}
\label{web3}
\end{figure} 

In Fig.~\ref{web3}, we plot the cocoon's shape and several trajectories for $\alpha=3$.  Compared to the previous cases, the trajectories are more curved.  Also, because of the steep density gradient, the density near the tip is much lower than the density near the base.  As a result, material originating near the axis wraps around and actually reaches the equator before material on the sides.  The path traced out by the bulge (dashed line) straightens out and never becomes parallel to the equator, as expected for an outflow which does not become spherical.

\begin{table*}
\centering
\begin{tabular}{| >{\centering\arraybackslash}m{2cm} || c | c | c | c | >{\centering\arraybackslash}m{2cm} |}
\hline 
\, & $\alpha \le 2b^2/a^2 $ & $2b^2/a^2<\alpha<2 $ & $\alpha=2$ & $2<\alpha<3 $ \\
\, & $(k_\alpha\ge1-b^2/a^2) $ & $(0<k_\alpha<1-b^2/a^2) $ & $ (k_\alpha=0)$ & $(-1/2<k_\alpha<0)$  \\ \hline  
$f_\alpha$ & \multicolumn{3}{  c | }{1} & $\left[\sin(\pi/\alpha)\right]^{\alpha/(\alpha-2)}$ \\ \hline
$A_\alpha$ & \multicolumn{2}{  c |}{$(2/k_\alpha^2) \sin^2(k_\alpha\pi/4)$} & $\pi^2/8$ & $(1/2k_\alpha^2) \cot^2(\pi/\alpha)$ \\ \hline
$\zeta$ & \multicolumn{2}{  c |}{$(1/k_\alpha)\left[(\zc/a)^{k_\alpha}-1\right]$} & $\ln(\zc/a)$ & $(-1/k_\alpha)\left[(\zc/a)^{-k_\alpha}-1\right]$ \\ \hline
bulge forms? & no & \multicolumn{3}{c |}{yes} \\ \hline
asymptotic shape & \multicolumn{3}{ c |}{sphere} & `peanut' \\ \hline
\end{tabular}
\caption{Asymptotic cocoon properties for the regimes discussed in Section~\ref{analytical}.}
\label{table1}
\end{table*}

We emphasize again that the above results are based on an idealized model that ignores heating due to collisions.  The effects of equatorial interactions will tend to make the outflow more spherical than what the analytical model predicts.  Interactions may be particularly important for steeper density profiles, because whereas in flat density profiles the equatorial collisions are glancing, in steep density profiles they can occur head-on (see Appendix~\ref{appendixc}).  This can be seen in Fig.~\ref{web3}: the first ejecta to reach the equator follow the dot-dashed path and collide fully head-on with material from the other side.    The other paths in Fig.~\ref{web3} also intersect the equator at a steep angle $\ga 45$ degrees.   In comparison, the trajectories in the $\alpha=1$ case (Fig.~\ref{web1}) approach the equator at a much shallower angle and interact more weakly.   

As a final note, we point out that for sufficiently steep $\alpha$, there can be a region inside the cocoon which is never shocked in the analytical model.  This is simply a mathematical feature of the idealized KA model, which is not present in numerical simulations (see Section 4), and which most likely does not exist in reality.  In a more realistic model taking interaction into account, we expect that a pressure gradient would develop due to collisions on either side of the unshocked region, which would then drive material into this region and shock it.  In any case, because the unshocked region is engulfed by the rest of the cocoon, its presence or absence does not affect our conclusions about the late-time behaviour. \\

\begin{figure} 
\centering
\includegraphics[width=\columnwidth,keepaspectratio]{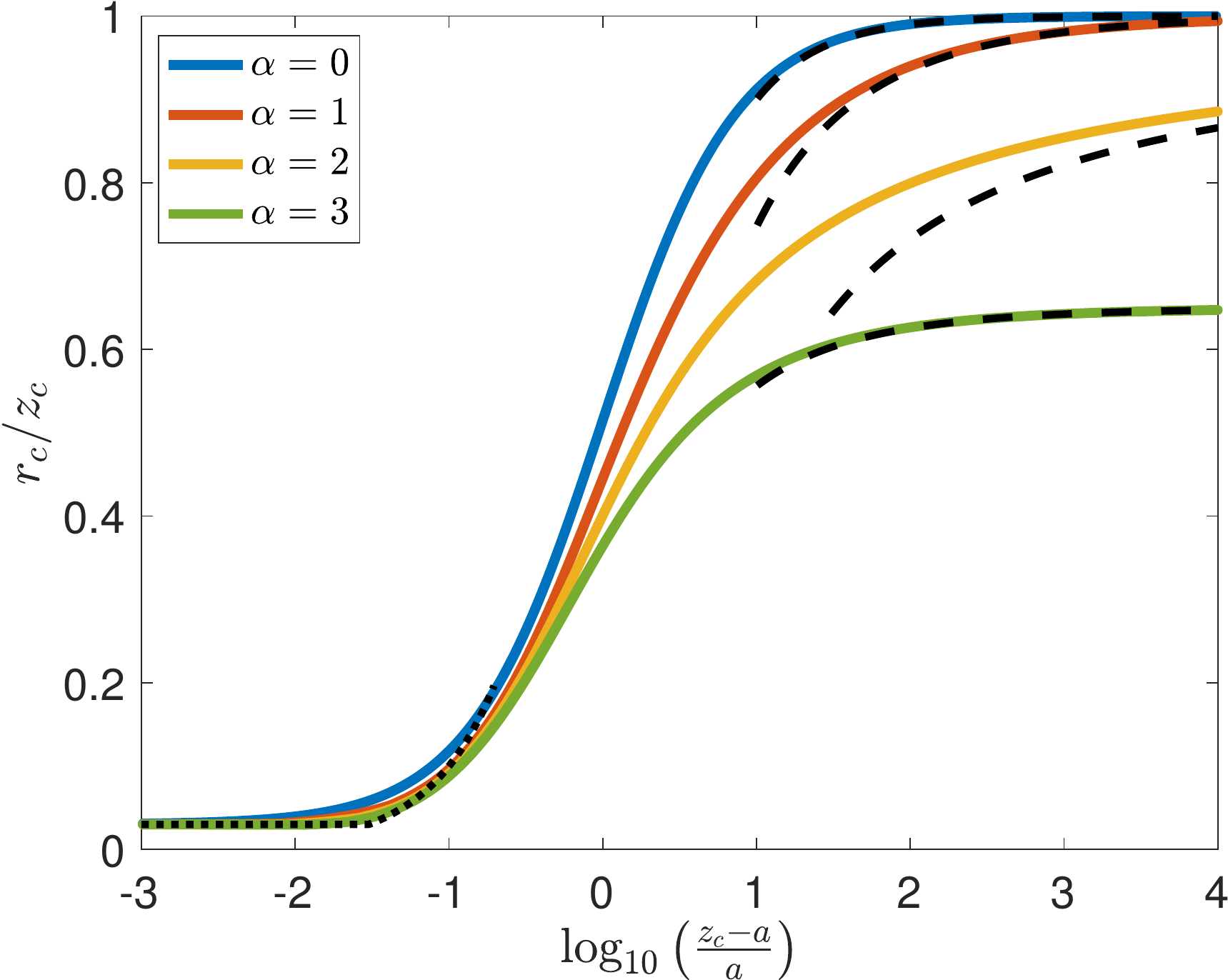}
\caption{The ratio $\rc/\zc$ as a function of cocoon height, for $b/a=0.03$ and $\alpha=0$ ,1, 2, and 3.  The dotted and dashed lines show the approximations given in equation~\ref{r_vs_z}, respectively in the limits $\zc-a \ll a$ and $\zc \gg 2a.$ }
\label{ratio_vs_z}
\end{figure}

Gathering results obtained so far, it is possible to succinctly express the evolution of the cocoon's width versus its height in each dynamical phase.  In the planar phase, the width and height remain roughly constant.  Once the width has doubled ($\rc \gg 2b$), the velocity becomes comparable in the forwards and sideways directions, and the change in width ($\rc-b \approx \rc$) is about the same as the change in height ($\zc-a$).  This approximation remains valid throughout the sideways expansion phase, i.e. while $\zc-a \ll a$.  Finally, in the quasi-spherical phase, the evolution depends on $\alpha$, with the ratio of width to height is given by equation~\ref{ratio1},~\ref{ratio2}, or~\ref{ratio3}, respectively for $\alpha < 2$, $\alpha = 2$, or $\alpha > 2$.  Combining these results, and assuming $b \ll a$, we then have
\begin{equation}
\label{r_vs_z}
\rc\simeq
\begin{cases} 
b, & \zc-a \ll b \\ 
\zc-a, & b \ll \zc-a \ll a \\ 
f_\alpha \zc\left(1-A_\alpha/\zeta\right), & \zc \gg 2a 
\end{cases}.
\end{equation}
Note that the intermediate regime $b \ll \zc-a \ll a$ may not be realized if $b/a$ is not particularly small.  In Table~\ref{table1}, we give the appropriate values of $f_\alpha$, $A_\alpha$, and $\zeta$ in the various regimes, and also summarize some important asymptotic characteristics of the cocoon.  

Fig.~\ref{ratio_vs_z} compares the limiting expressions in~\ref{r_vs_z} with the full solution for each of the values of $\alpha$ discussed above.  We see that when $\alpha$ is not too close to 2, the asymptotic approximation agrees well with the exact solution for $\zc \ga 10a$.  For $\alpha \approx 2$, however, the leading-order approximation given in equation~\ref{r_vs_z} becomes less accurate.  The reason for the slow convergence is that, since $\zeta \propto \ln \zc$ for $\alpha=2$, the contribution from higher-order terms in $\zeta$ is non-negligible.  If greater accuracy is required, the parametric equations presented in Appendix~\ref{appendixa} can be used to calculate $\rc/\zc$ instead. \\ \\

\subsection{The volume of the cocoon}
\label{volume}

In order to determine the expansion speed of the cocoon, or the time that has elapsed since energy deposition, it is necessary to know the cocoon's volume, $\Vc$.  However, due to the complicated shape of the boundary, it is not always possible to obtain an analytical expression for the volume.  To make the problem tractable, we consider the limiting behaviour of $\Vc$ for $t \ll t_a$ and $t \gg t_a$.  Assuming the initial shape is an ellipsoid, the initial volume is $V_0=(4/3)\pi ab^2$.  Subsequently, the volume grows as $\Vc \propto \rc^2 \zc$.  At early times ($t \ll t_a$), when the cocoon is still roughly elliptical, we approximate the volume as 
\begin{equation}
\label{Vplanar}
\Vc \approx V_0 (\zc/a) (\rc/b)^2.
\end{equation}  
At late times ($t \gg t_a$), we have $\rc \propto \zc$ and therefore $\Vc \propto \zc^3$.  In this case we write 
\begin{equation}
\label{Vspherical}
\Vc \approx (4/3) \pi \zc^3 C_\alpha,
\end{equation} 
where the constant of proportionality $C_\alpha$ is defined by
\begin{equation}
\label{Calpha}
C_\alpha \equiv \lim_{t\rightarrow \infty} \dfrac{3\Vc}{4 \pi \zc^3}.
\end{equation}
For $\alpha \le 2$, $C_\alpha=1$ since the outflow becomes spherical, while for steeper density profiles, $C_\alpha$ can be determined by integrating equation~\ref{shape3} over the enclosed volume.  We then have
\begin{equation}
\label{Calpha2}
C_\alpha = 
\begin{cases} 
1, & \alpha \le 2 \\
\int_0^{\pi/2} \left[ \cos(k_\alpha\theta)\right]^{3/|k_\alpha|} \sin \theta d\theta, & \alpha>2 
\end{cases}.
\end{equation}

\begin{figure} 
\centering
\includegraphics[width=\columnwidth,keepaspectratio]{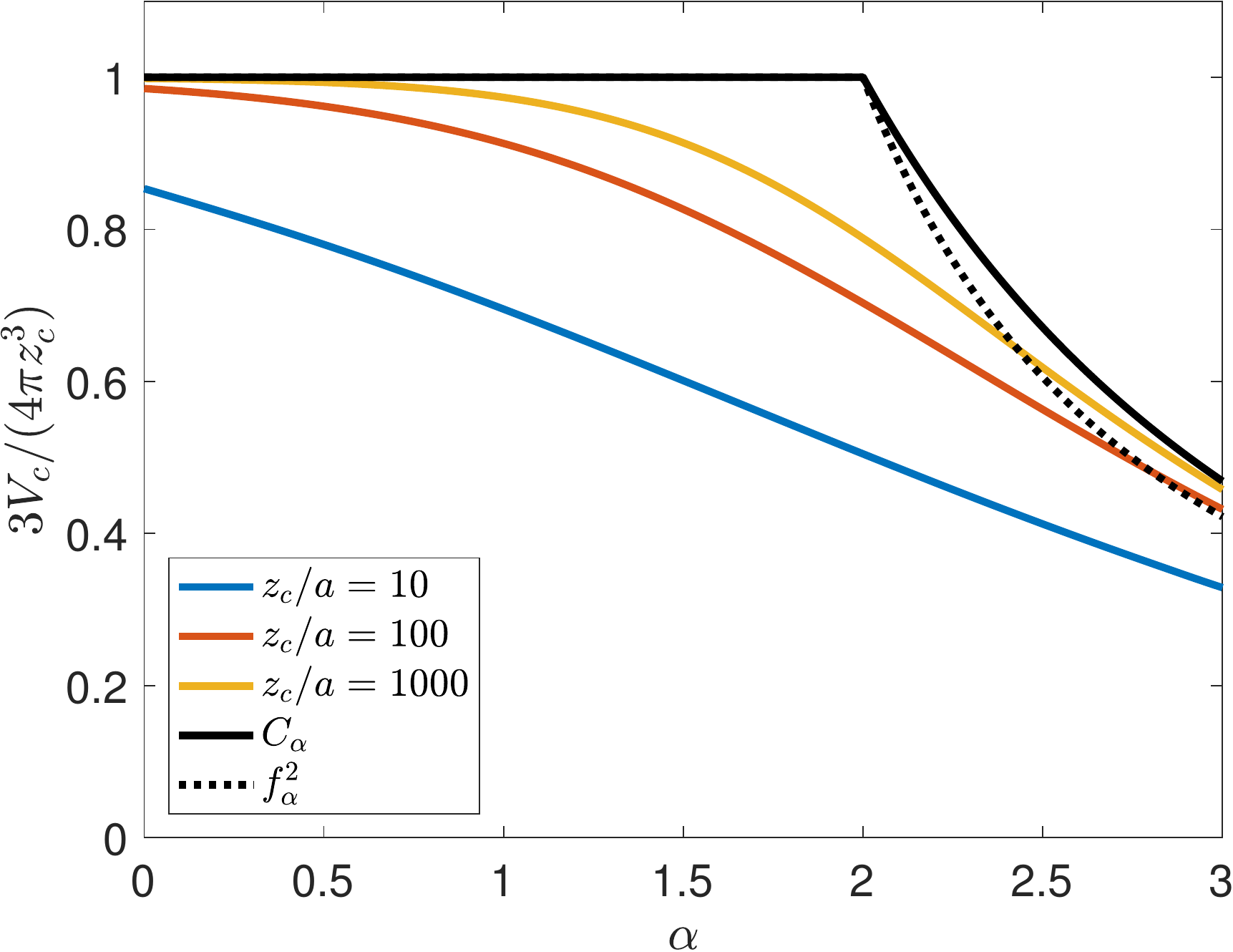}
\caption{The volume of the cocoon scaled to $4 \pi \zc^3/3$, as a function of $\alpha$ for $b/a=0.1$.  The blue, red, and yellow curves depict the volume when the cocoon has expanded to $10a$, $100a$, and $1000a$ respectively. The limiting value as $\zc \rightarrow \infty$ is shown in black: the solid line is the accurate value given by equation~\ref{Calpha}, while the dotted line is the simple approximation $C_\alpha \simeq f_\alpha^2$.}
\label{CV}
\end{figure}

The integral in equation~\ref{Calpha} can be evaluated analytically in limited cases, but it is generally more convenient to evaluate it numerically.  In Fig.~\ref{CV}, we plot $C_\alpha$ as a function of $\alpha$, and compare it to the value of $3\Vc/(4\pi \zc^3)$ found by direct numerical integration of equations~\ref{R1} and \ref{theta1} at $\zc/a=10$, 100, and 1000.  As expected from the results of Section~\ref{alphax}, the $\alpha=2$ case is slowest to converge to the limiting volume, while for $\alpha \la 1$, the asymptotic expression is already a reasonably good approximation for $\zc \ga 10a$.  Additionally, we plot the estimate $C_\alpha \simeq f_\alpha^2$ obtained by treating the cocoon as an ellipsoid with volume $(4/3)\pi \rc^2 \zc$.   Replacing $C_\alpha$ by $f_\alpha^2$ is reasonably accurate, and has the advantage of being easy to compute via eq.~\ref{falpha3}.

\subsection{The age of the cocoon}
\label{convert}

\begin{figure*} 
\centering
\includegraphics[width=\linewidth,keepaspectratio]{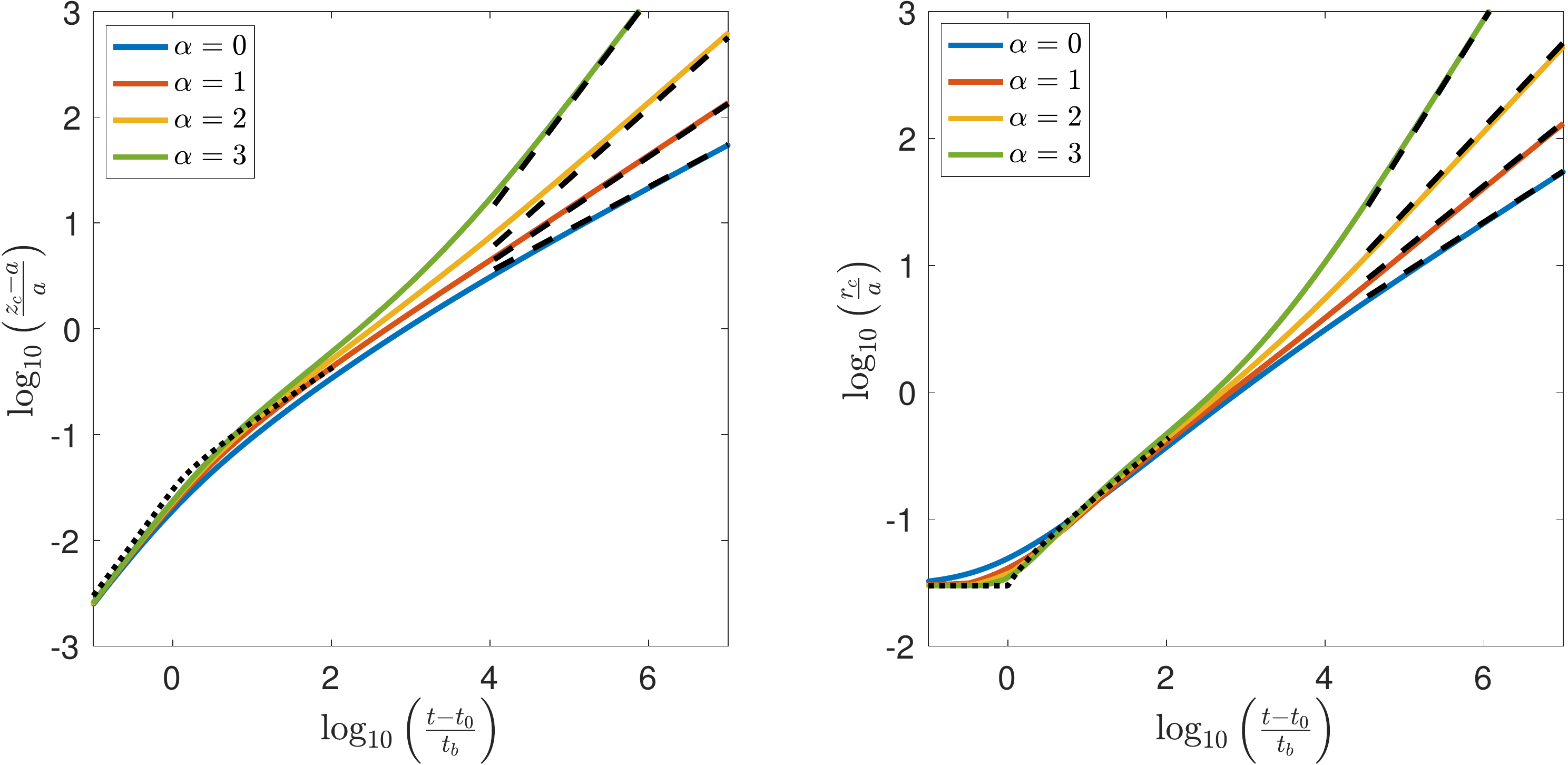}
\caption{The cocoon's height and width versus time, for $b/a=0.03$ and $\alpha=1$, 2, and 3.  The estimates from equations~\ref{z_vs_t} and~\ref{r_vs_t} are overlaid as dotted lines ($t/t_0 \ll a^2/b^2$) and dashed lines ($t/t_0 \gg a^2/b^2$).}
\label{rz_vs_t}
\end{figure*}

We now turn to the question of how $\zc$ and $\rc$ evolve in time.  To start, we rewrite equation~\ref{x} as 
\begin{equation}
\label{t_int}
\dfrac{t-t_0}{t_0} \simeq \dfrac{a}{b} \int_0^x \left(\dfrac{V(x')}{V_0} \right)^{1/2} \deriv x',
\end{equation}
where we used $P(x) V(x)=P_0 V_0$, $x(t_0)=0$, $\gamma=4/3$, and $t_0 \simeq t_b$ as given by equation~\ref{tb}.
Now, we differentiate equation~\ref{z1} to find $dx=(\zc/a)^{k-1} d(\zc/a)$ and change variables in the integral~\ref{t_int}:
\begin{equation}
\label{t_int2}
\dfrac{t-t_0}{t_0} \simeq \dfrac{a}{b}\int_a^{\zc} \left( \dfrac{V(\zc')}{V_0}\right)^{1/2} \left(\dfrac{\zc'}{a}\right)^{k-1} \dfrac{\deriv \zc'}{a}.
\end{equation}
Then, substituting for $\Vc(\zc)$ using equation~\ref{Vplanar} (for $\zc-a \ll a$) or equation~\ref{Vspherical} (for $\zc -a \gg a$), equation~\ref{t_int2} becomes
\begin{equation}
\label{t3}
\dfrac{t-t_0}{t_0} \simeq 
\begin{cases}
\left(\dfrac{a}{b} \right)^2 \int_a^{\zc} \left(\dfrac{\rc(\zc')}{a}\right) \dfrac{\deriv \zc'}{a}, & \zc -a \ll a \\
 \left(\dfrac{a}{b} \right)^2 C_\alpha^{1/2} \int_a^{\zc} \left(\dfrac{\zc'}{a} \right)^{(3-\alpha)/2} \dfrac{\deriv \zc'}{a}, & \zc -a \gg a \\
\end{cases}.
\end{equation}
Finally, we replace $\rc$ using equation~\ref{r_vs_z} and perform the integration to obtain
\begin{equation}
\label{t_vs_z}
\dfrac{t-t_0}{t_0} \simeq 
\begin{cases}
(a/b)(\zc/a-1), & \zc-a \la b \\
\frac{1}{2}+ \frac{1}{2}(a/b)^2(\zc/a-1)^2, & b \la \zc -a \ll a \\
\frac{2}{5-\alpha} C_\alpha^{1/2}  (a/b)^2 (\zc/a)^{(5-\alpha)/2}, & \zc \gg 2a
\end{cases}.
\end{equation}
The leading constant term in the second expression was chosen to ensure a smooth transition at $\zc-a = b$. 

Inverting equation~\ref{t_vs_z}, we find the height as a function of time:
\begin{equation}
\label{z_vs_t}
\zc \simeq 
\begin{cases}
a+b(t/t_0-1), & t \la 2 t_0\\
a+ b\left(2t/t_0-3\right)^{1/2}, & 2 t_0 \la t \ll t_a \\
a \left(\frac{5-\alpha}{2} C_\alpha^{-1/2} \frac{b^2}{a^2} \frac{t}{t_0}\right)^{2/(5-\alpha)}, & t \gg t_a
\end{cases}.
\end{equation}
As expected, the evolution becomes like an ST blast wave at late times with $\zc \propto t^{2/(5-\alpha)}$.    Finally, combining equations~\ref{r_vs_z} and ~\ref{z_vs_t} gives an approximation for the cocoon width,
\begin{equation}
\label{r_vs_t}
\rc \simeq 
\begin{cases}
b , & t - t_0 \ll t_0 \\
b\left(2t/t_0-3\right)^{1/2}, & t_0 \ll t-t_0 \ll t_a \\
a f_\alpha\left(\frac{5-\alpha}{2} C_\alpha^{-1/2} \frac{b^2}{a^2} \frac{t}{t_0}\right)^{2/(5-\alpha)}, & t \gg t_a
\end{cases}.
\end{equation}

In Fig.~\ref{rz_vs_t}, we plot the analytical estimates in equations~\ref{z_vs_t} and \ref{r_vs_t} respectively, compared to the accurate result obtained from numerical integration of the dynamical equations.  We choose a small value of ${b/a=0.03}$ so that the intermediate regime shows up clearly.  In general, the formulae in \ref{z_vs_t} and \ref{r_vs_t} do a good job at reproducing the full solution. 

\begin{figure}
\centering
\includegraphics[width=\columnwidth,keepaspectratio]{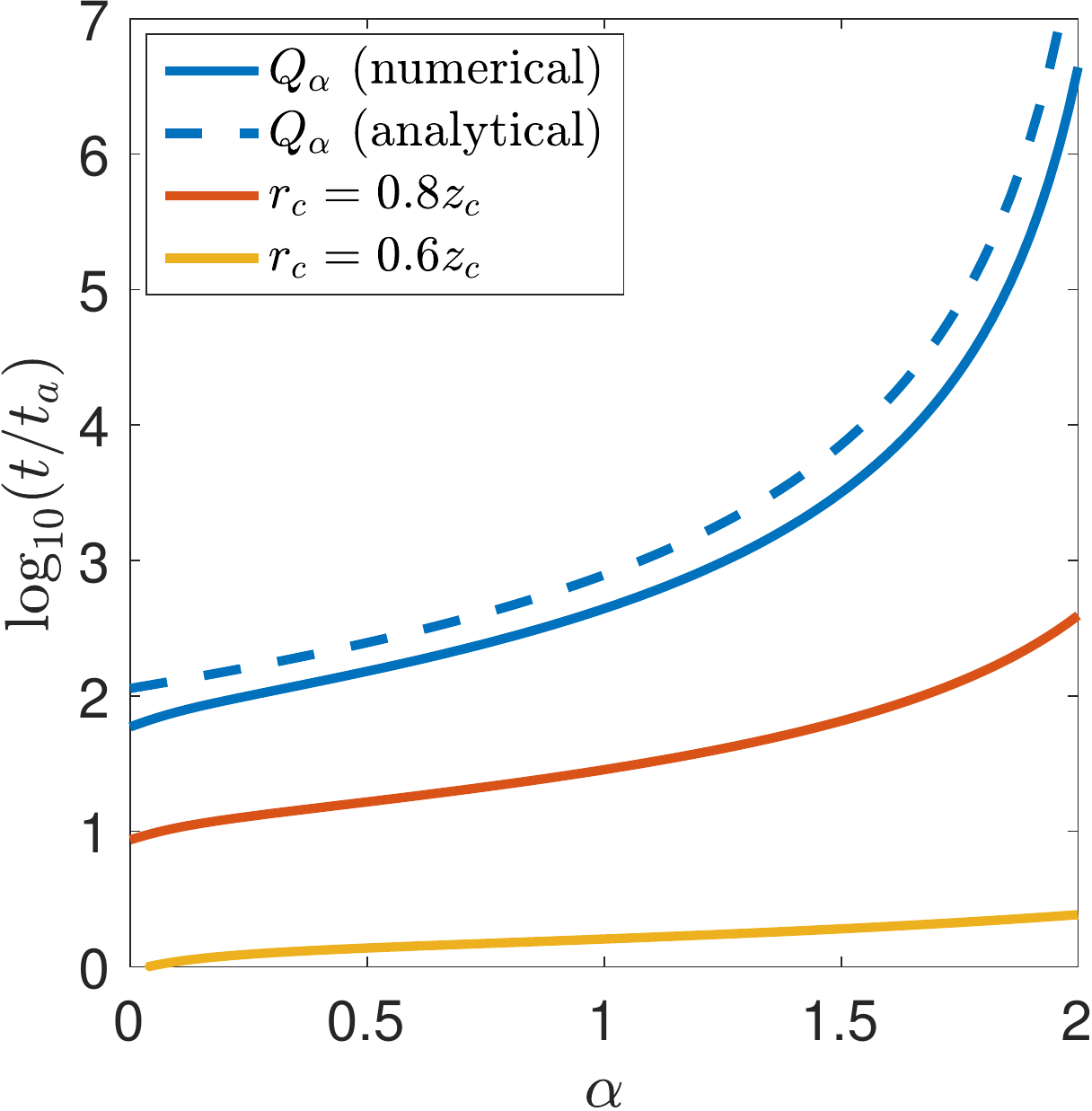}
\caption{The time it takes for the cocoon's width to become 60 (yellow), 80 (red), and 90 (blue) per cent of its height, as a function of $\alpha$, for an initially ellipsoidal cocoon with $a/b=10$.  All times are scaled to the time for the height to double, $t_a \simeq \sqrt{2} (a/b)^2 t_0$. The solid blue line shows the numerically-calculated value of $Q_\alpha$, while the dashed line represents the analytical estimate given by equation~\ref{tsph}. }
\label{Qalpha}
\end{figure}

We conclude this section with an estimate for the time-scale for the cocoon expansion to become spherical.  We define this time-scale, $\tsph$, as the time when the width equals 90 per cent of the height.  Note that $\tsph$ is only defined for $\alpha \le 2$, since for $\alpha >2 $ the outflow does not become spherical (as discussed in Section~\ref{alpha2-3}).  The spherization time can be written as the product of the time-scale for the height to double, $t_a$, and a scaling factor $Q_\alpha$ which depends on the density profile.  Since $t_a \simeq \sqrt{2} (a/b)^2 t_0$ (equations~\ref{tb} and \ref{ta}), we define $Q_\alpha$ via
\begin{equation}
\label{Qdef}
Q_\alpha \equiv (\sqrt{2} a^2/b^2)^{-1} \tsph/t_0 \simeq \tsph/t_a.
\end{equation}
The numerically-computed value of $Q_\alpha$ as a function $\alpha$ is plotted in Fig.~\ref{Qalpha}.  For reference, we also show the times when the width becomes 60 and 80 per cent of the height, as well as an analytical estimate for $Q_\alpha$ (see below).  The value of $Q_\alpha$ grows extremely fast with $\alpha$, with $Q_\alpha = 59$, 441, and $4.4 \times 10^6$ respectively for ${\alpha=0}$, 1, and 2.  Outflows in steep density profiles therefore take a considerably longer time to become spherical.

To obtain an analytical estimate for $Q_\alpha$, we start from equation~\ref{r_vs_z} and observe that $\rc/\zc = 0.9$ holds when ${\zeta \simeq 10 A_\alpha}$.  Using the definition of $\zeta$ (equation~\ref{Z}), we then determine that this occurs at ${\zc \simeq a(1+10 k_\alpha A_\alpha)^{1/k_\alpha}}$ when $\alpha<2$, or $\zc \simeq a{\rm e}^{5\pi^2/4}$ when $\alpha=2$.  Finally, we use equation~\ref{z_vs_t} to translate the limit on $\zc$ into a limit on the age, arriving at
\begin{equation}
\label{tsph}
Q_\alpha \simeq \dfrac{\tsph}{t_a} \simeq
\begin{cases}
\frac{\sqrt{2}}{5-\alpha} (1+10 k_\alpha A_\alpha )^{(5-\alpha)/(2-\alpha)} , & \alpha < 2 \\
(\sqrt{2}/3) {\rm e}^{15 \pi^2/8}, & \alpha = 2
\end{cases}.
\end{equation}
This approximation overestimates the true value of $Q_\alpha$, because equation~\ref{r_vs_z} underestimates the true value of $\rc/\zc$ at late times.  For $\alpha \la 1.5$, the analytical estimate of $Q_\alpha$ is roughly twice the numerically-determined value.  However, the approximation is less accurate when $\alpha$ is close to 2, because the series expansion of $\rc(\zc)$ converges slowly in that case, and we have used only the lowest-order term.  For $\alpha = 2$, equation~\ref{tsph} overestimates $Q_\alpha$ by a factor of 9.  None the less, Fig.~\ref{Qalpha} and equation~\ref{tsph} both clearly demonstrate that $\tsph$ is much larger than $t_a$, especially for $\alpha \sim 2$.  The long-lingering asphericity provides a useful means to constrain the jet properties, as we discuss in Section~\ref{jet}.  

\subsection{Eventual breakdown of the solution}
\label{breakdown}

In this section, we estimate the time at which the KA solution breaks down.  The KA model makes two key assumptions which may eventually become invalid.  First, the shock is assumed to be strong.  In a static medium, this condition holds until the outflow reaches pressure equilibrium with the surrounding medium, on a time-scale $\teq$.  In an expanding medium, on the other hand, the shock becomes weak once the shock velocity becomes comparable to the expansion velocity of the ambient gas, which takes a time $\tw$.  Second, the model assumes that the ambient medium extends to infinity.  If the extent is finite, then the solution is valid only up until the time $\tbo$ when the shock breaks out.  The KA model applies up until $t=\min(\teq,\tw,\tbo)$.

To make for an easy comparison, we will write each of the time-scales as a product of $t_a$ and a dimensionless factor $\eta$.  Also, we neglect order-unity constants which depend on the density profile.  We first consider $\teq$.  Pressure equilibrium is attained once the cocoon pressure, $\Pc$, becomes comparable to the ambient pressure, $P_a$.   We suppose that the ambient pressure obeys 
\begin{equation}
\label{Pa}
P_a = P_{a0} (\rho/\rho_0)^{\gamma_a} = P_{a0} (R/a)^{-\gamma_a \alpha},
\end{equation} 
where $\gamma_a$ is the adiabatic index of the ambient medium.\footnote{We note that pressure equilibrium can only be achieved if the cocoon pressure falls off more steeply than the ambient pressure; this is true only for $\alpha < 3/\gamma_a$.}  Furthermore, we assume that pressure upon choking satisfies $P_0 \gg P_{a0}$ (otherwise, the solution was not valid in the first place). In this case the shock remains strong throughout the planar phase.  

At the time $t_a$, when the outflow transitions from sideways expansion to quasi-spherical expansion, the cocoon pressure is $\Pc(t_a) \sim E_0/(4 \pi a^3)$.  Therefore, there are two possible scenarios: if $P_{a0} \ga \Pc(t_a)$, pressure equilibrium is reached during the sideways expansion phase, while if $P_{a0} \la \Pc(t_a)$, it happens in the quasi-spherical regime.  In the former case, the cocoon's height is given by $\zc \sim a$ and its volume is $\Vc \sim 4 \pi a \rc^2/3$.  The expansion occurs near the tip of the cocoon, where $P_a \sim P_{a0}$.  Pressure equilibrium is attained when the width is $\rc \sim a (E_0/4 \pi P_{a0} a^3)^{1/2}$.  In the latter case, the outflow is roughly spherical, and pressure balance is achieved when the shock's size satisfies $\zc \sim a (E_0/4 \pi P_a(\zc) a^3)^{1/3}$.  Substituting equation~\ref{Pa} for $\Pc(\zc)$ then leads to $\zc \sim a (E_0/4 \pi P_{a0} a^3)^{1/(3-\gamma_a \alpha)}$.  As a final step, we note that age of the system in the sideways expansion phase ($t_b \ll t \ll t_a$) is given by $t \sim t_0 (\rc/b)^2 \sim t_a (\rc/a)^2$ via equation~\ref{r_vs_t}, and the age in the quasi-spherical phase ($t \gg t_a$) is given by $t \sim t_a (\zc/a)^{(5-\alpha)/2}$ via equation~\ref{z_vs_t}.  We then arrive at the pressure equilibrium time-scale
\begin{equation}
\label{teq}
\teq \sim 
\begin{cases}
t_a \etaeq , & b^2/a^2 \ll \etaeq \la 1 \\
t_a \etaeq^{(5-\alpha)/2(3-\gamma_a \alpha)}, & \etaeq \ga 1 \\
\end{cases},
\end{equation}
where
\begin{equation}
\label{etaeq}
\etaeq \equiv \dfrac{E_0}{4\pi P_{a0} a^3}.
\end{equation}
The condition $\etaeq \gg (b/a)^2$ comes from our assumption that $P_0 \gg P_{a0}$.

The above analysis assumes a static medium.  An alternative possibility is that the cocoon propagates in a supersonically expanding medium with an expansion velocity $\vw$ exceeding the ambient sound speed.  Then, once $v \sim \vw$, the shock becomes weak and our solution no longer applies.  Let us assume a steady wind with $\rho \propto r^{-2}$ and $\vw=const$.  We also require that  $v \gg \vw$ initially.  Setting $v \sim (\Pc/\rho)^{1/2} \sim \vw$, the calculation then proceeds as before.  We ultimately arrive at
\begin{equation}
\label{tw}
\tw \sim
\begin{cases}
t_a \etaw , & b^2/a^2 \ll \etaw \la 1 \\
t_a \etaw^{3/2}, & \etaw \ga 1 \\
\end{cases},
\end{equation}
with $\etaw$ given by
\begin{equation}
\label{etaw}
\etaw \equiv \dfrac{E_0}{4\pi \rho_0 \vw^2 a^3} = \dfrac{E_0}{\dot{M} \vw a},
\end{equation}
where we used the relation $\dot{M} = 4 \pi R^2 \vw \rho(R)$ for a steady wind with a mass loss rate of $\dot{M}$.

Finally, we consider the break out of the shock from a density profile extending out to a radius of $R_*$.  The breakout time is sensitive to location where the jet is choked relative to $R_*$.  It is important to note that, due to the rapid sideways expansion and deceleration of a choked outflow, failed jets take considerably longer to break out than successful jets, unless the choking occurs close to the edge.  We suppose that the jet is choked sufficiently far from the edge (i.e., $R_* - a \gg b$), so that the breakout takes place in either the sideways expansion regime (if $b \ll R_*-a \ll a$) or in the quasi-spherical regime (if $R_*-a \gg a$).  Then, taking $\zc=R_*$ in equation~\ref{t_vs_z}, the breakout time is
\begin{equation}
\label{tbo}
\tbo \sim
\begin{cases}
\etabo^2 t_a, &b/a \ll \etabo \la 1 \\
\etabo^{(5-\alpha)/2} t_a, & \etabo \ga 1 \\
\end{cases},
\end{equation}
where
\begin{equation}
\label{etabo}
\etabo \equiv \dfrac{R_*-a}{a}.
\end{equation}

The eventual fate of the outflow depends on the ordering of the time-scales discussed above.  We investigate three astrophysically relevant scenarios: a GRB jet choked inside a WR star (i.e., a failed long GRB); a GRB jet choked within an extended, low-mass stellar envelope (as may be the case in $ll$GRBs); and an AGN jet in a cluster environment choked by the intracluster medium.  In each case, the ambient medium can be treated as static.  Therefore, the relevant time-scales are $t_a$, $\teq$, and $\tbo$.

In a typical long GRB, it takes the jet $\sim 15$ seconds to drill through the progenitor WR star \citep{bromberg12}.  Therefore, we expect jets with a duration of $\sim$ a few seconds to be choked at an appreciable fraction of $R_*$, i.e. we have $a/R_* \sim$ a few tenths.  In this case $\etabo$ is order-unity and $\tbo \sim t_a$.  A WR star can be approximated by an $n=3$ polytrope, with the ambient density and pressure scaling as $\rho \propto R^{-5/2}$ and ${P_a\propto \rho^{4/3} \propto R^{-10/3}}$ in the outer parts of the star.  As $P_a R^3 \propto R^{-1/3}$ changes slowly with radius, the quantity $P_{a0} a^3$ appearing in equation~\ref{etaeq} is roughly constant, regardless of the choking location.  Numerical modelling of pre-supernova stars \citep[e.g.,][]{heger05} suggests a pressure on the order of $2 \times 10^{18}\, \text{erg cm}^{-3}$ at $R=10^{10}$\,cm.  We then have $\etaeq \simeq 80 (E_0/10^{51}\,\text{erg})(P_{a0} a^3/10^{48}\,\text{erg})^{-1} \gg 1$.  Thus, $t_a \sim \tbo \ll \teq$, and the breakout occurs while the KA model is still valid.

The progenitors of $ll$GRBs, on the other hand, may have an extended optically thick envelope containing a mass $\sim 10^{-2} M_{\odot}$ within a radius $\sim 100-1000 R_{\odot}$ \citep{campana06,nakar15,ic16}.  The pressure in this envelope is unknown, but it must be smaller than in the WR star case, since the envelope is both cooler and less dense.  We then expect $\etaeq \ga 100$, as before.  However, because of the larger radius compared to the WR star case, a typical GRB jet is choked deep within the extended envelope, with $a/R_* \ll 1$.  We therefore have $\etabo \gg 1$ and a quasi-spherical breakout occurs at $\tbo \gg t_a$.  Assuming that the jet successfully escapes the star, $a$ is at least $\sim 10 R_{\odot}$, so $\etabo$ is at most $\sim 100$.  We then find $t_{a} \ll \tbo < \teq$, so in this case the KA model also applies up until breakout.

In AGNs, bubbles with a size of $\sim 10$\,kpc are sometimes observed in the centre of galaxy clusters \citep[e.g.,][]{birzan04,diehl08}.  If these bubbles were inflated by jets, the jets must have been choked at $a < 10$\,kpc.  However, the dense cores in clusters extend to much larger radii of ${R_* \sim 100}$\,kpc, so we have $a/R_* \la 0.1$ and $\etabo \ga 10$.  Thus, like in the $ll$GRB case, the breakout time is $\tbo \gg t_a$.  The difference is that, compared to the GRB case, the ambient pressure in an AGN environment is much more significant.  Taking a typical jet energy $E_0 \sim 10^{59}$\, erg and a typical ambient pressure of $P_a(10\,\text{kpc})\sim 0.1$\,keV\,cm$^{-3}$ at $R=10$\,kpc, we obtain  $\etaeq \ga 1.6 (E_0/10^{59}\,\text{erg})[P_a(10\,kpc)/0.1\,\text{keV\,cm}^{-3}]^{-1}$ for $a \la 10$\,kpc.  Assuming a flat ($\alpha \simeq 1$) and isothermal ($\gamma_a=1$) ambient medium, the lower limits on $\etabo$ and $\etaeq$ imply $\tbo \ga 100 t_a$ and $\teq \ga 2 t_a$, respectively.  Furthermore, in the isothermal case, the ratio $\tbo/\teq$ is independent of $a$.  Dividing equation~\ref{tbo} by equation~\ref{teq} and applying $P_{a0} = P(10\,\text{kpc})(a/10\,\text{kpc})^{-1}$, we obtain
\begin{equation}
\dfrac{\tbo}{\teq} \sim 60 \left(\dfrac{E_0}{10^{59}\,\text{erg}}\right)^{-1} \left(\dfrac{R_*}{100\,\text{kpc}}\right)^2 \left(\dfrac{P_a(10\,\text{kpc})}{0.1\,\text{keV\,cm}^{-3}}\right).
\end{equation}
Therefore, AGN jets choked in the core of galaxy clusters satisfy $t_a \la \teq \ll \tbo$.  Pressure equilibrium is achieved early in the quasi-spherical phase, and the KA approximation breaks down long before the flow crosses the cluster core.

In AGNs, buoyancy will also become important at late times.  The effects of buoyancy are discussed in more detail in a separate paper (Irwin et al., in prep.), where we apply the model to young bubbles inflated by jets in galaxy clusters.  Here, we simply remark that if the ambient medium is in hydrostatic balance, and the size of the bubbles is comparable to their distance from the central source, the buoyancy time-scale $t_{buoy}$ is comparable to $\teq$ \citep[][see also Irwin et al., in prep]{churazov01}.

\section{Inferring the jet parameters from the cocoon properties}
\label{jet}

For a cocoon formed by a choked jet expanding with a non-relativistic velocity, the initial conditions of the cocoon ($a$, $b$, and $E_0$) are related to the parameters of the jet (the luminosity $\Lj$, the duration $\tj$, and the initial opening angle $\thetaop$) in a straightforward way.  In this section, we consider how to work backwards from the observed characteristics of the cocoon to obtain useful constraints on the jet properties.  We will assume that it is possible to estimate the cocoon's height ($\zc$), width ($\rc$), and energy ($E_0$) from observations, and additionally that the density profile is known.  The calculation then proceeds in two steps.  First, we use the KA model to estimate the age of the system and determine the height and width upon choking ($a$ and $b$).  Then, we apply the jet propagation model of \citet{b11} to find the jet properties needed to match this initial shape.    Throughout this discussion, we ignore order-unity prefactors for simplicity.  

We caution that our analysis applies specifically to outflows which are initially narrow with $b \ll a$, as expected when the cocoon is inflated by a relativistic jet.  When the observed aspect ratio is $\zc/\rc \gg 1$, the association with a jet is not controversial.  However, when $\zc/\rc$ is of order unity, the alternative possibility that the outflow was powered by a wide-angle wind cannot be ruled out.  

The first step is to determine which dynamical phase is appropriate.  Roughly speaking, the outflow is in the quasi-spherical phase when $\rc/\zc \ga 0.5$, and in the planar or sideways expansion phase when $\rc/\zc \la 0.5$.  From equations~\ref{r_vs_t} and~\ref{z_vs_t}, along with the definitions of $t_b$ and $t_a$, we deduce that the age in each case is approximately
\begin{equation}
\label{tage}
t \sim 
\begin{cases}
\left(\dfrac{\rho(\zc) \zc \rc^4}{E_0} \right)^{1/2}, & \zc/\rc - 1 \gg 1 \\
\left(\dfrac{\rho(\zc) \zc^5}{E_0}\right)^{1/2}, & \zc/\rc - 1 \ll 1 \\
\end{cases},
\end{equation}
where $\rho(\zc)$ is the observed density at $\zc$.

Next, we wish to estimate $a$ and $b$.  Because the outflow is slow to become spherical (see Section~\ref{convert}), the dependence of the aspect ratio on $a$ persists until well after $t_a$.  The observed value of $\rc/\zc$ is then directly related to the ratio $\zc/a$, as shown in Fig.~\ref{ratio_vs_z}.  In the quasi-spherical case there is also a significant dependence on the density profile.  An analytical approximation for $a$ is obtained by inverting equation~\ref{r_vs_z}:
\begin{equation}
\label{a}
a \simeq
\begin{cases}
\zc-\rc, & \zc/\rc - 1 \gg 1 \\
\zc\left(1+\dfrac{A_\alpha |k_\alpha|}{1-\rc/\zc}\right)^{-1/|k_\alpha|}, & \zc/\rc - 1 \ll 1 \\
\end{cases}.
\end{equation}
For elongated outflows in the planar or sideways expansion phase, we find that the choking occurred closer to the edge of the outflow than the centre, while in quasi-spherical outflows the opposite holds true.

The initial width, however, is more difficult to determine.  In the quasi-spherical regime, the shape has little to no dependence on the initial width, so $b$ cannot be recovered.  In the sideways expansion phase, there is a weak dependence on $b$, but there is also another issue, which is that the sideways expansion and planar cases are difficult to distinguish observationally.  One indication that the cocoon is in the sideways expansion phase is a bulging out towards the tip, as seen in Fig.~\ref{comparison}. However, without knowing the initial shape, this method is unreliable, and moreover it only applies when the density profile is fairly steep.  With good observations, a better way to test which case is relevant is to compare the strength of the shock along the axis to the strength near the widest part of the cocoon.  If the shock is significantly stronger towards the axis, it would be good evidence for a recently quenched jet, suggesting a planar evolution with $t \sim t_0$ and $b \sim \rc$.

Statistically speaking, though, it is much more likely to catch the system during the sideways expansion phase, since it lasts roughly $(a/b)^2$ times longer than the planar phase.  In this case it is hard to tell if the jet was choked somewhat recently, with a width comparable to the observed width, or if it was choked farther in the past, but was initially narrower.  All we can say for certain is that $b < \rc$ and $t_0 < t$.  For a known choking radius $a$ given by equation~\ref{a}, the degeneracy between the initial width and the choking time is captured by
\begin{equation}
\label{degen}
t_0 b^{-2} \sim \left(\dfrac{\rho_0 a}{E_0}\right)^{1/2} = \left(\dfrac{\rho(\zc) \zc}{E_0}\right)^{1/2} \left(\dfrac{a}{\zc}\right)^{(1-\alpha)/2},
\end{equation}
where in the second equality we used $\rho_0 a^\alpha = \rho(\zc) \zc^\alpha$ to replace $\rho_0$ by observable quantities.

Now that we have established how $a$ and $b$ depend on the observables, let us consider how they relate to the jet properties.  Suppose the jet is injected relativistically with luminosity $\Lj$ and opening angle $\thetaop$, and operates for a duration $\tj$.  \citet{b11} showed that the dynamics of the jet are governed by a dimensionless quantity $\tilde{L}$.  At the time $t=t_0$ when the jet is choked, $\tilde{L} = \text{max}\{(\frac{\Lj}{\rho_0 t_0^2 \thetaop^4 c^5})^{2/5}, \frac{\Lj}{\rho_0 t_0^2 \thetaop^2 c^5} \}$.  The jet is collimated with a non-relativistic head when $\tilde{L} \ll 1$; collimated with a relativistic head when $1 \ll \tilde{L} \ll \thetaop^{-4/3}$; and uncollimated with a relativistic head when $\tilde{L} \gg \thetaop^{-4/3}$.   $\tilde{L}$ is closely related to the head velocity $\betah$, with $\betah \approx \tilde{L}^{1/2} $ for a non-relativistic jet, and $2 \Gamma_h^2=\tilde{L}^{1/2}$ for a relativistic head, where $\Gamma_h \approx (1-\betah^2)^{-1/2}$.  

The case of an uncollimated jet can be further subdivided depending on the relative values of $\tilde{L}$ and $\thetaop^{-4}$.  If $\tilde{L} \ll 4 \thetaop^{-4}$, the jet and cocoon are in causal contact and the pressure in the cocoon remains more or less uniform.  However, if $\tilde{L} \gg 4 \thetaop^{-4}$, causal connection is lost and the approximation of uniform pressure breaks down.  We therefore restrict our discussion to the $\tilde{L} \ll 4 \thetaop^{-4}$ case in which the KA remains valid.

After the jet is switched off at $\tj$, how long does it take to be choked?  Material continues to flow into the cocoon until $t_0$, when the last bit of material launched by the jet catches up to the head.  For a non-relativistic head, this time is negligible compared to $\tj$, and therefore $t_0 \simeq \tj$.   If the head is relativistic instead, choking takes longer, and $t_0 = \tj/(1-\betah) \approx 2 \Gamma_h^2 \tj$ \citep{nakar15}.  Thus we have
\begin{equation}
\label{t_jet}
t_0 \simeq \begin{cases}
\tj, & \tilde{L} \ll 1 \\
\tilde{L}^{1/2} \tj , & \tilde{L} \gg 1
\end{cases}.
\end{equation}  
The distance the head travels in time $t_0$ is $\betah c \tj$ in the former case, and $2 \Gamma_h^2 c \tj$ in the latter; in either regime, we can write 
\begin{equation}
\label{a_jet}
a \simeq \tilde{L}^{1/2} c \tj.
\end{equation}
Then, estimating the cocoon width at the time of choking as ${b \simeq (\betac/\betah) a}$, where the sideways expansion speed is $\betac \simeq \tilde{L}^{1/2} \thetaop$ ($\tilde{L} \ll \thetaop^{-4/3}$), or $\betac \simeq \tilde{L}^{1/8} \thetaop^{1/2}$ ($\thetaop^{-4/3} \ll \tilde{L} \ll 4 \thetaop^{-4}$)\citep{b11}, we arrive at
\begin{equation}
\label{b_jet}
b \simeq \begin{cases}
\tilde{L}^{1/2} \thetaop c \tj , & \tilde{L} \ll 1 \\
\tilde{L} \thetaop c \tj , & 1 \ll \tilde{L} \ll \thetaop^{-4/3} \\
\tilde{L}^{5/8} \thetaop^{1/2} c \tj , & \thetaop^{-4/3} \ll \tilde{L} \ll 4\thetaop^{-4}
\end{cases}.
\end{equation}

We now have three equations (\ref{t_jet}, \ref{a_jet}, and \ref{b_jet}) relating the initial conditions of the cocoon ($a$, $b$, and $t_0$) to three quantities describing the jet ($\tilde{L}$, $\tj$, and $\thetaop$).  When the choking radius satisfies $a \ll c t_0$, we find that the head was non-relativistic, with $\tilde{L}$ given by 
\begin{equation}
\label{ltilde_nr} 
\tilde{L} \simeq \left(\dfrac{a}{ct_0}\right)^2 \simeq \dfrac{E_0a}{\rho_0 c^2 b^4} \ll 1.
\end{equation}
In this case, the three jet parameters are uniquely determined by the cocoon properties:
\begin{equation}
\label{nr_coll}
\begin{array}{lll}
\Lj & \simeq  E_0/t_0 & \simeq   E_0^{3/2} \rho_0^{-1/2} a^{-1/2} b^{-2} \\
\tj & \simeq   t_0 & \simeq  E_0^{-1/2} \rho_0^{1/2} a^{1/2} b^{2} \\
\thetaop &  \simeq   b/a & \,  \,\\
\end{array}.
\end{equation}

When the jet head is relativistic, however, a degeneracy arises because $a$ and $t_0$ are related trivially by $a=ct_0$.  In this case only two of the cocoon parameters are independent and $\tilde{L}$ is not uniquely determined by the cocoon properties.  Instead, $\tilde{L}$ and $\thetaop$ follow a closure relation given by
\begin{equation}
\label{ltilde_r}
b/a \simeq
\begin{cases}
\tilde{L}^{1/2} \thetaop, & 1 \ll \tilde{L} \ll \theta^{-4/3} \\
\tilde{L}^{1/8} \thetaop^{1/2}, & \theta^{-4/3} \ll \tilde{L} \ll 4\theta^{-4} \\
\end{cases}.
\end{equation}
None the less, the relationship between the jet parameters can still be constrained.  In either case, we have the same constraint on the total injected energy as before, i.e. $\Lj \tj \simeq E_0$.  For a collimated jet with a relativistic head ($1 \ll \tilde{L} \ll \theta^{-4/3}$), we obtain the additional constraint
\begin{equation}
\label{rel_coll}
\tj \thetaop^{-1}  \simeq (a/b) t_0 \simeq E_0^{-1/2} \rho_0^{1/2} a^{3/2} b,
\end{equation}
while for an uncollimated jet in causal contact with its cocoon ($\theta^{-4/3} \ll \tilde{L} \ll 4\theta^{-4}$), we find
\begin{equation}
\label{rel_uncoll}
\tj \thetaop^{-2}  \simeq (a/b)^4 t_0 \simeq E_0^{-1/2} \rho_0^{1/2} a^{9/2} b^{-2}.
\end{equation}

In both GRBs and AGNs, the usual case is that the jet head is Newtonian \citep[see, for example, the discussion in][]{b11}.  The jet properties are then given by equation~\ref{nr_coll}.  However, a precise calculation requires knowledge of the initial width, which may be difficult to extract from observations for the reasons discussed above.   In cases where $E_0$ and $a$ are known, but $b$ is unconstrained, we can only learn about the quantities $\Lj \tj$ and $\tj \thetaop^{-2}$, which are independent of $b$.  From equation~\ref{nr_coll} we have $\Lj \tj = E_0$ (as expected), and $\tj \thetaop^{-2} \simeq (a/b)^2 t_0 \simeq t_a$.  We then find that $\tj \thetaop^{-2}$ satisfies a relation resembling equation~\ref{degen}:
\begin{equation}
\label{degen2}
\tj \thetaop^{-2} \sim a^2 (t_0 b^{-2}) = \left(\dfrac{\rho(\zc) \zc^5}{E_0}\right)^{1/2} \left(\dfrac{a}{\zc}\right)^{(5-\alpha)/2}.
\end{equation}
The reason this degeneracy comes about is that the choking radius scales as $a \propto \tilde{L}^{1/2} \tj$ via equation~\ref{a_jet}.  In the non-relativistic regime, this becomes $a  \propto E_0^{1/5} (\tj/\thetaop^2)^{1/5}$.  Therefore, two jets with the same energy and the same $\tj \thetaop^{-2}$ are choked at the same location.  Even if the two jets had different widths upon choking, after several $t_0$ the initial width becomes unimportant, and the resulting outflows look about the same.

To conclude this section, we consider how the jet properties influence the late-time evolution.  The characteristic time-scale for the cocoon's height to double is related to the jet parameters by
\begin{equation}
\label{td} 
t_a \simeq (a/b)^2 t_0 \simeq
\begin{cases}
\thetaop^{-2} \tj, & \tilde{L} \ll 1 \\
\tilde{L}^{-1/2} \thetaop^{-2} \tj, & 1 \ll \tilde{L} \ll \thetaop^{-4/3} \\
\tilde{L}^{1/4} \thetaop^{-1} \tj, & \thetaop^{-4/3} \ll \tilde{L} \ll 4\thetaop^{-4}
\end{cases}.
\end{equation}
Unsurprisingly, $t_a$ is larger for narrow or long-lived jets. Interestingly, however, for jets of a given opening angle and duration, $t_a$ has a minimum with respect to $\tilde{L}$, with $\min(t_a) \simeq \thetaop^{-4/3} \tj$ occurring at $\tilde{L} \simeq \thetaop^{-4/3}$. This suggests that barely collimated jets become spherical on the shortest time-scale compared to the jet working time, which makes sense given that the time to become spherical scales as $\tsph \propto (a/b)^2 t_0$.  As $\tilde{L}$ increases, $a/b$ decreases while $t_0$ grows.  A tightly collimated jet is easy to suffocate, but takes a longer time to spherize because it leaves behind a narrow cocoon.  On the other hand, a powerful uncollimated jet produces a wider cocoon, but takes a longer time to choke in the first place because the jet head is highly relativistic.  Mildly collimated jets balance these two effects to minimize the spherization time.  The radius where the flow becomes spherical, however, is $\propto a \propto \tilde{L}^{1/2} \tj$ and is always larger for more powerful jets.

\section{Comparison with Numerical Simulations}
\label{numerical}

To check the validity of the analytical results, we compared them to numerical simulations carried out by a collaborator, O. Gottlieb, using the publicly available hydrodynamics code PLUTO \citep{mignone07}.  The axisymmetric simulation setup is akin to previous studies of GRB jet propagation \citep[e.g., ][]{gottlieb18,harrison18}.  The jet is injected with a two-sided luminosity of $10^{51}$\,erg\,s$^{-1}$ and a Lorentz factor of 5 into a nozzle of width $8\times10^7$\,cm.   An adiabatic index of 4/3 is used in all cases.  The lower boundary of the simulation is placed at $z_{inj}=10^9$\,cm, and the external density is defined by $\rho(R)=6\times10^4 (R/z_{inj})^{-\alpha}$\,g\,cm$^{-3}$.  The ambient pressure is set to a negligibly low value. 

\begin{figure*} 
\centering
\includegraphics[width=2.1\columnwidth]{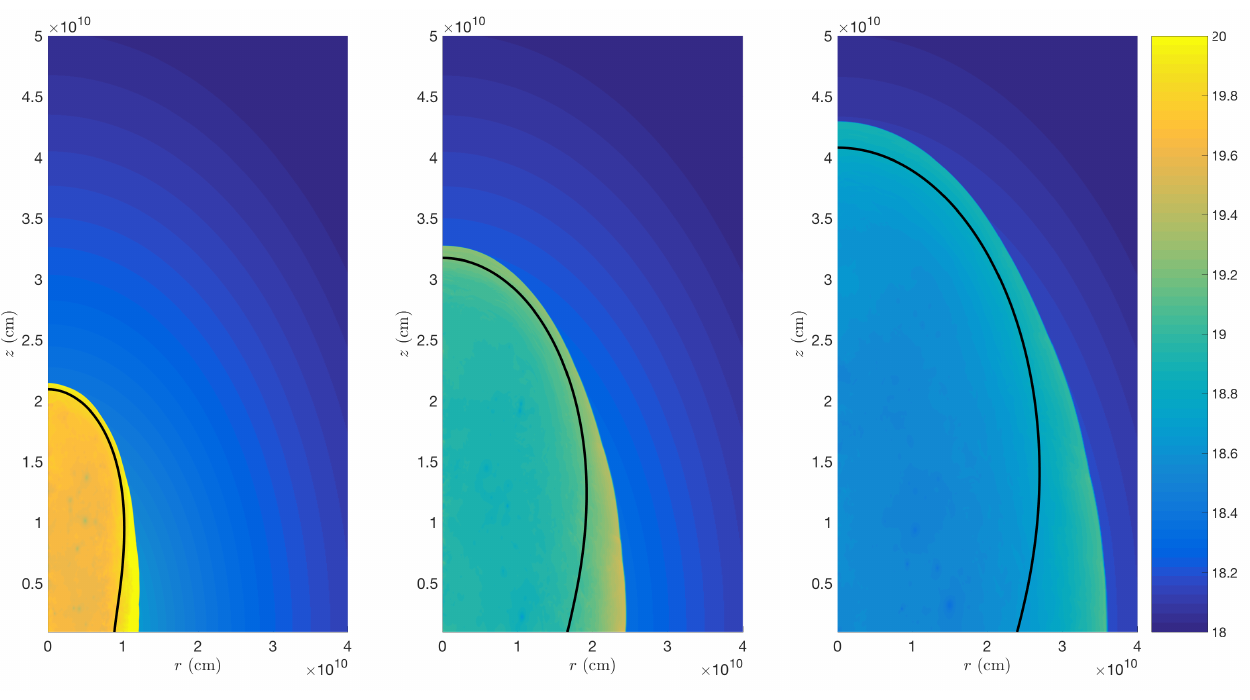}
\caption{PLUTO simulations of a choked relativistic jet in a $\rho \propto R^{-1}$ density profile.  Snapshots are shown at $t=36$, 116, and 220\,s, when the outflow has reached heights of roughly 2, 3, and $4 \times 10^{10}$\,cm.  The colour scale indicates log pressure in cgs units.  The analytical solution obtained from the KA is shown for comparison as a heavy black line.}
\label{sim_alpha1}
\end{figure*}

\begin{figure*}
\centering
\includegraphics[width=2.1\columnwidth,keepaspectratio]{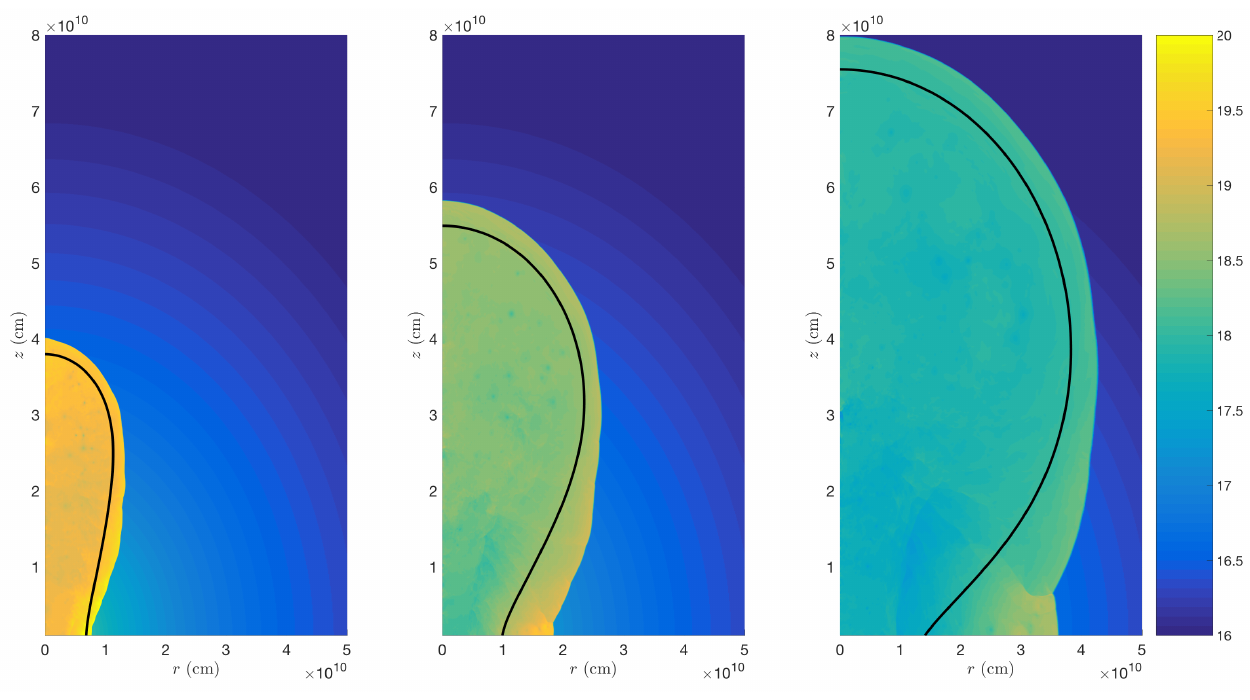}
\caption{As in Fig.~\ref{sim_alpha1}, except for $\rho \propto R^{-2}$ at $t=13$, 37, and 81\,s.}
\label{sim_alpha2}
\end{figure*}

\begin{figure*} 
\centering
\includegraphics[width=2.1\columnwidth,keepaspectratio]{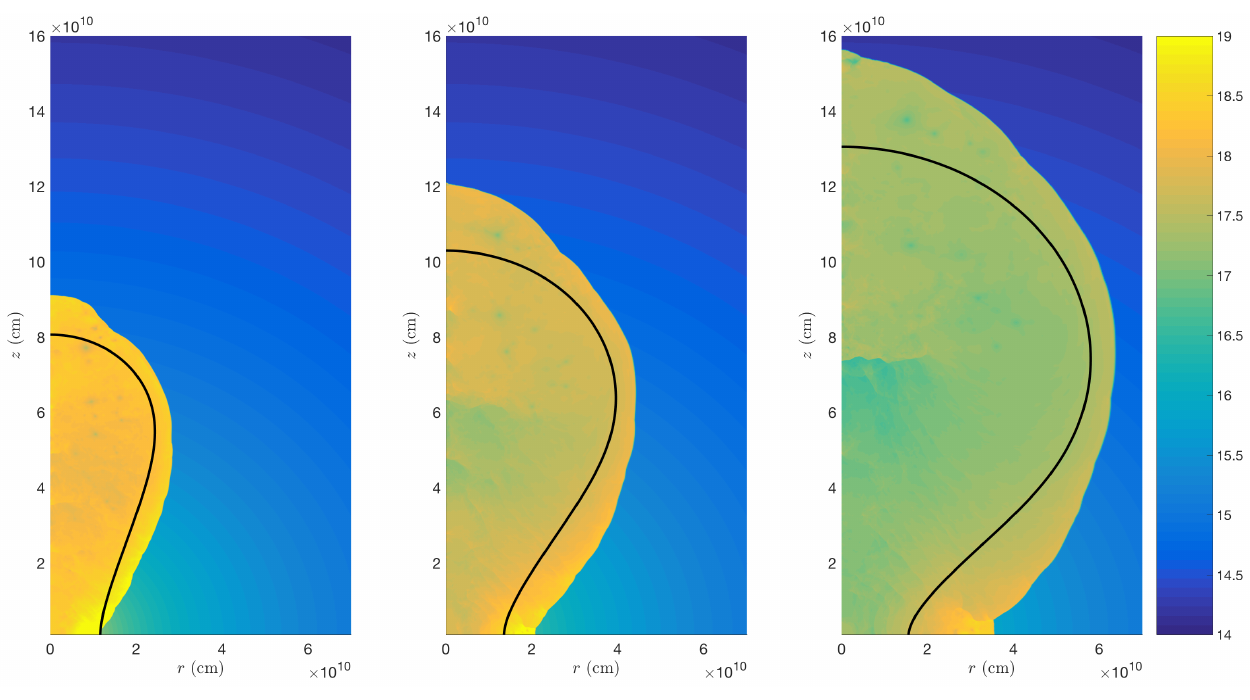}
\caption{As in Fig.~\ref{sim_alpha1}, but for $\rho \propto R^{-2.5}$ at $t=16$, 30, and 52\,s.}
\label{sim_alpha25}
\end{figure*}

We compared our results with three simulations, for three values of $\alpha$ relevant to astrophysical jets: $\alpha=1$, as is typical for the dense core of a galaxy cluster; $\alpha=2$, as expected for a dense circumstellar wind; and $\alpha=2.5$, as appropriate in the outer parts of a WR star.  Tracer particles are included in the unshocked jet to determine when all of the jet material has flowed into the cocoon. Once all the jet has been shocked, the maximum speed of the tracers drops.  In each simulation, we locate the time step where this drop occurs and call this the choking time.   We then find the height, width, and volume-averaged pressure of the outflow at the choking time, and use these as initial conditions for the analytical model.  From there we evolve the analytical cocoon model according to Section~\ref{analytical}, assuming a nominal value of $\lambda=1$ in equation~\ref{P}.

The simulation results are shown in figures~\ref{sim_alpha1}, \ref{sim_alpha2}, and \ref{sim_alpha25}, respectively for $\alpha=1$, 2, and 2.5.  The analytical solution obtained from the KA at each time is also shown in a heavy black line.  In the $\alpha=1$ and $\alpha=2$ cases, there is good agreement (within 10--15 per cent) between the analytical and numerical results out to about 45 degrees from the axis.  The slight disagreement is due to the fact that the pressure behind the shock in the simulations is slightly larger than assumed by the KA.  The shapes can be made to agree by adjusting the parameter $\lambda$, with $\lambda\approx 1.2$ giving an improved fit.

Beyond 45 degrees, the shapes start to deviate, and as expected the agreement becomes worse near the equator.  Typically, the analytical model underestimates the equatorial radius by a factor of 2--3.  As discussed above, one possible source of the discrepancy at the equator is the increased pressure due to collisions.  However, another reason that the equatorial radius is larger in the simulations is that the analytical model assumes, following the KA, that the \textit{initial} pressure is uniform.  In the simulations, this is not the case; rather, there is a clear gradient in pressure and velocity at the time of choking, with both being larger towards the equator.  One possible reason why the shock might be stronger at large angles is that the ejecta near the equator were shocked at earlier times when the sideways expansion was faster and the density was higher.  An alternative possibility is that some interaction at the equator already took place while the jet was active.  

In the $\alpha=2.5$ case, there are also discrepancies near the equator, for the same reasons discussed above.  In addition, the extent along the axis is somewhat larger in the numerical simulation compared to before.  This comes about because the analytical model assumes that the outflow has already decelerated to a steady state where the velocity is given by $\sim (P/\rho)^{1/2}$, or in other words, that the ram pressure of the jet suddenly becomes unimportant at the choking time.  It seems that this assumption is reasonable when $\alpha = 1$ or 2, but not for $\alpha=2.5$.  This makes sense considering that, prior to choking, the jet head is accelerating in the $\alpha=2.5$ case, whereas in the other cases it is is decelerating ($\alpha=1$) or has a constant velocity ($\alpha=2$).  Therefore, for $\alpha=2.5$, it takes longer for the on-axis ejecta to decelerate and approach the KA prediction.    In fact, a shock is clearly visible in Fig.~\ref{sim_alpha25}, indicating that deceleration is ongoing.    Despite the differences on the axis and at the equator, the analytical solution still does a good job at predicting the cocoon's width to within roughly 10 per cent.

In addition to the jet simulations described above, we also tested our conclusion that the outflow does not become spherical for $\alpha>2$.  To do so, we compare with the results of a different simulation, in which an ellipsoid filled with a uniform high pressure is placed into a $\rho \propto R^{-2.5}$ density profile and allowed to evolve.   Since there is no need to resolve a jet in this case, the flow can be tracked to much larger radii, until it becomes approximately self-similar.  The initial aspect ratio of the outflow is $a/b=70/3 \gg 1$, and the ambient pressure is set to a small value ($10^{-8}$ times the initial cocoon pressure) so that it does not affect the evolution. 

\begin{figure} 
\centering
\includegraphics[width=\columnwidth,keepaspectratio]{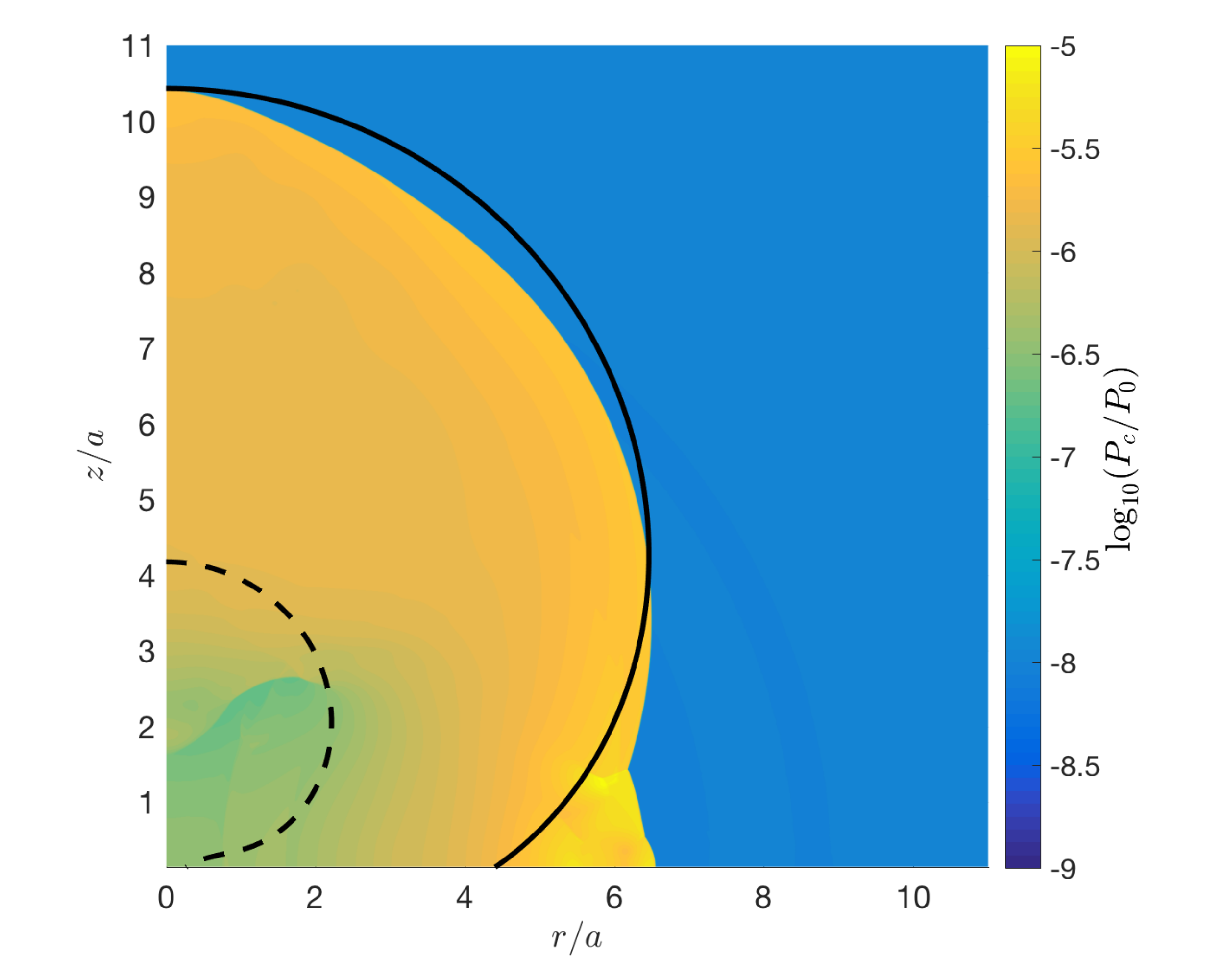}
\caption{The shape of the cocoon for a $\rho \propto R^{-2.5}$ density profile and an initial aspect ratio $a/b=70/3$, at a time $t/t_0=840$ when the height is $\zc/a=10.4$.  The colour scale indicates the ratio of the pressure to the initial cocoon pressure.  The black lines show the shape of the shock in the analytical model, scaled in two different ways.  The solid line is scaled to the same $\zc/a$ as the simulation, while the dashed line is scaled to the same $t/t_0$.} 
\label{asymptoticfig}
\end{figure}

In Fig.~\ref{asymptoticfig}, we compare the shape of the cocoon seen in the simulation to the prediction of the analytical model.  The figure shows a snapshot of the simulation at $t/t_0 = 840$, when the height is $\zc/a = 10.4$.  As expected, we find that the outflow does not become spherical when $t$ grows large.  The shape in the analytic model, using the same values of $\zc/a$ and $b/a$, is drawn as a solid black line.  The shape observed in the simulations is slightly narrower towards the axis and wider towards the base, but otherwise agrees well with the analytical prediction.  In particular, the ratio of the cocoon's width to its height is extremely close to the analytically-predicted value of 0.62.

However, although the agreement \textit{at a given height} is very good, the agreement \textit{at a given time} is noticeably worse.  Although the cocoon's size grows as $\propto t^{2/(5-\alpha)}$ in the simulations (as expected), the simulation reaches the quasi-spherical phase more quickly than anticipated analytically.  As a result, the size of the cocoon in the simulation is larger than the analytically-estimated size at a given time.  At late times, the ratio of the sizes approaches a constant value of $\approx 2.5$.  For example, whereas the simulations reach a height of $\zc/a = 10.4$ at a time $t/t_0=840$, the analytical model predicts a height of only $\zc/a = 4.2$ at this time, as shown by the dashed line in Fig.~\ref{asymptoticfig}.  One possible reason for the discrepancy is the fact that, as discussed above, there is a clear pressure gradient along the cocoon surface in the simulations, with the pressure increasing by a factor of a few towards the equator.  The higher pressure towards the sides enhances the lateral expansion at early times, causing the outflow to transition to the quasi-spherical regime sooner than it would if the pressure were constant across the surface.

From these considerations, we conclude that our results describing the shock's shape as a function of height (i.e., eq.~\ref{r_vs_z} and Fig.~\ref{ratio_vs_z}) are rather accurate, while our expressions relating the height and width to time (i.e., eqs.~\ref{z_vs_t}--\ref{r_vs_t}, and Fig.~\ref{rz_vs_t}) are only reliable to within a factor of a few.  An accurate calibration of the analytical model to more extensive numerical simulations is reserved as an interesting topic for future study. \\

\section{Discussion and Conclusions}
\label{conclusions}

Jets which fail to penetrate the surrounding medium are expected to leave behind a narrow cocoon filled with hot gas of almost uniform pressure.  To describe the evolution of these systems, we consider an axisymmetric generalization of the ST solution in a power-law ambient density profile $\rho \propto R^{-\alpha}$, in which the explosion energy is deposited throughout an elongated region along the polar axis.  Applying the Kompaneets approximation (KA), we derive a differential equation governing the shock motion, and solve it to yield analytical expressions for the shape of the shock.  
For a given curve $R_i(\theta_i)$ describing the initial shock shape, we obtain exact parametric solutions for the shock coordinates $R(\theta_i,\zc)$ and $\theta(\theta_i,\zc)$ as functions of $\zc$, the shock height along the axis.
The solutions vary in form depending on $\alpha$. In addition, we provide simplified expressions for the height, width, and volume of the shock versus time in the case of an initially ellipsoidal cocoon with an aspect ratio $a/b \gg 1$, as expected for a jet.  Our solution extends the previous work of B11, which describes the dynamics of the system while the jet is active, to times after energy injection has ended.

We find that the evolution of the outflow is characterized by three dynamical regimes, which we call the planar regime, the sideways expansion regime, and the quasi-spherical regime.  Two important dynamical time-scales are the time for the outflow's width to double ($t_b$) and the time for its height to double ($t_a$). For a jet choked at time $t_0$, the planar regime persists until the age is $\sim 2 t_0$.  During this time, dynamical quantities remain roughly constant, and the outflow more or less keeps its original shape.  Once $t \sim 2 t_b \sim 2t_0$, sideways expansion starts to become important; the pressure in the cocoon drops and the shock starts to decelerate.  The shape also changes, with the cocoon bulging out towards the tip.  This is especially pronounced in steep density gradients where there is a large density contrast between the base and the tip.  Most of the sideways expansion occurs over the time for the cocoon's height to double, which is $t \sim t_a \sim (a/b)^2 t_0$, where $a/b$ is the aspect ratio of the outflow upon choking.  After this, the evolution becomes quasi-spherical, the dependence on the initial shape becomes weak, and the shock's shape is governed primarily by the density profile.  As $a/b$ approaches unity, the two characteristic time-scales ($t_b$ and $t_a$) become the same and the usual spherical blast wave solution is recovered.

Although the height ($\zc$) and width ($\rc$) of the cocoon become comparable around $t_a$, the time for them to become equal (to within, say, 10 per cent) can be considerably longer.  In general we find that the asymptotic behaviour depends strongly on the density profile.  For $\alpha \le 2$, the flow becomes spherical.  The convergence to a sphere is slower for steeper density profiles, with the quantity $1 - \rc/\zc$  going to zero as $\propto \zc^{(\alpha-2)/2}$ for $\alpha < 2$, or as $\propto (\ln \zc)^{-1}$ for $\alpha=2$.  An approximation for the time it takes for the outflow to become spherical in different density profiles is given in eq.~\ref{tsph}.  We find that in a constant-density medium, the flow spherizes after about $\sim 10  t_a$, while in a wind-like medium it takes $\sim 1000 t_a$.  

In steep density profiles with $\alpha > 2$, the flow does not become spherical (at least in the idealized case where the KA applies).  Instead, the ratio $\rc/\zc$ converges to a constant value of $[\sin(\pi/\alpha)]^{\alpha/(\alpha-2)}$ as ${\propto \zc^{(2-\alpha)/4}}$.  As before, the convergence becomes slower as $\alpha \rightarrow 2$.  For $2 < \alpha < 3$, the shape of the shock at infinity is narrower towards the equator, like a peanut.

Comparing the analytical solution with numerical hydrodynamics simulations, we find a generally good agreement for $\zc/a \la $ a few, even without calibration.  We compared our results with realistic simulations of a choked jet for three different density profiles of astrophysical interest, with $\alpha=1$, 2, and 2.5.  In each case, the height and the width of the outflow estimated analytically match the numerically determined value to within about 15 per cent.  The agreement is not so good near the equator, but the KA is known to have issues in that region.  In addition, we explored the late-time behaviour in the $\alpha=2.5$ case through comparison with a simplified simulation in which an elliptical region is filled with an initially high pressure and left to evolve for $840 t_0$.  For $\zc/a \gg 1$, we find that the size of the simulated outflow is larger than predicted analytically by a factor of $\approx 2.5$.  However, when rescaled to be the same size, the analytical and numerical shapes are in close agreement.

The fact that the time-scale to become quasi-spherical is long compared to the working time of the jet is good news when it comes to observational applications, since it means that it is not necessary to catch the system shortly after jet activity ceases in order to meaningfully constrain the jet.  The relationship between the parameters of the choked jet model (initial height $a$, initial width $b$, and the energy $E_0$) and the parameters characterizing the jet (luminosity $\Lj$, duration $\tj$, and opening angle $\thetaop$) is discussed in Section~\ref{jet}.  By measuring the size and shape of a choked jet outflow, along with its energy, it is in principle possible to work backwards to obtain the initial conditions at the moment of choking, and therefore derive the jet properties.  However, because the system quickly becomes insensitive to the initial width, constraining $b$ is difficult in practice, unless we happen to catch the system early on.  The situation where only $E_0$ and $a$ can be estimated from observations is expected to be more typical.  In this case, the jet parameters are subject to degeneracy and cannot be determined uniquely.  None the less, knowing the location where the jet was choked is a powerful constraint, and one which is not attainable in spherically symmetric models.

We conclude by discussing several possible future applications of our results.  The first involves quenched AGN jets in galaxy clusters.  These jets inflate bubbles filled with shocked jet debris that initially expand supersonically, driving a shock into the ambient medium, and later drift away due to buoyancy.  Our model can be used to describe the shape of the forward shock surrounding the bubbles during the early evolution when the shocked jet material has not yet reached pressure equilibrium with the intracluster medium or been deformed by buoyancy.   The size, shape, and speed of the forward shock can be directly measured from X-ray observations, and then compared with the predictions of the model to derive constraints on the jet properties.   This idea is developed further and applied to 5 AGNs with suitably young bubbles in a separate paper (Irwin et al., in preparation).

Another possible application deals with observations of SNe that may have harboured a jet.  We envision a scenario where the jet is choked before breaking out of its progenitor star, at a radius $a < R_*$ where $R_*$ is the star's radius.  The cocoon continues to propagate, eventually breaking out of the star when its height is $\zc = R_*$.  If the choking does not occur too deep inside the star, the width of the outflow upon breakout, $\rc$, will be small compared to $R_*$, and the mass swept up by the cocoon will also be small compared to the star's mass $M_*$.   Once outside the star, the cocoon is expected to quickly expand sideways and engulf the star due to the much lower density.  However, since a typical GRB jet imparts an energy of $10^{51}$ erg which is comparable to the SN energy, the cocoon will expand faster than the SN due to its lower mass.  Therefore, a prediction for the choked jet scenario is the presence of a small amount of mass ($0.01$--$0.1 M_*$) surrounding the supernova that is moving with a speed several times greater than the bulk SN velocity. 

This fast-moving ejecta, which is only visible at early times, may have already been observed in some objects \citep[e.g.,][]{piran19,izzo19}.  Early-time spectroscopy makes it possible to estimate the total mass ($M_{\rm ej}$) and energy ($E_{\rm ej}$) contained in this high-velocity component, and can even constrain the distribution of mass with velocity.  $E_{\rm ej}$ corresponds to the cocoon energy $E_0$ in our model, while $M_{\rm ej}$ serves as a proxy for the width at breakout, with $M_{\rm ej} \sim M_* (\rc/R_*)^2$.   By applying our results, this information can be translated into an estimate of the choking radius, which provides significant insight into the jet properties.  Whether the choked jet model can reproduce the observed mass-velocity distribution in \citet{piran19} is an interesting question for future study.
 
The asphericity of the cocoon also has important implications for shock breakout, where even a small departure from spherical flow can lead to oblique effects that dramatically alter the observed signature \citep[e.g.,][]{matzner13}.  Therefore, we typically expect the breakout of a choked jet to look markedly different than a fully spherical shock breakout.  At the least, an aspherical breakout will last longer and be fainter than a spherical breakout of the same energy, because it takes some time for the breakout to wrap around from the pole to the equator.  The expected changes in the breakout emission when it is powered by a choked jet, rather than a spherical SN explosion, will be addressed in future work.

As these examples show, a 2D analytical model for the shape of the shock in choked jet outflows has a wide variety of astrophysical applications, and offers significant advantages over the standard simplifying assumption of spherical symmetry.  The Kompaneets approximation model developed here is a useful tool which is straightforward and inexpensive to implement, without sacrificing too much accuracy.  Bridging the gap between the jet phase and the spherical phase of evolution is an important step towards a complete picture of jet dynamics, which will remain important as observations continue to reveal evidence for suffocated jets in diverse astrophysical systems.

\section*{Acknowledgements}

We thank O. Gottlieb, who conducted the numerical simulations discussed in Section 5 and graciously shared his data with us, for his generous assistance; his contributions greatly strengthened the final paper.
This research is supported by the CHE-ISF I-Core Center for Excellence in Astrophysics. 
TP is supported by an advanced ERC grant TReX. 
This work was supported in part by the Zuckerman STEM Leadership Program.
CI and EN were partially supported by consolidator ERC grant JetNS and by an ISF grant 1114/17.

\appendix
\onecolumn
\section{Glossary of symbols}
\label{appendix0}

Many of the quantities appearing in our problem evolve in time.  Additionally, at a given time, certain quantities such as the shock velocity vary along the cocoon surface.  Since each point on this surface had a unique initial angular position, $\theta_i$, at the moment the jet was choked (see Section~\ref{analytical}), we use $\theta_i$ to parametrize the relative position on the shock front.  In Table~\ref{table2}, we list the symbols defined throughout the paper, explicitly including any dependence on $t$ or $\theta_i$ in parentheses following each symbol.

\begin{table*}
\centering
\caption{Glossary of symbols.}
\begin{tabular}[t]{l l l}
\textbf{Symbol} & \textbf{Meaning} & \textbf{Introduced} \\ \hline
\multicolumn{3}{c}{Coordinates}  \\ \hline
$R$, $\theta$ & polar coordinates & \,   \\
$r$, $z$ & cylindrical coordinates & \,  \\ \hline
\multicolumn{3}{c}{Initial conditions}  \\ \hline
$E_0$ & energy of the cocoon & Section~\ref{overview}  \\
$a$ & initial height of the cocoon & Section~\ref{overview}   \\
$b$ & initial width of the cocoon & Section~\ref{overview}   \\  
$P_0$ & initial pressure in the cocoon & Section~\ref{overview} \\
$V_0$ & initial volume in the cocoon & Section~\ref{overview} \\
$\rho_0$ & ambient density at $R=a$ & eq.~\ref{rho}   \\
$\alpha$ & power-law index of the density profile & eq.~\ref{rho}  \\ 
$k_\alpha$ & frequently appearing quantity defined by $k_\alpha \equiv 1-\alpha/2$ & eq.~\ref{kalpha} \\
$R_i(\theta_i)$ & curve describing the shock's initial shape & eq.~\ref{R_i} \\
$\chi_i(\theta_i)$ & angle between the axis and the initial direction of motion & eqs.~\ref{chi_i} and~\ref{thetaN} \\ 
$\omega_i(\theta_i)$ & angle between the initial direction of motion and the radial direction & eq.~\ref{omega} \\ \hline
\multicolumn{3}{c}{Time-scales} \\ \hline
$t_0$ & time when the jet is choked & Section~\ref{overview} \\
$t_b$ & time it takes for the cocoon's width to double & eq.~\ref{tb} \\
$t_a$ & time it takes for the cocoon's height to double & eq.~\ref{ta} \\
$\tsph$ & time for the cocoon's width to become 90 per cent of its height & eqs.~\ref{Qdef} and~\ref{tsph} \\ 
$\teq$ & time to reach pressure equilibrium with the ambient medium & eqs.~\ref{teq} and~\ref{etaeq}  \\
$\tw$ & time when the solution breaks down in an expanding medium & eqs.~\ref{tw} and~\ref{etaw} \\
$\tbo(R_*)$ & time when the cocoon breaks out of an ambient medium extending to $R_*$ & eqs.~\ref{tbo} and~\ref{etabo} \\ \hline
\multicolumn{3}{c}{Cocoon properties}  \\ \hline
$\Vc(t)$ & volume of the cocoon & Section~\ref{overview} \\
$\Pc(t)$ & average pressure in the cocoon & Section~\ref{overview}  \\ 
$\lambda$ & dimensionless ratio of postshock pressure to volume-averaged pressure & eq.~\ref{P} \\
$x(t)$ & dimensionless time parameter used to solve for the shock's shape & eq.~\ref{x} \\
$\zc(t)$ & height of the cocoon & eq.~\ref{zcdef}  \\ 
$\rc(t)$ & width of the cocoon & eq.~\ref{rcdef} \\
$\req(t)$ & width of the cocoon at the equator \, & see Fig.~\ref{illustration} \\ 
$\epsiloneff(t)$ & effective eccentricity, i.e. $\epsiloneff\equiv \sqrt{1-\rc^2/\zc^2}$ & Sections~\ref{alpha0} and~\ref{alphax} \\ 
$R(\theta_i,t)$, $\theta(\theta_i,t)$ & Parametric equations for the shape of the shock at time $t$ & Sections~\ref{alpha0} and~\ref{alphax} \\ \hline
\multicolumn{3}{c}{Quantities related to the bulge} \\ \hline
$\Rb(t)$, $\thetab(t)$ & polar coordinates of the bulge & eq.~\ref{thetab} and~\ref{Rb}  \\ 
$\rb(t)$, $\zb(t)$ & cylindrical coordinates of the bulge & see Fig.~\ref{illustration}  \\ 
$\thetabo(t)$, $\Rbo(t)$, $\chibo(t)$ & initial values of $\theta_i$, $R_i$, and $\chi_i$ for a trajectory passing through ($\Rb,\thetab$) & see Fig.~\ref{bulgeparams} \\ \hline
\multicolumn{3}{c}{Asymptotic quantities} \\ \hline
$\zeta(t)$ & quantity important in the expansion of $\rc(\zc)$ at infinity & eq.~\ref{Z} \\
$A_\alpha$ & coefficient appearing in the expansion of $\rc(\zc)$ at infinity & eq.~\ref{Aalpha1} and ~\ref{Aalpha3} \\
$f_\alpha$, $g_\alpha$ & ratios of $\rc/\zc$ and $\req/\zc$ respectively, as $t \rightarrow \infty$ & eq.~\ref{fdefine} \\
$C_\alpha$ & ratio of $\Vc$ to $4\pi \zc^3/3$ as $t \rightarrow \infty$ & eq.~\ref{Calpha} and~\ref{Calpha2} \\
$\thetabinf$ & value of $\thetab(t)$ as $t \rightarrow \infty$ & eq.~\ref{thetabmax} \\ 
$\thetabomin$ & value of $\thetabo(t)$ as $t \rightarrow \infty$ & Section~\ref{alphax} \\ \hline
\multicolumn{3}{c}{Jet properties}  \\ \hline
$\Lj$ & jet luminosity & Section~\ref{jet}  \\ 
$\tj$ & jet duration & Section~\ref{jet}  \\ 
$\thetaop$ & jet opening angle at injection & Section~\ref{jet}  \\ 
$\betah$, $\Gamma_h$ & velocity and Lorentz factor of the jet head upon choking & Section~\ref{jet}  \\ 
$\betac$ & sideways velocity of the cocoon upon choking & Section~\ref{jet}  \\ 
$\tilde{L}$ & dimensionless parameter governing jet propagation & Section~\ref{jet}  \\ \hline
\end{tabular}
\label{table2}
\end{table*}

\section{Parametric equations for the cocoon shape}
\label{appendixa}

In this appendix, we derive parametric equations for the height and width of the cocoon as a function of the parameter $\thetabo$.  As discussed in Section~\ref{alphax}, a trajectory originating from $\thetabo$ becomes parallel to the equator at the point ($\Rb,\thetab$), with $\thetab$ and $\Rb$ given by equations~\ref{thetab} and \ref{Rb} respectively.  How long does it take a particle at $(\Rbo,\thetabo)$ to travel to $(\Rb,\thetab)$?  By substituting $\theta(x(\thetabo),\thetabo)=\thetab$ into equation~\ref{theta1} or \ref{theta2}, as appropriate, we obtain 
\begin{equation}
\label{x0}
x(\thetabo) = 
\begin{cases}
\dfrac{1}{k_\alpha} \left(\dfrac{\Rbo}{a}\right)^{k_\alpha} \dfrac{\sin(k_\alpha \Deltab)}{\sin(\omegabo-k_\alpha \Deltab)} & \alpha \ne 2 \\
\dfrac{\Deltab}{\sin\omegabo} & \alpha=2 
\end{cases} ,
\end{equation}
where we have defined
\begin{equation} 
\label{Deltab}
\Deltab(\thetabo) \equiv \thetab-\thetabo = \dfrac{2}{\alpha}\left(\pi/2-\chibo\right).
\end{equation}
and
\begin{equation}
\label{omegab0}
\omegabo(\thetabo) \equiv \chibo-\thetabo
\end{equation}
to make the formulae more compact.  The values of $\Rbo$ and $\chibo$ are given by setting $\theta_i=\thetabo$ in equations~\ref{R_i} and \ref{thetaN}, respectively. 

If the initial shape is an ellipsoid with $\Rbo$ and $\chibo$ given by equations~\ref{R_i} and~\ref{thetaN} at $\theta_i=\thetabo$, then equation~\ref{x0} only has a solution when $x$ is greater than a critical value,
\begin{equation}
\label{xb}
x_b = \lim_{\thetabo\rightarrow \pi/2} x(\thetabo) = \left(\dfrac{b}{a}\right)^{k_\alpha} \dfrac{2 b^2/\alpha a^2}{1-2b^2/\alpha a^2}.
\end{equation}
Physically, this corresponds to the time when the bulge width becomes equal to the equatorial width, i.e. we have $\rb=\req$ at the time $t=t(x_b)$.  The relation between $\rc$ and $\zc$ is different before and after this transition, with $\rc=\req$ up to $x = x_b$, and $\rc=\rb$ afterwards.  Note that if $\alpha \le 2b^2/a^2$, $x_b$ is negative or undefined.  Since $x$ can only be positive and finite, this means that when $\alpha \le 2b^2/a^2$, the transition at $x=x_b$ never occurs. The bulge width never exceeds the equatorial width in this case, and $\rc=\req$ at all times.  However, as long as $b/a$ is small, this is only relevant to very flat density profiles with $\alpha \approx 0$.  In the case where $b \ll a$ and $\alpha \gg 2b^2/a^2$, the equatorial phase lasts only briefly, until the height has increased by $ ax_b \approx (2/\alpha)(b/a)^{2+k_\alpha} a \ll a$, and the bulge-dominated case is the relevant one for most of the evolution.

When the width is determined by equatorial expansion (i.e. when $\alpha \le 2b^2/a^2$, or when $\alpha > 2b^2/a^2$ but $x\le x_b$), the cocoon shape parameters are simple functions of the parameter $x$.  The height is given by equation~\ref{z1} or~\ref{z2}, and the equatorial radius is found by taking $\thetabo=\chibo=\pi/2$ and $\Rbo=b$.  We then have $\rc=\req$, and
\begin{equation}
\label{system_equatorial}
\begin{array}{ll}
\zc(x) = a(1+k_\alpha x)^{1/k_\alpha} \\
\rc(x) = b\left[1+(a/b)^{k_\alpha} k_\alpha x\right]^{1/k_\alpha}\\
\end{array}
\end{equation}
In this regime, it is possible to express $\rc$ as an explicit function of $\zc$:
\begin{equation}
\label{r_z}
\rc(\zc) = \req(\zc) =
\begin{cases}
b \left\{1 + \left(\dfrac{a}{b}\right)^{k_\alpha} \left[\left(\dfrac{\zc}{a}\right)^{k_\alpha}-1\right]\right\}^{1/k_\alpha} & \alpha \ne 2 \\
b \left(\dfrac{\zc}{a}\right) & \alpha =2
\end{cases} .
\end{equation}
For $k_\alpha=1$ (constant density), we reproduce equation~\ref{ratio0}.  

When $\alpha > 2b^2/a^2$ and $x>x_b$, however, the width is governed by the bulge, and the relation between $\rc$ and $\zc$ becomes more nuanced.  It is no longer possible to express $\rc$ as an explicit function of $\zc$, but the width and height can be parametrized in terms of $\thetabo$.  The height of the cocoon when $x=x(\thetabo)$ is, combining equations~\ref{z1} or \ref{z2} with \ref{x0},
\begin{equation}
\label{z_th0}
\zc (\thetabo) = 
\begin{cases}
a \left[1 + \left(\dfrac{\Rbo}{a}\right)^{k_\alpha} \dfrac{\sin(k_\alpha \Deltab)}{\cos \thetab} \right]^{1/k_\alpha} & \alpha \ne 2 \\
a {\rm e}^{\Deltab \csc\omegabo} & \alpha=2
\end{cases},
\end{equation}
with $\thetab$ given by equation~\ref{thetab}.  The width, on the other hand, is $\rc=\rb=\Rb \sin \thetab$; from equations~\ref{thetab} and \ref{Rb}, we have
\begin{equation}
\label{r_th0}
\rc(\thetabo) =
\begin{cases}
\Rbo \sin\thetab \left(\dfrac{\sin \omegabo}{\cos\thetab}\right)^{1/k_\alpha} & \alpha \ne 2 \\
\Rbo \sin\thetab {\rm e}^{\Deltab \cot \omegabo} & \alpha = 2
\end{cases}.
\end{equation}
Similarly, the height of the bulge above the equatorial plane is $\zb=\Rb \cos \thetab$, or
\begin{equation}
\label{zb_th0}
\zb(\thetabo) = 
\begin{cases}
\Rbo \cos\thetab \left(\dfrac{\sin \omegabo}{\cos\thetab}\right)^{1/k_\alpha} & \alpha \ne 2 \\
\Rbo \cos\thetab {\rm e}^{\Deltab \cot \omegabo} & \alpha = 2
\end{cases}.
\end{equation}
The equatorial width in this case is discussed separately, in Appendix~\ref{appendixc}.
In the first part of equations~\ref{z_th0}--\ref{zb_th0}, we simplified the denominator by using the identity $\omegabo - k_\alpha \Deltab = \pi/2 - \thetab$, which can be derived from equations~\ref{Deltab}--\ref{omegab0}.  

Since we expect $\rc \propto \zc$ to hold approximately at late times, it is sometimes more convenient to work with the dimensionless ratio $\rc/\zc$.  Dividing equation~\ref{r_th0} by equation~\ref{z_th0}, we obtain
\begin{equation}
\label{ratio_th0}
\dfrac{\rc}{\zc} =
\begin{cases}
\left(\dfrac{\Rbo}{a}\right) \sin \thetab \left[ \dfrac{\sin \omegabo}{\cos \thetab + (\Rbo/a)^{k_\alpha} \sin(k_\alpha \Deltab)} \right]^{1/k_\alpha} & \alpha \ne 2 \\
\left(\dfrac{\Rbo}{a}\right) \sin \thetab  {\rm e}^{\Deltab(\cot \omegabo-\csc \omegabo)} & \alpha=2
\end{cases} .
\end{equation}

\section{Evolution of the cocoon width versus height}
\label{appendixb}

\begin{figure}
\centering
\includegraphics[width=\textwidth]{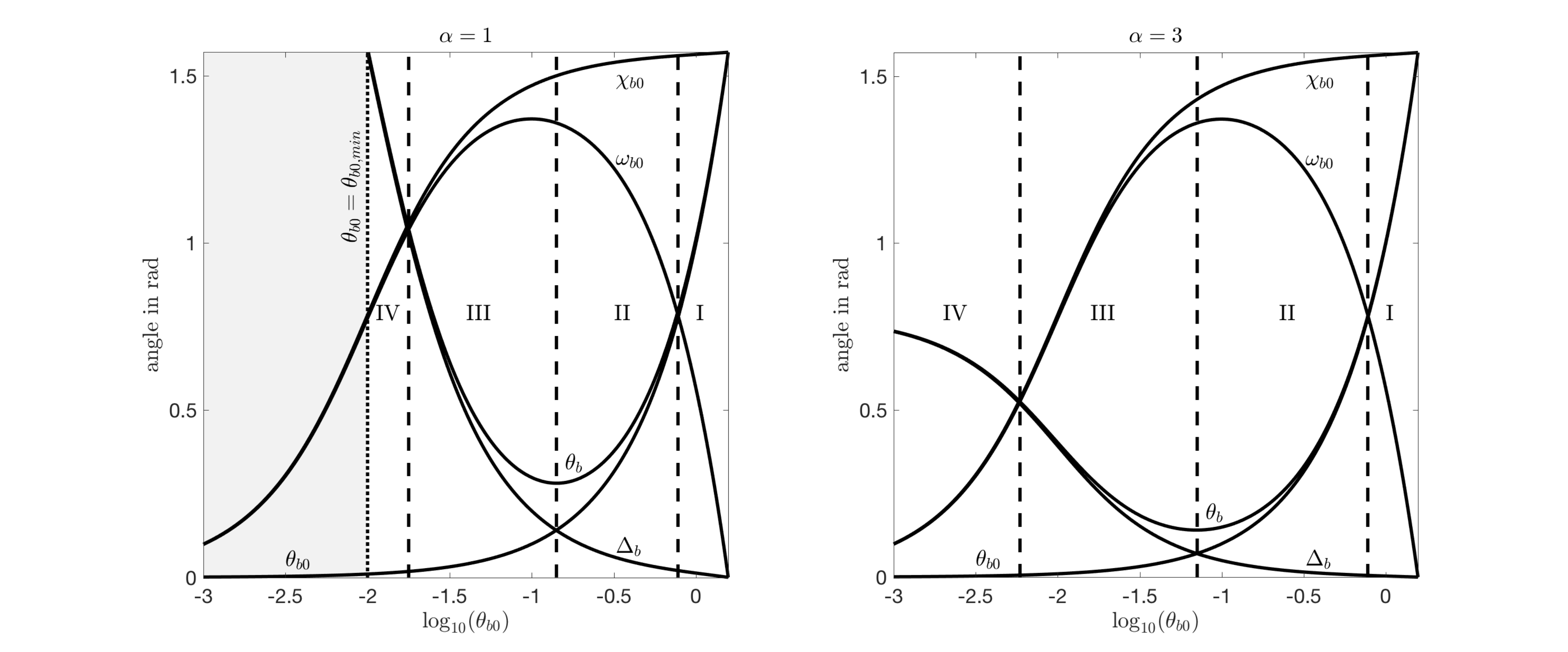}
\caption{The angular scales $\thetabo$, $\chibo$, $\thetab$, $\omegabo$, and $\Deltab$ as functions of $\thetabo$, for an initial ellipsoid with $b/a=0.1$ and $\alpha=1$ (left panel) or $\alpha=3$ (right panel).  The four dynamical phases described in the text are labeled with Roman numerals and separated by vertical dashed lines.  The critical value $\thetabomin$ (which is only defined for $\alpha<2$) is indicated by a dotted line.  The greyed-out region has $\thetab>\pi/2$ and does not affect the evolution of $\rc/\zc$.}
\label{phases}
\end{figure}

Approximate formulae for $\rc$ versus $\zc$ can be obtained in certain limiting situations where equations~\ref{z_th0} and~\ref{r_th0} simplify.  To understand the interesting limiting cases, let us investigate how the three angular scales appearing in equation~\ref{ratio_th0} ($\thetab$, $\omegabo$, and $\Deltab$) vary with $\thetabo$.  (We restrict the discussion here to an initially ellipsoidal cocoon, but the conclusions can be generalized to other initial shapes.)  To guide the discussion, we plot $\chibo$, $\omegabo\equiv\chibo-\thetabo$, $\thetab$, and $\Deltab\equiv\thetab-\thetabo$ versus $\thetabo$ in Fig.~\ref{phases} for $\alpha=1$ (left panel) or 3 (right panel) and $a/b=10$.  Focusing now on the $\thetab$ and $\omegabo$ curves, the key features are as follows:
\begin{itemize}
\item Differentiation of equation~\ref{thetab} implies that $\thetab$ has a minimum if $\deriv\chibo/\deriv\thetabo=\alpha/2$ is satisfied.  Applying equation~\ref{thetaN}, we see that this condition is only fulfilled when $\frac{2b^2}{a^2} < \alpha < \frac{2a^2}{b^2}$.  If we assume that $b\ll a$, and that $\frac{2b^2}{a^2} \ll \alpha \ll \frac{2a^2}{b^2}$ holds as well,\footnote{The case of $\alpha\le\frac{2b^2}{a^2}$ describes the scenario where the cocoon always remains widest in the equatorial plane, as discussed above.  The case of $\alpha \ge \frac{2 a^2}{b^2}$ we do not discuss further, because if $a \gg b$, it requires an extreme value of $\alpha$.} then $\thetab$ has a minimum value of $\approx \frac{2b}{a} \sqrt{\frac{2}{\alpha}}$ at $\thetabo \approx \frac{b}{a} \sqrt{\frac{2}{\alpha}}$.  Away from the minimum, $\thetab$ rises to $\pi/2$ at $\thetabo=\pi/2$, and to $\pi/\alpha$ at $\thetabo=0$.\footnote{If $\thetab >\pi/2$ for some $\thetabo$, the physical interpretation is that trajectories stemming from $\thetabo$ never become parallel to the equator.}  If $\alpha<2$, $\thetab$ becomes equal to $\pi/2$ at $\thetabo=\thetabomin$ (see below).
\item $\omegabo$ is determined from equation~\ref{thetaN}, and its behaviour is straightforward since there is no dependence on $\alpha$: $\omegabo$ is 0 when $\thetabo=0$, rises to a maximum value of $\omegabo\approx \pi/2-b/a$ at $\thetabo \approx b/a$, and then goes to zero again as $\thetabo \rightarrow \pi/2$.
\item The $\thetab$ and $\omegabo$ curves have two crossings, one at $\thetabo \sim b^2/a^2$, and another at $\thetabo \sim \pi/4$.  Over the range $b^2/a^2 \la \thetabo \la \pi/4$, $\thetab<\omegabo$; otherwise $\thetab>\omegabo$.
\end{itemize}

As discussed in Section~\ref{alpha0-2}, $\thetabomin$ corresponds to a special trajectory separating the ejecta that eventually interact at the equator from those that do not. To find its value, we look for a trajectory that has $\theta \rightarrow \pi/2$ as $x \rightarrow \infty$.  Then, from equation~\ref{theta1}, we get an implicit equation,
\begin{equation}
\label{thetac1}
\thetabomin = \dfrac{2}{\alpha}\left(\chibomin-k_\alpha\pi/2\right),
\end{equation}
where $\chibomin$ is related to $\thetabomin$ by equation~\ref{thetaN}.  Alternatively, formula~\ref{thetac1} may be obtained by taking $\thetab \rightarrow \pi/2$ in ~\ref{thetab}.  Note that when $\alpha = 2b^2/a^2$, the solution to equation~\ref{thetac1} is $\thetabomin = \chibomin= \pi/2$.  This is another way to see that there is no interaction in the $\alpha \le 2b^2/a^2$ case.  As $\alpha$ increases from this value, $\thetabomin$ decreases, and a larger portion of the cocoon becomes subject to collisions.  When $\alpha \rightarrow 2$, $\thetabomin \rightarrow 0$, and all parts of the surface will interact given sufficient time.  As long as $\alpha \gg 2b^2/a^2$, $\thetabomin$ is small and can be approximated by
\begin{equation}
\label{thetac1_approx}
\thetabomin \approx (b^2/a^2) \tan \left(k_\alpha \pi/2\right).
\end{equation}

\renewcommand{\labelenumi}{\Roman{enumi}. }
The outflow's shape evolves in a different way depending on the relative size of the angular parameters.  There are four distinct evolutionary phases (labeled by Roman numerals in the figure) as $\thetabo$ is decreased from $\pi/2$ to $\thetabomin$ (when $\alpha<2$) or to $0$ (when $\alpha \ge 2$).  These four phases are closely related to the dynamical phases discussed in Section~\ref{overview}.  The relationship between the angular scales in each phase is as follows:
\begin{enumerate}
\item $\thetabo \approx \pi/2$ and $\chibo > \thetabo \approx \thetab \gg \omegabo \gg \Deltab$.  This corresponds to the early planar phase, when $x \sim x_b$ (see also Appendix~\ref{appendixa}), and the bulge is still located close to the equator.
\item $b/a \ll \thetabo \ll \pi/4$ and $\chibo > \omegabo \gg \thetabo \approx \thetab \gg \Deltab$.  This applies during the later part of the planar phase, once $x \gg x_b$ and the bulge is closer to the axis than to the equator.
\item $b^2/a^2 \ll \thetabo \ll b/a$ and $\chibo \simeq \omegabo \gg \thetab \simeq \Deltab \gg \thetabo$.  This holds during the sideways expansion phase.
\item $\thetabo \rightarrow \thetabomin$ or $\thetabo \rightarrow 0$ (respectively for $\alpha <2$ and $\alpha \ge 2$), and $\thetab \approx \Deltab > \chibo \approx \omegabo \gg \thetabo$.  This is appropriate in the quasi-spherical phase.
\end{enumerate}
We consider each phase in turn below.

\textbf{Phase I ($\thetabo \approx \pi/2$:)} Let $\delta \equiv \pi/2-\thetabo$ be a small quantity.  Then, expanding equations~\ref{thetaN} and \ref{thetab} in terms of $\delta$, we find $\chibo\approx \pi/2-(b^2/a^2)\delta$, $\omegabo \approx (1-b^2/a^2)\delta$, $\thetab \approx \pi/2-(1-2 b^2/\alpha a^2)\delta$, and $\Deltab \approx (2b^2/\alpha a^2) \delta$.   In this case, the term in brackets in equation~\ref{ratio_th0} is unity, up to a small correction of order $b^2/a^2$.  Therefore we can write $\rc/\zc \approx (\Rbo/a) \sin \thetab \approx b/a$.  

\textbf{Phase II ($b/a \la \thetabo \la \pi/4$):} In this limit $\delta=\pi/2-\thetabo \ge \pi/4$ is no longer a small quantity.  However, $(b^2/a^2) \delta$ is still small.  So, to a reasonable approximation $\thetab \approx \thetabo$, $\chibo\approx \pi/2$, and $\omegabo \approx \pi/2-\thetabo$.  Thus the bracketed part of equation~\ref{r_z} is still order-unity, and $\rc/\zc \approx (\Rbo/a) \sin \thetab$ remains valid.  Although $\Rb$ and $\thetab$ both vary with $\thetabo$ in this case, one decreases while the other increases.  The effects roughly cancel so that $\rb$ turns out to be nearly constant with respect to $\thetabo$.  When phase II ends, $\thetab$ is near its minimum value of  $\approx (2b/a)\sqrt{2/\alpha}$.  At that time, $\Rb \approx \Rbo \approx a/\sqrt{1+2/\alpha}$ via equations~\ref{R_i} and~\ref{Rb}, and $\rc/\zc \approx \Rb \thetab/a \approx (2b/a)/ \sqrt{1+\alpha/2}$.  We therefore find that $\rc/\zc \simeq b/a$ throughout this phase, to within a factor of two.  This suggests that the end of phase II roughly coincides with the end of the planar phase.

\textbf{Phase III ($b^2/a^2 \la \thetabo \la b/a$):} In phases III and IV, $\thetabo$ is the smallest angular scale in the problem.  In the limit $\thetabo\rightarrow 0$, the shock's shape is similar to that of an on-axis point explosion at $z=a$.  Since $\thetabo$ can be neglected, we have $\thetab \approx \Deltab$ and $\chibo \approx \omegabo \approx \pi/2-(\alpha/2)\thetab$.  Equations~\ref{r_z} and \ref{z_th0} then simplify to functions of $\thetab$ alone:
\begin{equation}
\label{z34}
\zc(\thetab) \approx 
\begin{cases}
a \left[1+ \left(\dfrac{\Rbo}{a}\right)^{k_\alpha} \dfrac{\sin(k_\alpha\thetab)}{\cos \thetab}\right]^{1/k_\alpha} & \alpha \ne 2 \\
a {\rm e}^{\thetab/\cos( \alpha \thetab/2)} & \alpha=2
\end{cases}
\end{equation}
and
\begin{equation}
\label{ratio34}
\dfrac{\rc}{\zc} \approx 
\begin{cases}
\left(\dfrac{\Rbo}{a}\right) \sin \thetab \left[ \dfrac{\cos( \alpha \thetab/2)}{\cos \thetab + (\Rbo/a)^{k_\alpha} \sin(k_\alpha\thetab)} \right]^{1/k_\alpha} & \alpha \ne 2 \\
\left(\dfrac{\Rbo}{a}\right) \sin \thetab {\rm e}^{\thetab(\tan\thetab-\sec\thetab)} & \alpha=2
\end{cases} ;
\end{equation}
these expressions are valid for both phase III and phase IV, since so far we have only assumed $\chibo \gg \thetabo$ and $\thetab \gg \thetabo$.  In what follows it will also be useful to know how the parameter $\zeta$ varies with $\thetab$.  Combining equation~\ref{z34} with the definition of $\zeta$ from equation~\ref{Z}, we obtain
\begin{equation}
\label{Zb}
\zeta \approx 
\begin{cases}
\left(\dfrac{\Rbo}{a}\right)^{k_\alpha} \dfrac{\sin(k_\alpha \thetab)}{k_\alpha \cos \thetab} & \alpha < 2 \\
\dfrac{\thetab}{\cos \thetab} & \alpha=2 \\
\dfrac{\sin(k_\alpha \thetab)}{k_\alpha[(a/\Rbo)^{k_\alpha} \cos \thetab+\sin (k_\alpha \thetab) ]} & \alpha>2
\end{cases}.
\end{equation}

For phase III specifically, we want $\thetab \ll \chibo$, and since $\chibo \le \pi/2$ this implies $\thetab$ must be small.  With the approximation $\Rbo \approx a$, expanding~\ref{z34} and \ref{ratio34} to leading order for $\thetab \ll 1$ yields simply $\rc/\zc \simeq \thetab$ and $\zc \simeq a(1+\thetab)$.  Therefore,
\begin{equation}
\label{r_z3{r_z3}}
\dfrac{\rc}{\zc} \simeq \dfrac{\zc}{a}-1.
\end{equation}
We caution that this approximation is coarse and is only relevant for sufficiently large $a/b$. The $\rc \propto \zc-a$ behaviour can be understood through analogy with a point explosion at $z=a$.  Within a small neighbourhood of size $\ll a$ near the tip of the cocoon, the density is nearly constant.  Since parts of the cocoon moving in different directions encounter the same density, the shocked region near $\zc=a$ is a roughly sphere of radius $\zc -a$.  As phase III ends when $\thetab \sim 1$ and $\zc \sim a$, it can be identified with the sideways expansion phase discussed in Section~\ref{overview}.

\textbf{Phase IV ($\thetabo \la b^2/a^2$):} Finally, we address the case where $\thetab > \chibo \gg \thetabo$.  This limit describes the behaviour as the outflow's size tends to infinity.  In contrast to the previous phases, the evolution in phase IV depends strongly on $\alpha$.  Below, we separately discuss the cases $\alpha<2$, $\alpha=2$, and $\alpha >2$.  The basic solution method is the same in each case.  First, we create some small quantity $\delta \equiv \thetabinf - \thetab$, with $\thetabinf$ defined as in equation~\ref{thetabmax}.  Next, we replace $\thetab$ by $\delta$ in equation~\ref{ratio34}, and expand to leading order in $\delta$.  We then do the same for equation~\ref{Zb} to obtain $\zeta$ in terms of $\delta$.  Finally, we combine the two expressions to determine $\rc/\zc$ to first order in $\zeta$.

For $\alpha<2$, we have $\thetab \rightarrow \pi/2$ as $\thetabo \rightarrow \thetabomin$, so we let $\delta = \pi/2-\thetab$.  Then, ignoring terms of order $\delta^2$ or higher, we find $\cos \thetab \approx \delta$; $\sin \thetab \approx 1$; $\cos(\alpha \thetab/2) \approx \sin(k_\alpha \pi/2) [1+ (\alpha/2)\delta \cot (k_\alpha\pi/2)]$; and $\sin(k_\alpha\thetab) \approx \sin(k_\alpha \pi/2)[1 - k_\alpha\delta \cot(k_\alpha \pi/2)]$.  Applying equation~\ref{Zb}, we obtain 
\begin{equation}
\label{Z1_approx}
\zeta^{-1} \approx \left[(\Rbomin/a)^{k_\alpha} \sin(k_\alpha \pi/2)\right]^{-1} k_\alpha \delta \ll 1,
\end{equation}
where $\Rbomin\equiv R_i(\theta_i=\thetabomin)$.  Now, expanding equation~\ref{ratio34} and using~\ref{Z1_approx} to replace $\delta$ by $\zeta$, we find
\begin{equation}
\label{ratiom0-2}
\dfrac{\rc}{\zc} \approx 1 - \dfrac{1}{k_\alpha^2}\left[1-\left(\dfrac{\Rbomin}{a}\right)^{k_\alpha} \cos\left(\dfrac{k_\alpha \pi}{2}\right) \right] \zeta^{-1}.
\end{equation}
If we further assume that $\thetabomin$ is small (i.e., that $\alpha \gg 2b^2/a^2$), then $\Rbomin \approx a$ and equation~\ref{ratiom0-2} reduces to the form in equations~\ref{ratio1}--\ref{Aalpha1}.

For $\alpha=2$, we can repeat the same procedure using the second expressions in~\ref{ratio34} and~\ref{Zb}, or simply take the limit $k_\alpha\rightarrow0$ in formula~\ref{ratiom0-2}.  Either way, we recover equation~\ref{ratio2}.  

Lastly, we treat the case of $\alpha>2$, for which $\thetabo \rightarrow 0$ and $\thetab \rightarrow \thetabinf= \pi/\alpha$.  This time, we suppose $\delta = \pi/\alpha-\thetab$ is small.  In this limit, it can be shown from equations~\ref{R0} and \ref{thetab} that $\Rbo/a = 1 - \mathcal{O}(\delta^2)$, so we ignore corrections arising from the $\Rbo/a$ factors.  Using $k_\alpha=1-\alpha/2$, the denominator appearing in equations~\ref{ratio34} and~\ref{Zb} can be rewritten as: 
\begin{equation}
\cos \thetab+\sin (k_\alpha\thetab) =  \sin\thetab \cos(\alpha \thetab/2) + \cos \thetab\left[1- \cos(\delta \alpha/2) \right].
\end{equation}  
The term in brackets can be neglected, since it vanishes as $\delta^2$ when $\delta \rightarrow 0$.  Therefore, equations~\ref{ratio34} and~\ref{Zb} simplify to
\begin{equation}
\label{ratiom_simp}
\dfrac{\rc}{\zc} \approx \left[\sin \thetab\right]^{1-1/k_\alpha}
\end{equation}
and
\begin{equation}
\label{Z_simp}
\zeta \approx \dfrac{\sin(k_\alpha \thetab)}{k_\alpha \sin \thetab \cos (\alpha \thetab/2)}.
\end{equation}
When $\thetab \rightarrow \pi/\alpha$, we see that $\rc/\zc$ approaches the value $f_\alpha$ given by equation~\ref{falpha3}.

Expanding once more in terms of $\delta$ yields $\sin \thetab \approx \sin(\pi/\alpha) [1-\delta \cot(\pi/\alpha)]$; $\sin (k_\alpha\thetab) \approx -\cos(\pi/\alpha) { [1-k_\alpha\delta \cot(k_\alpha \pi/\alpha)]}$; and $\cos(\alpha \thetab/2) \approx (\alpha/2)\delta$.  So, to leading order,
\begin{equation}
\label{Z3_approx}
\zeta^{-1} \approx (\alpha/2) |k_\alpha| \delta \tan(\pi/\alpha).
\end{equation}
Expanding formula~\ref{ratiom_simp} up to first order in $\delta$, and then substituting $\delta$ for $\zeta$ according to equation~\ref{Z3_approx}, we arrive at equations~\ref{ratio3}--\ref{Aalpha3}.
  
\section{Collisions and the evolution of the equatorial width versus height}
\label{appendixc}

Just like the other shape parameters, the equatorial radius $\req$ can also be parametrized in terms of $\thetabo$.  Before continuing, we stress again that the presence of shocks near the equator (formed either by the jet's passage or the collision of cocoon material) may significantly alter the evolution of $\req$.  As compared to numerical simulations (see Section~\ref{numerical}), the equations presented here are only accurate to within a factor of a few.  However, the true equatorial radius cannot be less than the analytical estimate which ignores interactions.

Consider the ejecta that are just arriving to the equator at the time $x=x(t)$.  These ejecta must satisfy $\theta(x(\thetaeqo),\thetaeqo)=\pi/2$, where $\thetaeqo$ is their initial angle.  Similar to the way that the width of the bulge can be parametrized in terms of $\thetabo$ (see Appendix~\ref{appendixa}), the width at the equator can be parametrized in terms of $\thetaeqo$. Equation~\ref{theta1} then leads to the condition
\begin{equation}
\label{thetaeq}
x(\thetaeqo) = \dfrac{1}{k_\alpha} \left(\dfrac{\Reqo}{a}\right)^{k_\alpha} \dfrac{\sin (k_\alpha \Deltaeq)}{\sin(\omegaeqo-k_\alpha \Deltaeq)}, 
\end{equation}
where (in analogy with Appendix~\ref{appendixa}) we defined
\begin{equation}
\label{deltaeq}
\Deltaeq \equiv \pi/2-\thetaeqo
\end{equation}
and 
\begin{equation}
\label{omegaeq0}
\omegaeqo \equiv \chieqo-\thetaeqo,
\end{equation}
and $\Reqo$ and $\chieqo$ are now given by taking $\theta_i=\thetaeqo$ in equations~\ref{R_i} and~\ref{thetaN}.  

Depending on the density profile and how much time has elapsed since choking, there can be 0, 1, 2, or 3 values of $\thetaeqo$ which satisfy $x(t)=x(\thetaeqo)$ on the interval $0 < \thetaeqo < \pi/2$.  The number of solutions for $\thetaeqo$ corresponds to the number of places where the cocoon surface intersects the equator (not counting the point $\thetaeqo=\pi/2$).  If there are no solutions for $\thetaeqo$, the equatorial radius $\req$ is determined by the ejecta which started at $\theta_i=\pi/2$, and $\req$ is described by equation~\ref{r_z}.  Otherwise, the value of $\req$ is determined by the intersection point farthest from the axis, which is set by the smallest solution for $\thetaeqo$.  We then have $\req = R(x(\thetaeqo),\thetaeqo)$.  Inserting $x=x(\thetaeqo)$ into equation~\ref{R1} and simplifying, we obtain
\begin{equation}
\label{req}
\req(\thetaeqo) = \Reqo \left(\dfrac{\sin \omegaeqo}{\sin(\omegaeqo-k_\alpha \Deltaeq)} \right)^{1/k_\alpha}.
\end{equation}
To compare $\req$ to the values of $\zc$, $\rc$, and $\zb$ given in Appendix~\ref{appendixa}, it is necessary to determine the relationship between $\thetaeqo$ and $\thetabo$.  This is done by equating the two expressions for $x$, ~\ref{x0} and~\ref{thetaeq}, and solving numerically to obtain $\thetaeqo$ in terms of $\thetabo$, or vice versa.

Equation~\ref{thetaeq} is useful for determining if and when different parts of the surface reach the equator and undergo a collision.  In general, there are two ways that collisions can occur.  The first way involves material just adjacent to the equator that gets pushed down into the equatorial plane as the cocoon expands.  This material tends to impact the equator at a relatively shallow angle; therefore, we call collisions that happen in this way `glancing collisions.'    Another type of collision can also occur in steep density profiles, because the trajectories are highly curved, and due to the large density contrast, material at large radii travels considerably faster than material at small radii.  If $\alpha$ is sufficiently large, material originating near the tip of the cocoon wraps around along a curved trajectory and impacts the equator before material that started near the equator.  The first material to reach the equator in this way has a velocity perpendicular to the equator at the moment of impact.  We therefore call this a `head-on collision.'    As we will see, the relationship between $\alpha$ and the ratio $b/a$ governs whether or not collisions occur, and also which type of collisions is relevant.  In addition, the behaviour is different for $\alpha \le 2$ and $2 < \alpha < 3$.  We discuss the various possibilities in detail below.

Let us first consider the case $\alpha \le 2$. In this scenario, we find that there is no solution to equation~\ref{thetaeq} when $\alpha \le 2b^2/a^2$, indicating that no collisions occur in this case.  When $\alpha > 2b^2/a^2$, we find that glancing collisions start to occur at $x=x_{eq}$, where 
\begin{equation}
\label{xeq}
x_{eq} \equiv \lim_{\thetaeqo\rightarrow\pi/2} x(\thetaeqo) = \left(\dfrac{b}{a}\right)^{k_\alpha} \dfrac{2/\alpha}{1-2b^2/\alpha a^2}=(a^2/b^2)x_b
\end{equation}
for an initially ellipsoidal cocoon.  Just as $\rc$ has a transition at $x=x_b$ (as discussed in Appendix~\ref{appendixa}), $\req$ breaks to a different behaviour at $x=x_{eq}$.  While $x<x_{eq}$, there is no solution for $\thetaeqo$ and $\req$ is given by equation~\ref{r_z}.  Once $x>x_{eq}$, ejecta from $\theta_i  < \pi/2$ start to arrive at the equator, and there is one solution for $\thetaeqo$.  Thereafter, $\req$ is determined by equation~\ref{req}.  Head-on collisions never occur for $\alpha \le 2$ (see below).

When $\alpha>2$, the situation is more complicated, because both glancing and head-on collisions are possible.  To determine if and when a head-on collision occurs, we look for trajectories that intersect the equatorial plane at right angles.  Such trajectories satisfy $dR/d\theta = 0$ at the point $(R,\theta)=(\req,\pi/2)$.  From equation~\ref{traj1}, we find that ejecta originating from $\theta_i$ undergo a head-on collision if the condition
\begin{equation}
\label{theta0col}
\theta_i = \dfrac{2}{\alpha}\left[\chi_i-(1+k_\alpha)\pi/2\right],
\end{equation}
is satisfied.  
Solving for $\alpha$ yields
\begin{equation}
\label{alphacol0}
\alpha = \dfrac{2(\pi-\chi_i)}{\pi/2-\theta_i}.
\end{equation}
For an initially ellipsoidal shape, the right hand side of equation~\ref{alphacol0} has a minimum value. Therefore, equation~\ref{alphacol0} only has a solution if $\alpha \ge \alpha_{col,ho}$, where 
\begin{equation}
\label{alphacol}
\alpha_{col,ho} = \min\left(\dfrac{2[\pi-\chi_i]}{\pi/2-\theta_{i}}\right) \approx 2\left(1+\dfrac{4b}{\pi a}\right),
\end{equation}
to leading order in $b/a$.  If $\alpha < \alpha_{col,ho}$, equation~\ref{theta0col} does not have a solution, and no head-on collision occurs.  (Because $\alpha_{col,ho}$ is always greater than 2, this also implies that head-on collisions are impossible for $\alpha <2$.)  

In the $\alpha \ge \alpha_{col,ho}$ case, equation~\ref{theta0col} admits two solutions.  The relevant one for determining $\req$ is approximated by
\begin{equation}
\label{thetacol1}
\theta_i \approx (b^2/a^2)\tan[\pi(1-\alpha/4)]
\end{equation}
for $b/a \ll 1$.  Eq.~\ref{thetacol1} pertains to the path taken by the first ejecta to experience a head-on collision (e.g., the dot-dashed line in Fig.~\ref{web3}).  If we set $\theta_i = \thetaeqo$, equation~\ref{theta0col} implies $\omegaeqo \approx \pi(1-\alpha/4)$, $\Deltaeq \approx \pi/2$, and $\Reqo \approx a$.  Inserting these values into equations~\ref{thetaeq} and~\ref{req},  we find that the first head-on collision occurs at a time
\begin{equation}
\label{xcol}
x_{col} \approx \dfrac{1}{k_\alpha} \sin(k_\alpha\pi/2)
\end{equation}
and a radius 
\begin{equation}
\label{rcol}
r_{col} \approx a \left(\sin[(1+k_\alpha)\pi/2]\right)^{1/k_\alpha}.
\end{equation}

The evolution of $\req$ in the $2< \alpha < 3$ case depends on how $\alpha$ compares with $\alpha_{col,ho}$.  If $2 < \alpha < \alpha_{col,ho}$, material just adjacent to the equator starts to undergo collisions at $x=x_{eq}$, and from then on there are continuous glancing collisions, as in the case $2b^2/a^2 < \alpha \le 2$ discussed above.  In this case, the equatorial radius $\req$ grows continuously with $x$.  However, if $\alpha_{col,ho} \le \alpha < 3$, there is a head-on collision that occurs at $x=x_{col}$. The head-on collision causes the equatorial radius to suddenly increase to $\approx r_{col}$, so that $\req$ has a discontinuity at $x_{col}$.  The value of $\req$ increases smoothly up for $x<x_{col}$, then jumps to a higher value at $x=x_{col}$, and afterwards continues to increase gradually.

\begin{figure} 
\centering
\includegraphics[width=.5\textwidth]{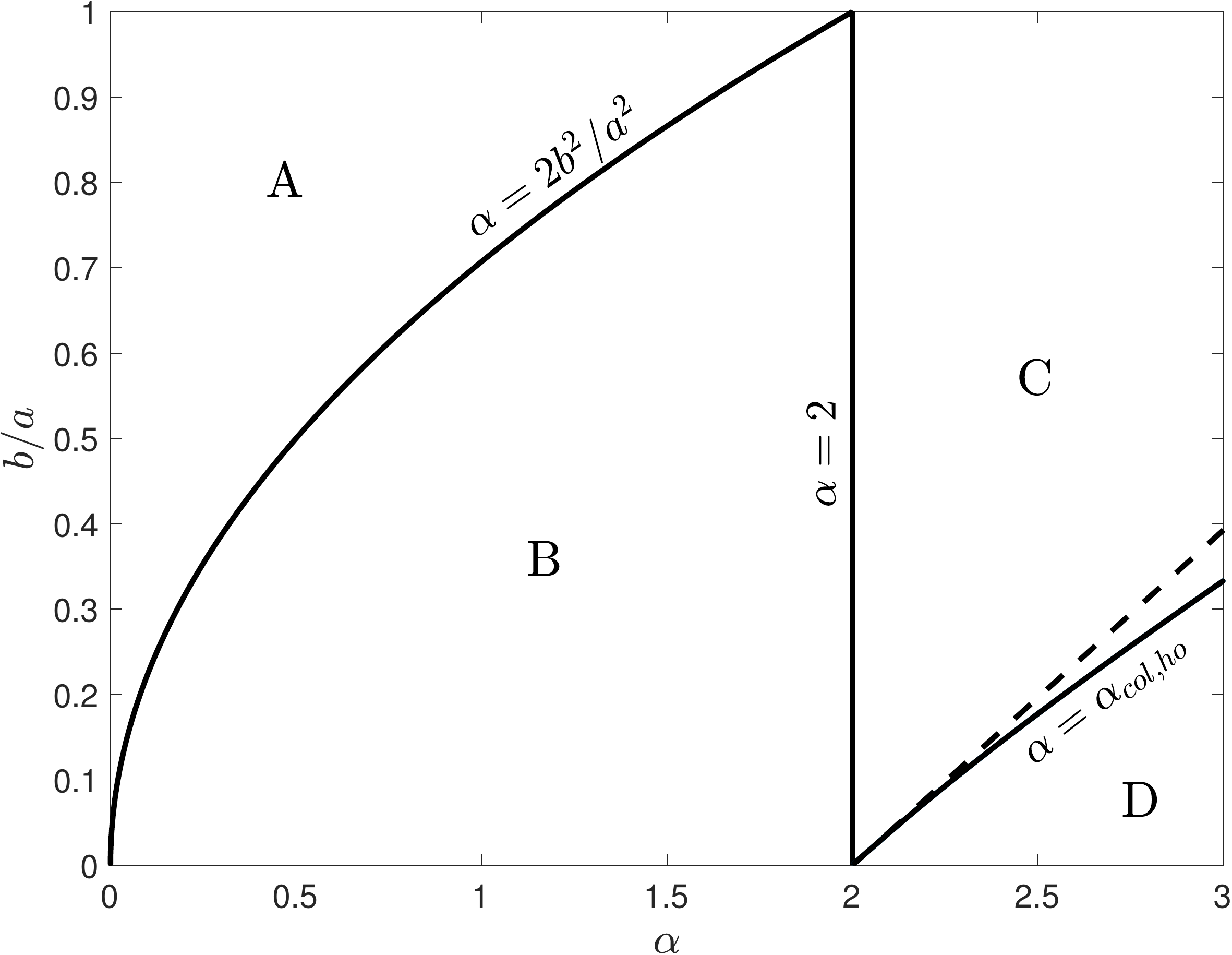}
\caption{The critical values of $\alpha=2b^2/a^2$ and $\alpha=\alpha_{col,ho}$ versus $b/a$ for an initially ellipsoidal cocoon.  The dashed line shows the approximation for $\alpha_{col,ho}$ for $b/a \ll 1$, as given by equation~\ref{alphacol}.  The regions labeled A through D are discussed in the text.}
\label{collisions}
\end{figure}

The results of this section are summarized by Fig.~\ref{collisions}, where we plot the critical values of $\alpha$ as functions of $b/a$.  The phase space of $\alpha$ versus $b/a$ can be divided into the four regions labeled A through D in the figure, each with a distinct behaviour:
\begin{itemize}
\item A ($\alpha \le 2b^2/a^2$): There are no collisions.  
\item B ($2b^2/a^2 < \alpha \le 2$): The part of the surface with $\theta_i > \thetabomin$ eventually experiences glancing collisions.  The rest of the surface never reaches the equator.
\item C ($2 < \alpha < \alpha_{col,ho}$): All the ejecta ultimately undergo a glancing collision.
\item D ($\alpha_{col,ho} \le \alpha < 3$):  All the ejecta eventually experience a collision.  For material starting out near the equator, the collision is glancing; for material originating near the axis, the collision is head-on.
\end{itemize}

\begin{thebibliography}{99}
\bibitem[Abbott et al.(2017)]{abbott17} 
	Abbott, B.~P., Abbott, R., Abbott, T.~D., et al.\ 2017, Physical Review Letters, 119, 161101 
\bibitem[Aloy et al.(2000)]{aloy00} 
	Aloy, M.~A., M{\"u}ller, E., Ib{\'a}{\~n}ez, J.~M., Mart{\'{\i}}, J.~M., \& MacFadyen, A.\ 2000, \apjl, 531, L119 
\bibitem[Ayal \& Piran(2001)]{ap01} 
	Ayal, S., \& Piran, T.\ 2001, \apj, 555, 23 
\bibitem[Bannikova et al.(2012)]{bannikova12} 
	Bannikova, E.~Y., Karnaushenko, A.~V., Kontorovich, V.~M., \& Shulga, V.~M.\ 2012, Astronomy Reports, 56, 496 
\bibitem[Begelman \& Cioffi(1989)]{bc89} 
	Begelman, M.~C., \& Cioffi, D.~F.\ 1989, \apjl, 345, L21 
\bibitem[B{\^i}rzan et al.(2004)]{birzan04} 
	B{\^i}rzan, L., Rafferty, D.~A., McNamara, B.~R., Wise, M.~W., \& Nulsen, P.~E.~J.\ 2004, \apj, 607, 800 
\bibitem[Bisnovatyj-Kogan \& Blinnikov(1982)]{bb82} 
	Bisnovatyj-Kogan, G.~S., \& Blinnikov, S.~I.\ 1982, \azh, 59, 876 
\bibitem[Bisnovatyi-Kogan \& Silich(1995)]{bs95} 
	Bisnovatyi-Kogan, G.~S., \& Silich, S.~A.\ 1995, Reviews of Modern Physics, 67, 661 
\bibitem[Blandford et al.(2018)]{blandford19} 
	Blandford, R., Meier, D., \& Readhead, A.\ 2018, arXiv:1812.06025 
\bibitem[Bromberg et al.(2011)]{b11} 
	Bromberg, O., Nakar, E., Piran, T., \& Sari, R.\ 2011, \apj, 740, 100 
\bibitem[Bromberg et al.(2012)]{bromberg12} 
	Bromberg, O., Nakar, E., Piran, T., \& Sari, R.\ 2012, \apj, 749, 110 
\bibitem[Campana et al.(2006)]{campana06} 
	Campana, S., Mangano, V., Blustin, A.~J., et al.\ 2006, \nat, 442, 1008 
\bibitem[Churazov et al.(2000)]{churazov00} 
	Churazov, E., Forman, W., Jones, C., \& B{\"o}hringer, H.\ 2000, \aap, 356, 788 
\bibitem[Churazov et al.(2001)]{churazov01} 
	Churazov, E., Br{\"u}ggen, M., Kaiser, C.~R., B{\"o}hringer, H., \& Forman, W.\ 2001, \apj, 554, 261 
\bibitem[Diehl et al.(2008)]{diehl08} 
	Diehl, S., Li, H., Fryer, C.~L., \& Rafferty, D.\ 2008, \apj, 687, 173 
\bibitem[Fabian et al.(2002)]{fabian02} 
	Fabian, A.~C., Celotti, A., Blundell, K.~M., Kassim, N.~E., \& Perley, R.~A.\ 2002, \mnras, 331, 369 
\bibitem[Gottlieb et al.(2018)]{gottlieb18} 
	Gottlieb, O., Nakar, E., \& Piran, T.\ 2018, \mnras, 473, 576 
\bibitem[Harrison et al.(2018)]{harrison18} 
	Harrison, R., Gottlieb, O., \& Nakar, E.\ 2018, \mnras, 477, 2128 
\bibitem[Heger et al.(2005)]{heger05} 
	Heger, A., Woosley, S.~E., \& Spruit, H.~C.\ 2005, \apj, 626, 350 
\bibitem[Irwin \& Chevalier(2016)]{ic16} 
	Irwin, C.~M., \& Chevalier, R.~A.\ 2016, \mnras, 460, 1680 
\bibitem[Ito et al.(2015)]{ito15} 
	Ito, H., Matsumoto, J., Nagataki, S., et al.\ 2015, \apj, 814, L29.
\bibitem[Izzo et al.(2019)]{izzo19} 
	Izzo, L., de Ugarte Postigo, A., Maeda, K., et al.\ 2019, \nat, 565, 324 
\bibitem[Kompaneets(1960)]{kompaneets60}
	Kompaneets, A. 1960, Soviet Phys. Doklady, 130, 46
\bibitem[Koo \& McKee(1990)]{koo90} 
	Koo, B.-C., \& McKee, C.~F.\ 1990, \apj, 354, 513 
\bibitem[Korycansky(1992)]{k92}
	Korycansky, D.~G. 1992, \apj, 398, 184
\bibitem[Kulkarni et al.(1998)]{kulkarni98} 
	Kulkarni, S.~R., Frail, D.~A., Wieringa, M.~H., et al.\ 1998, \nat, 395, 663 
\bibitem[Lazzati \& Begelman(2005)]{lb05} 
	Lazzati, D., \& Begelman, M.~C.\ 2005, \apj, 629, 903 
\bibitem[Lazzati et al.(2009)]{lmb09} 
	Lazzati, D., Morsony, B.~J., \& Begelman, M.~C.\ 2009, \apj, 700, L47
\bibitem[Lazzati et al.(2012)]{lazzati12} 
	Lazzati, D., Morsony, B.~J., Blackwell, C.~H., \& Begelman, M.~C.\ 2012, \apj, 750, 68 
\bibitem[L{\'o}pez-C{\'a}mara et al.(2013)]{lopez13} 
	L{\'o}pez-C{\'a}mara, D., Morsony, B.~J., Begelman, M.~C., \& Lazzati, D.\ 2013, \apj, 767, 19 
\bibitem[Lyutikov(2011)]{lyutikov11} 
	Lyutikov, M.\ 2011, \mnras, 411, 2054 
\bibitem[MacFadyen et al.(2001)]{mwh01} 
	MacFadyen, A.~I., Woosley, S.~E., \& Heger, A.\ 2001, \apj, 550, 410 
\bibitem[Mart{\'{\i}} et al.(1997)]{marti97} 
	Mart{\'{\i}}, J.~M., M{\"u}ller, E., Font, J.~A., Ib{\'a}{\~n}ez, J.~M.~Z., \& Marquina, A.\ 1997, \apj, 479, 151 
\bibitem[Matzner(2003)]{matzner03} 
	Matzner, C.~D.\ 2003, \mnras, 345, 575 
\bibitem[Matzner et al.(2013)]{matzner13} 
	Matzner, C.~D., Levin, Y., \& Ro, S.\ 2013, \apj, 779, 60
\bibitem[McNamara et al.(2000)]{mcnamara00} 
	McNamara, B.~R., Wise, M., Nulsen, P.~E.~J., et al.\ 2000, \apjl, 534, L135 
\bibitem[McNamara et al.(2005)]{mcnamara05} 
	McNamara, B.~R., Nulsen, P.~E.~J., Wise, M.~W., et al.\ 2005, \nat, 433, 45 
\bibitem[M{\'e}sz{\'a}ros \& Waxman(2001)]{mw01} 
	M{\'e}sz{\'a}ros, P., \& Waxman, E.\ 2001, Physical Review Letters, 87, 171102 
\bibitem[Mignone et al.(2007)]{mignone07} 
	Mignone, A., Bodo, G., Massaglia, S., et al.\ 2007, \apjs, 170, 228 
\bibitem[Mizuta et al.(2006)]{mizuta06} 
	Mizuta, A., Yamasaki, T., Nagataki, S., \& Mineshige, S.\ 2006, \apj, 651, 960 
\bibitem[Mizuta \& Aloy(2009)]{mizuta09} 
	Mizuta, A., \& Aloy, M.~A.\ 2009, \apj, 699, 1261 
\bibitem[Mizuta \& Ioka(2013)]{mizuta13} 
	Mizuta, A., \& Ioka, K.\ 2013, \apj, 777, 162 
\bibitem[Moharana \& Piran(2017)]{moharana17} 
	Moharana, R., \& Piran, T.\ 2017, \mnras, 472, L55  
\bibitem[Morsony et al.(2007)]{mlb07} 
	Morsony, B.~J., Lazzati, D., \& Begelman, M.~C.\ 2007, \apj, 665, 569.
\bibitem[Nagakura et al.(2011)]{nagakura11} 
	Nagakura, H., Ito, H., Kiuchi, K., \& Yamada, S.\ 2011, \apj, 731, 80 
\bibitem[Nakar(2015)]{nakar15} 
	Nakar, E.\ 2015, \apj, 807, 172 
\bibitem[Nakar \& Piran(2017)]{np17} 
	Nakar, E., \& Piran, T.\ 2017, \apj, 834, 28 
\bibitem[Nakar \& Sari(2012)]{ns12} 
	Nakar, E., \& Sari, R.\ 2012, \apj, 747, 88 
\bibitem[Piran et al.(2019)]{piran19} 
	Piran, T., Nakar, E., Mazzali, P., \& Pian, E.\ 2019, \apjl, 871, L25 
\bibitem[Rees(1984)]{rees84} 
	Rees, M.~J.\ 1984, \araa, 22, 471
\bibitem[Rimoldi et al.(2015)]{rimoldi15} 
	Rimoldi, A., Rossi, E.~M., Piran, T., \& Portegies Zwart, S.\ 2015, \mnras, 447, 3096 
\bibitem[Sedov(1959)]{sedov59} 
	Sedov, L.~I.\ 1959, Similarity and Dimensional Methods in Mechanics, New York: Academic Press, 1959, 
\bibitem[Soderberg et al.(2006)]{soderberg06} 
	Soderberg, A.~M., Kulkarni, S.~R., Nakar, E., et al.\ 2006, \nat, 442, 1014 
\bibitem[Tang \& Churazov(2017)]{tc17} 
	Tang, X., \& Churazov, E.\ 2017, \mnras, 468, 3516 
\bibitem[Taylor(1950)]{taylor50} 
	Taylor, G.\ 1950, Proceedings of the Royal Society of London Series A, 201, 159 
\bibitem[Wang et al.(2008)]{waz08} 
	Wang, P., Abel, T., \& Zhang, W.\ 2008, \apjs, 176, 467.
\bibitem[Waxman \& Shvarts(1993)]{waxman93}
	Waxman, E., \& Shvarts, D. 1993, Physics of Fluids A: Fluid Dynamics, 5, 1035
\bibitem[Woosley \& Bloom(2006)]{wb06} 
	Woosley, S.~E., \& Bloom, J.~S.\ 2006, \araa, 44, 507 
\bibitem[Zhang et al.(2004)]{zwh04} 
	Zhang, W., Woosley, S.~E., \& Heger, A.\ 2004, \apj, 608, 365 
\end{thebibliography}
\end{document}